\documentclass[fleqn,usenatbib]{mnras}
\usepackage{deluxetable}
\usepackage{graphicx}
\usepackage{color,soul}
\usepackage{newtxtext,newtxmath}
\usepackage[T1]{fontenc}
\usepackage{ae,aecompl}

\begin{document}
\label{firstpage}
\pagerange{\pageref{firstpage}--\pageref{lastpage}}

\title[EDGES: Description and S.B. properties]{The Extended Disk Galaxy Exploration Science Survey: Description and Surface Brightness Profile Properties}
\author[Staudaher et al.]{
Shawn M. Staudaher,$^{1,2}$\thanks{E-mail: sstaudaher@gatech.edu}
Daniel A. Dale,$^{1}$\thanks{E-mail: ddale@uwyo.edu}
Liese van Zee$^{3}$
\\
$^{1}$Physics and Astronomy Department, University of Wyoming, WY 82071\\
$^{2}$Center for 21$^{\rm st}$ Century Universities, The Georgia Institute of Technology, GA 30332\\
$^{3}$Astronomy Department, Indiana University, IN 47405
}

% These dates will be filled out by the publisher
\date{Accepted 2019 March 27. Received 2019 March 27; in original form 2018 September 25}

% Enter the current year, for the copyright statements etc.
\pubyear{2019}

\maketitle

\def\VegaSB{${\rm \mu~(Vega~mag/\square \arcsec)}$}
\def\SB{${\rm \mu~(mag/\square \arcsec)}$}
\def\um{${\rm \mu}$m}

\begin{abstract}
The survey description and near-infrared properties for 92 galaxies are presented for the Extended Disc Galaxy Exploration Science (EDGES) Survey, along with an investigation into the properties of the stellar halos of these galaxies. EDGES is a {\it Spitzer} Space Telescope Warm Mission program designed to reach the surface brightness limit ($\sim$0.5~kJy~sr$^{-1}$ or 29~AB~mag~arcsec$^{-2}$) of the Infrared Array Camera (IRAC) 3.6 and 4.5~\um\ bands for a wide range of galaxy types found within the local volume. The surface brightness profiles exhibit a large range in disc scalelength, with breaks more frequently seen than in previous studies, owing in large part to the extremely deep near-infrared imaging. A number of these surface brightness profile breaks may be due to stellar halos, up to 7 galaxies out of the full sample of 92 galaxies, and we explore these implications in relation to current cosmological models. We also report the discovery of a new tidal stream near NGC~3953.
\end{abstract}


\section{INTRODUCTION}

A significant hurdle of $\Lambda$CDM-based cosmological simulations has been the so-called ``missing satellite problem''. Simply put, early $\Lambda$CDM-based simulations \citep{Kauffmann+93, Navarro+96} found an order of magnitude more satellite galaxies than observed orbiting the Milky Way \citep{Klypin+99, Moore+99, Bullock+00}. The degree of this problem has since diminished due to a number of new findings which include: models with more detailed physics \citep[e.g.][]{Guo+11, Wetzel+16}, a more complete understanding of the satellite mass function \citep[e.g.][]{Bullock+10, Guo+15}, and the finding that observed satellites are less massive than they were originally due to tidal stripping \citep{Kravtsov+04}. Also, more low-mass galaxies have recently been discovered near the Milky Way \citep[e.g.][]{Belokurov+06, Zucker+06, Martin+07, Koposov+15}, the Andromeda galaxy \citep[e.g.][]{Zucker+07, Irwin+08}, and beyond \citep[e.g.][]{Javanmardi+16, Munoz+15} since the ``missing satellite problem'' was originally posited.

An alternative test of $\Lambda$CDM-based simulations is the comparison of measured to predicted stellar halo masses. Stellar halos are diffuse spherical distributions of stars which are thought to surround most galaxies larger than dwarfs. These halos are predominantly made up of stars which have been stripped via tidal forces from satellite galaxies (which are often destroyed in this process) and deposited onto a larger host galaxy. The number and size of satellite galaxies available to be tidally disrupted is determined by cosmological parameters. It is therefore possible to use cosmological simulations to determine the predicted distribution of stellar halo masses, and then test these predictions with observations.

The relation between stellar halo mass and total stellar mass has been simulated by a few teams thus far \citep{Johnston+08, Purcell+11, Cooper+13, Rodriguez-Gomez+15,Rodriguez-Gomez+16, Elias+18}, who find that the stellar halo mass fraction increases with total stellar mass, but with a high amount of scatter. This scatter is likely due to the finding that the majority of stellar halo mass is accreted from a single, large, satellite \citep{Bullock+05, DSouza+Bell18}. Observational confirmation of these results has proven to be difficult. Stellar halos are amongst the dimmest stellar components of a galaxy at 28~AB~mag~arcsec$^{-2}$ \citep{Cooper+13}, where slight deviations from perfect flat-fielding of images, the extended wings of point spread functions (PSFs) \citep{Sandin14, Sandin15}, and the time required to reach these depths, hinders detections. Nevertheless, several stellar halo masses have been measured in individual galaxies. These include the Milky Way \citep{Carollo+10},  NGC253 \citep{Bailin+11}, M31 \citep{Courteau+11}, M101 \citep{vanDokkum+14}, NGC3115 \citep{Peacock+15}, M63 \citep{Staudaher+15}, and UGC00180 \citep{Trujillo+16}. Stellar halos have also been found within small targeted surveys including GHOSTS \citep{Harmsen+17} and Dragonfly \citep{Merritt+16}. The consensus from these results is that the simulations find more massive stellar halos than observed. However, the methodologies used to calculate stellar halo masses vary greatly: from counting stars in the Milky Way \citep{Carollo+10}, to optical colors \citep{vanDokkum+14, Trujillo+16}, to near-infrared {\it Spitzer Space Telescope} 3.6~\um\ imaging \citep{Courteau+11, Staudaher+15}, to {\it Hubble Space Telescope} color-magnitude diagrams which resolve red giant stars \citep{Bailin+11, Peacock+15, Harmsen+17}. A large sample size is also important due to the aformentioned stochastic accretion process of stellar halo populations. To observationally test the predicted halo mass relation requires a large robust sample with consistent measurement techniques.

\begin{figure*}
    \includegraphics[width=140mm]{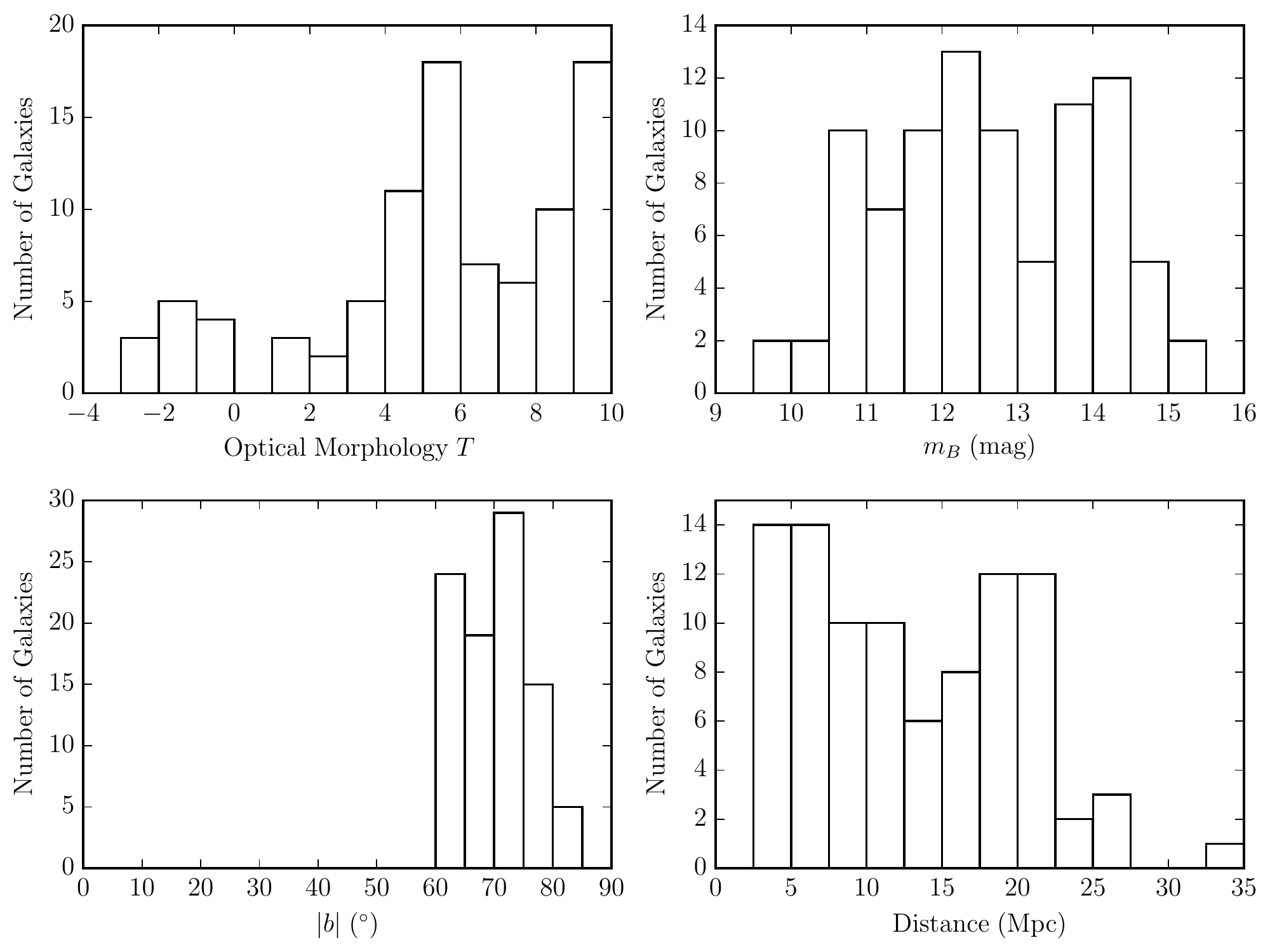}
    \caption{The distribution of optical morphology (top left), apparent $B$-band magnitude (top right), Galactic latitude, $|b|$ (bottom left), and distance (bottom right).}
    \label{fig:sample_hist}
\end{figure*}

The first large scale observational examination \citep{DSouza+14} of stellar halos was done by grouping $>$45,000 Sloan Digital Sky Survey (SDSS) galaxies by mass bins, and then stacking every image in a given bin to produce an extremely deep image ($\mu_r\sim32$ mag~arcsec$^{-2}$) of an average galaxy. They found that the average stellar halo for galaxies from 10$^{10}$-10$^{11.4}$~M$_{\odot}$ (split into 6 mass bins) are in general agreement with simulations. The Dragonfly group uses a robotic array of refractive telescopes which helps to minimize scattered light and thus leads to improved flat-fielding \citep{Abraham+14}.  They are in the process of conducting a survey of stellar halos \citep{Merritt+16} and have successfully measured the stellar halo mass fraction for five galaxies, with three stellar halos remaining undetected down to $\mu_g\sim31$~mag~arcsec$^{-2}$. The Dragonfly results systematically find stellar halo mass fractions lower than predicted by simulations. A promising survey, MADCASH (Magellanic Analog Dwarf Companions And Stellar Halos) has discovered a faint dwarf galaxy \citep{Carlin+16} and will examine resolved stars within stellar halos for several nearby LMC analogs. The first large scale survey of individually resolved stellar halos was completed with data from the Hyper Suprime-Cam-based Subaru Strategic Program \citep{Aihara+17} in \cite{Huang+17}. They studied $\sim$3,000 galaxies between $0.3~<~z~<0.5$ using high quality data that reaches $i~>~28.5$ mag~arcsec$^{-2}$ allowing measurements of stellar halos to 100~kpc. They find an astounding agreement with the Illustris-based simulations of \cite{Rodriguez-Gomez+16}. The majority of their galaxies fit within the 1-$\sigma$ envelope of the predicted relation between the fraction of stellar halo mass and total stellar mass from $11-12$ log($M_{\star}/M_{\odot}$).

We pursue an observational approach to quantifying the stellar halos in nearby galaxies, one that is based on extremely deep and spatially wide near-infrared imaging.  The Extended Disc Galaxy Exploration Science (EDGES) program is a Warm Spitzer survey of 92 galaxies designed to measure the dimmest stellar structures in nearby galaxies. In this paper we first discuss the sample selection and explain our general observational strategy. We detail the processing techniques which generate our mosaics and surface brightness profiles. We then show how the surface brightness profiles are decomposed into individual components and how the mass of these components is measured. From these measurements the statistics of up-bending, and down-bending breaks are examined, and these findings are compared to predictions from $\Lambda$CDM-based simulations.

\section{SAMPLE}
\label{sec:sample}

EDGES consists of 92 nearby galaxies spanning a wide range of physical properties and possible formation histories (as shown in \cite{Dale+16} and Dale (2019, in prep.)). Due to the exploratory nature of the project, several galaxies of each morphological type, luminosity, and inclination angle are included in order to explore the relationships between the extended stellar distribution and global properties of galaxies in general. The selected galaxies are drawn from a parent sample of all galaxies listed in the NASA/IPAC Extragalactic Database (NED) with velocities less than 3000 km~s$^{-1}$ and $|b|~>20^{\circ}$. The primary selection criteria include both a minimum and maximum distance cut, a minimum and maximum angular size cut ($2\arcmin~<D_{25}~<13\arcmin)$, an apparent magnitude limit ($m_B~<~16$), and a very strict Galactic latitude constraint ($|b|~>~60^{\circ}$). The latter criterion is imposed based on our experience with data from our pilot projects in Cycle 6 (PIDs 60094 and 60116, see \mbox{\cite{Barnes+14}}), where even moderate galactic latitude fields have significant contamination by foreground stars at the depths of these images. In addition, to minimise confusion with background structure, we exclude all potential targets within 20$^{\circ}$ of the center of the Virgo cluster. Note that the combination of the distance cut and apparent magnitude limit results in a nearly complete parent sample for galaxies brighter than $M_B$ of $-15$; thus, in order to maintain a nearly complete parent sample while still including a representative sample of low mass galaxies, we further limit the sample to those galaxies brighter than $M_B$ of $-14$. This final selection criterion ensures that the majority of galaxies in this sample are representative of systems that encompass the majority of baryonic mass in this volume element.

The volume element for this survey was selected to optimise our ability to identify and trace extended stellar components within the observational constraints of Spitzer and other telescopes. Specifically, we excluded galaxies within 2~Mpc of the Milky Way since both the primary target and any associated stellar streams are too large to map efficiently with IRAC, with angular sizes up to tens of degrees. We also imposed a maximum redshift constraint ($v < 1100$ km s$^{-1}$) to facilitate targeted follow-up observations of stellar features identified in the survey; this outer distance limit enables ancillary observations designed to investigate the resolved stellar populations of selected small fields to further explore their origin and structure. The redshift limit allows such observations to probe the tip of the red giant branch of these targeted fields with reasonable integration times using existing instrumentation available on ground-based telescopes and the Hubble Space Telescope.

Minimum and maximum angular size constraints were imposed to best match our observational mapping strategy to the areal coverage required to trace the extended stellar population ($5\times R_{25}$). We excluded small angular extent targets from the sample in order to best match the minimum primary field-of-view (FOV) from our mapping strategy ($10 \arcmin \times 10 \arcmin$ for an $8 \times 8$ grid) with the known optical extent. While this criterion preferentially excluded low mass galaxies from the final sample, our science goals only require a representative sample, with well defined selection criteria, which can then be compared to a sample selected in a similar manner from $\Lambda$CDM simulations; thus, a slight over-representation of massive galaxies (relative to the parent sample) is acceptable. The maximum size limit was imposed so that the maximum FOV to be mapped could be accomplished within a single Spitzer Astronomical Observing Request (AOR). While it was nominally possible to split observations into multiple AORs to map the large FOVs required for the 7 extremely large nearby galaxies in this survey volume, the potential systematic errors associated with matching sky levels for data taken at different times would have posed significant challenges to meeting our science goals. 

The above constraints resulted in a potential sample of 122 galaxies. We examined each field visually to eliminate galaxies that would require masking a significant fraction of the FOV due to either numerous conspicuous foreground or background source contamination (Milky Way stars, galaxies, and galaxy clusters). We also examined each potential field with the Spitzer planning software package SPOT to verify that each target has low infrared sky levels at both 3.6~\um\ and 4.5~\um. As expected given the location of these galaxies, the typical infrared sky levels for this sample are less than 0.1~MJy~sr$^{-1}$ at 3.6~\um.

The final galaxy sample is a well defined and statistically representative sample of normal galaxies within this volume (see Figure~\ref{fig:sample_hist} for the general properties of this sample and Table~\ref{table:general} for the properties of individual galaxies). Our deep observations of a large FOV (at least $5~R_{25}$) around this representative sample of 92 galaxies provides an unprecedented view of the faint stellar populations associated with nearby galaxies.

\section{OBSERVATIONAL STRATEGY AND DATA PROCESSING}

\subsection{Mapping strategy}

Our observations are designed to map a field of view that corresponds to at least a factor of 5 beyond a galaxy's bright optical component. The need for wide field mapping observations is driven by the estimated sizes of dark matter halos relative to their bright baryonic components and by the observed extent of the largest gaseous discs currently known. Kinematic tracers indicate that the dark matter halos extend at least a factor of 5 beyond the high surface brightness component for both early- \citep[e.g][]{Rhode+07} and late-type galaxies \citep[e.g.][]{Zaritsky+93, Christlein+Zaritsky08}. Further, extremely large atomic gas discs (HI-to-optical size ratios of 7-10) have been identified in several low mass dwarf irregular galaxies \citep[e.g. NGC~3741][]{Begum+08}, indicating that the baryonic component may extend well beyond the high surface brightness regime even in isolated galaxies. Finally, \cite{Regan+06} find smoothly varying stellar surface brightnesses at 3.6~\um\ out to at least 2~$R_{25}$, the limit of the SINGS and S$^4$G data \citep{Kennicutt+03, Sheth+10}, indicating that deeper observations would reveal a more extended stellar component in most spiral galaxies. Thus, while our observations are exploratory in nature, our observing strategy samples the faint extended stellar distributions for every galaxy.

Astronomical Observing Requests (AORs) were constructed using the successful strategy employed for our Cycle~6 pilot studies (see \cite{Barnes+14}). Mosaics are built upon a grid of 100$\arcsec$ spacings ($\sim$one-third the IRAC FOV). Two sets of maps are obtained for each source to enable asteroid removal and to enhance map sensitivity and redundancy. At any given location within the map cores there are a total of 18 100~s frames resulting in a net integration per sky position of 1800~s (along with a 1200~s, 100$\arcsec$-wide ``inner periphery'' and a 600~s, 100$\arcsec$-wide ``outer periphery''). The mosaics for 3.6~\um\ observations are centered on the target galaxies, but even our smallest maps have sufficient sky coverage that the FOV of the corresponding 4.5~\um\ mosaic includes the galaxy as well. The smallest maps are $8 \times 8$ mosaics, providing 10$\arcmin$ map cores at the deepest 1800~s effective integration. For highly inclined galaxies, the mosaic pattern was modified to generate long rectangular strips for 31 fields (such restrictions were not required for the face-on galaxies for which there is no preferred direction for the map).

\subsection{Pre-processing and mosaicking}

The EDGES observations took place from 2011-06-20 to 2013-03-13 and are currently freely available at the Spitzer Heritage Archive under PID:80025, and we plan to make the fully processed mosaics available via NED. The raw data of EDGES are a massive trove of 33,434 individual pointings totaling over a month of exposure time (1005.3~hours). In order to analyse these data we have created mosaics using the MOPEX pipeline with the BCD images including standard pre- and post-processing fixes for Warm Mission data. The pre-processing steps include the removal of a frame-wide bias caused by ``stuck'' pixels, the so-called ``column-pulldown effect'', the ``first-frame effect'', and a dependence of the bias on frame number. A more detailed description of this process is found in \cite{Staudaher+15}, and see also \cite{Krick+11} for an analysis of the effects removed in pre-processing.

\subsection{Post-processing}

Post-processing of the mosaics begins with the removal of the 3.6~\um\ sky by subtracting a fitted plane gradient to a suite of individually defined rectangular sky regions with the IDL function {\tt MPFITFUN}. The subset of galaxies with saturated pixels at their centers are corrected with shallower archival data. Archival mosaics are generated from BCD-level data using our pipeline with our pixel mapping solution (the Fiducial Image Frame), and saturated pixels in EDGES mosaics are replaced with the unsaturated values from the newly generated archival mosaics. Also, as a further check, the pixel values within the archival mosaics are compared to the pixel values in the EDGES mosaics.

Foreground stars and background galaxies are the largest contaminants in the EDGES mosaics due to the extremely low background noise level of $\sim2.5$~kJy~sr$^{-1}$. To remove these contaminants a SExtractor \citep{Bertin+96} catalog of sources is generated for every mosaic. The SExtractor algorithm correctly identifies stars and background galaxies beyond the extent of EDGES galaxies. However, within the diffuse emission of our targets SExtractor overestimates the extent of point sources and misidentifies spiral structure as independent galaxies. To remedy these issues the extent of point sources within extended structure are adjusted by hand and misidentified spiral structures are removed. Once the catalogs have been properly edited they are used to generate masks for each mosaic.

\begin{figure}
    \includegraphics[width=\columnwidth]{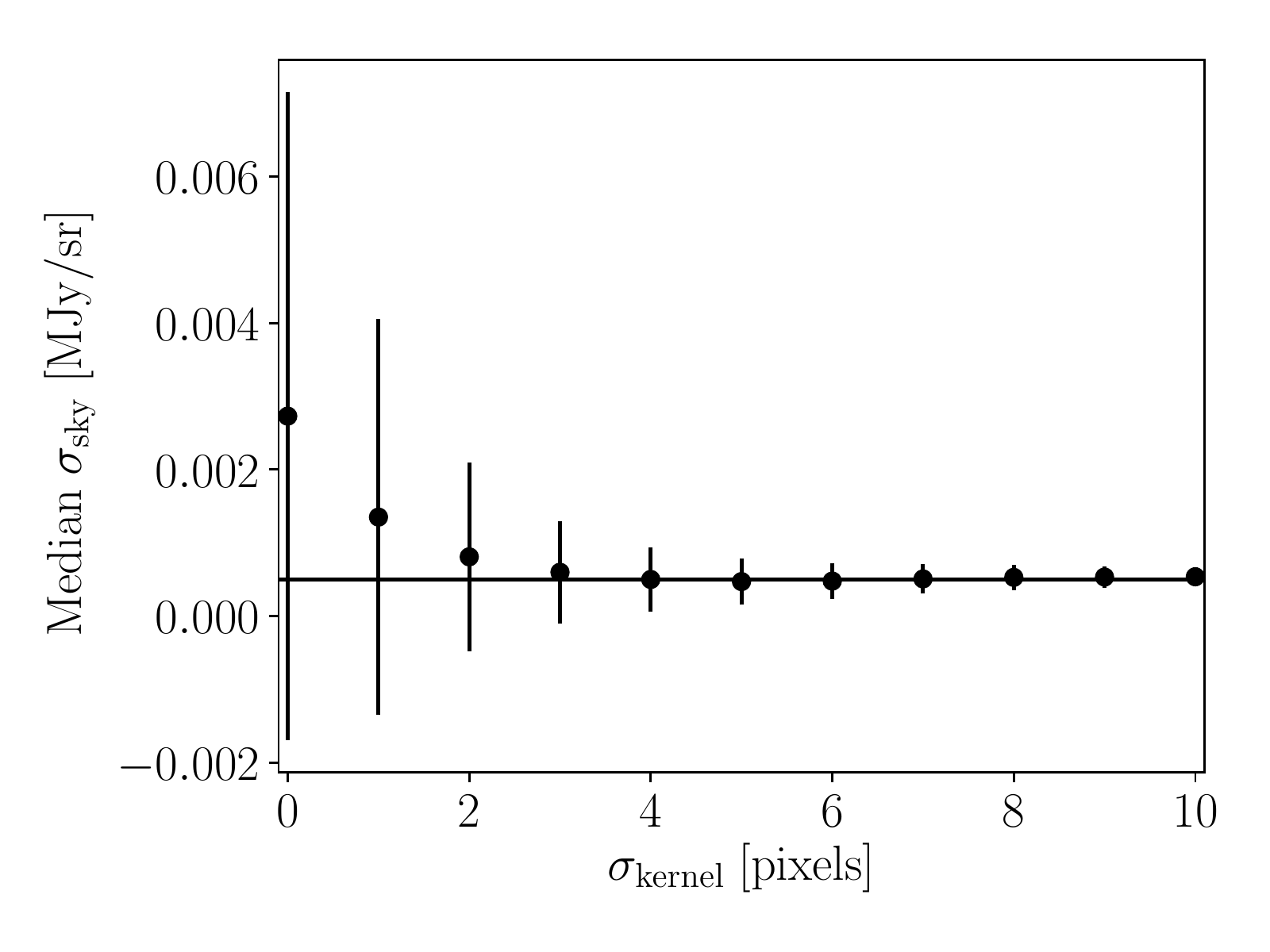}
    \caption{The dependence of the standard deviation within sky apertures on smoothing kernel size.}
    \label{fig:smooth_test}
\end{figure}

\begin{figure*}
    \centering
    \begin{tabular}{cc}
        \includegraphics[width=75mm]{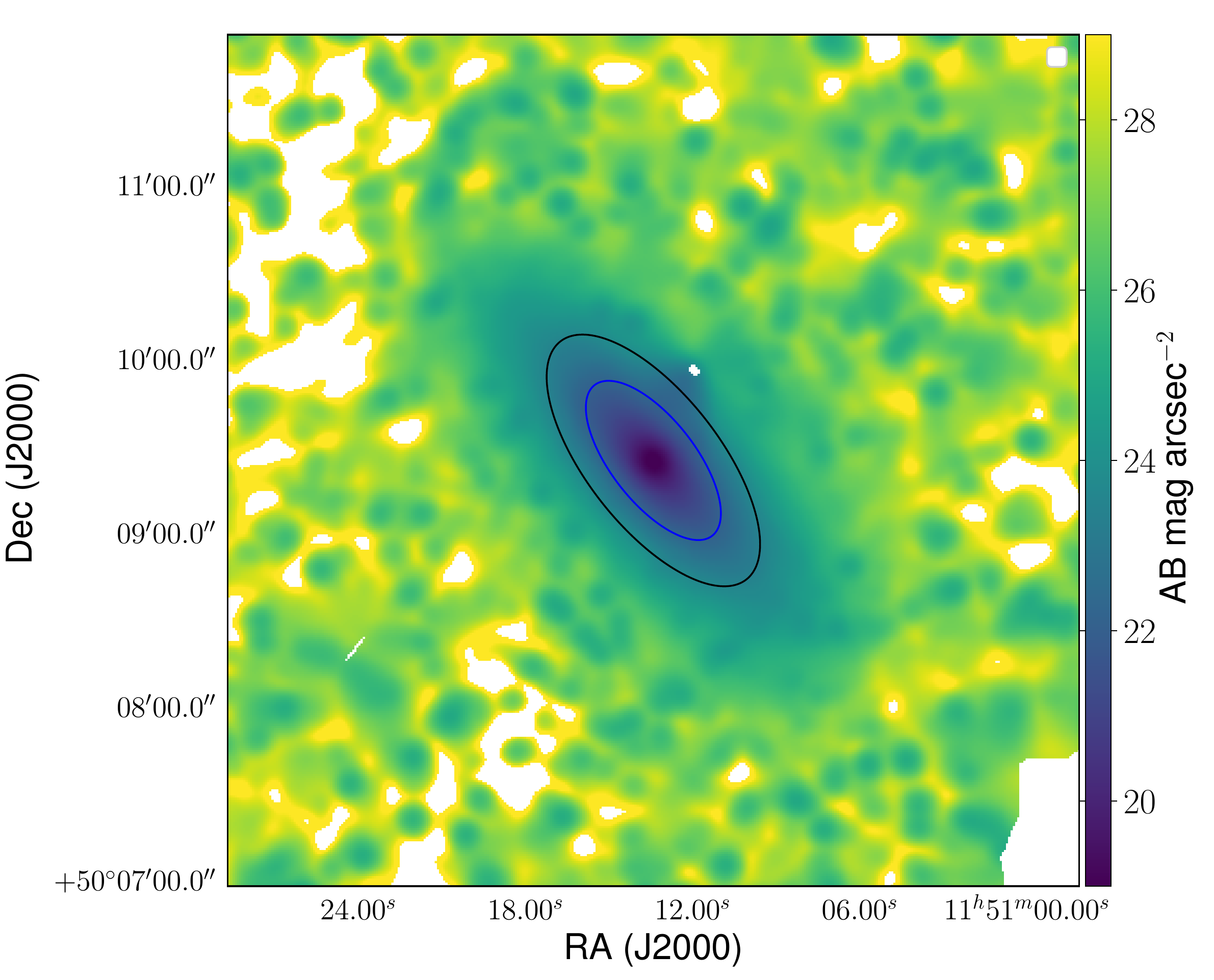} &
        \includegraphics[width=75mm]{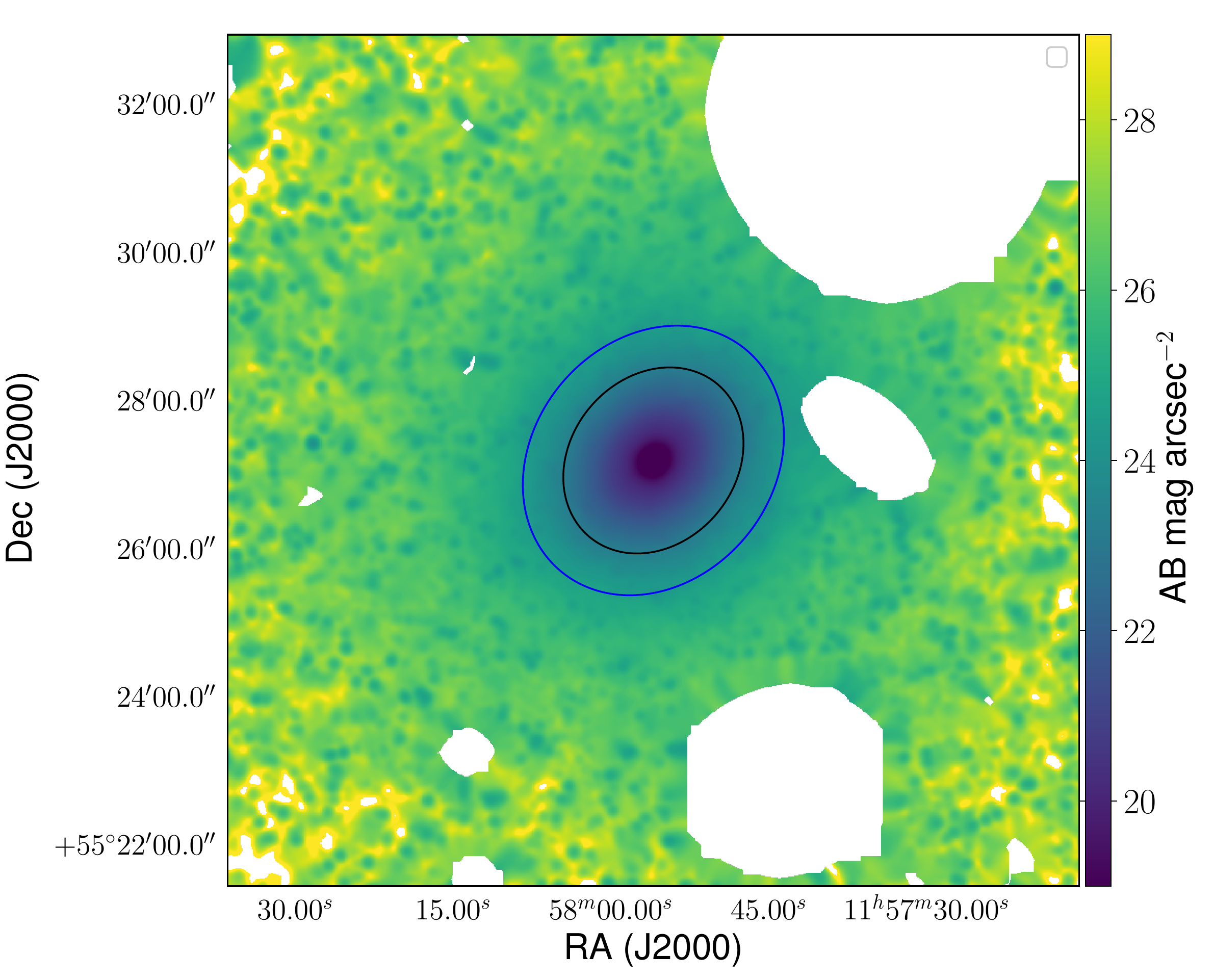} \\
        (a) NGC 3922 & (b) NGC 3998 \\[6pt]
        \includegraphics[width=75mm]{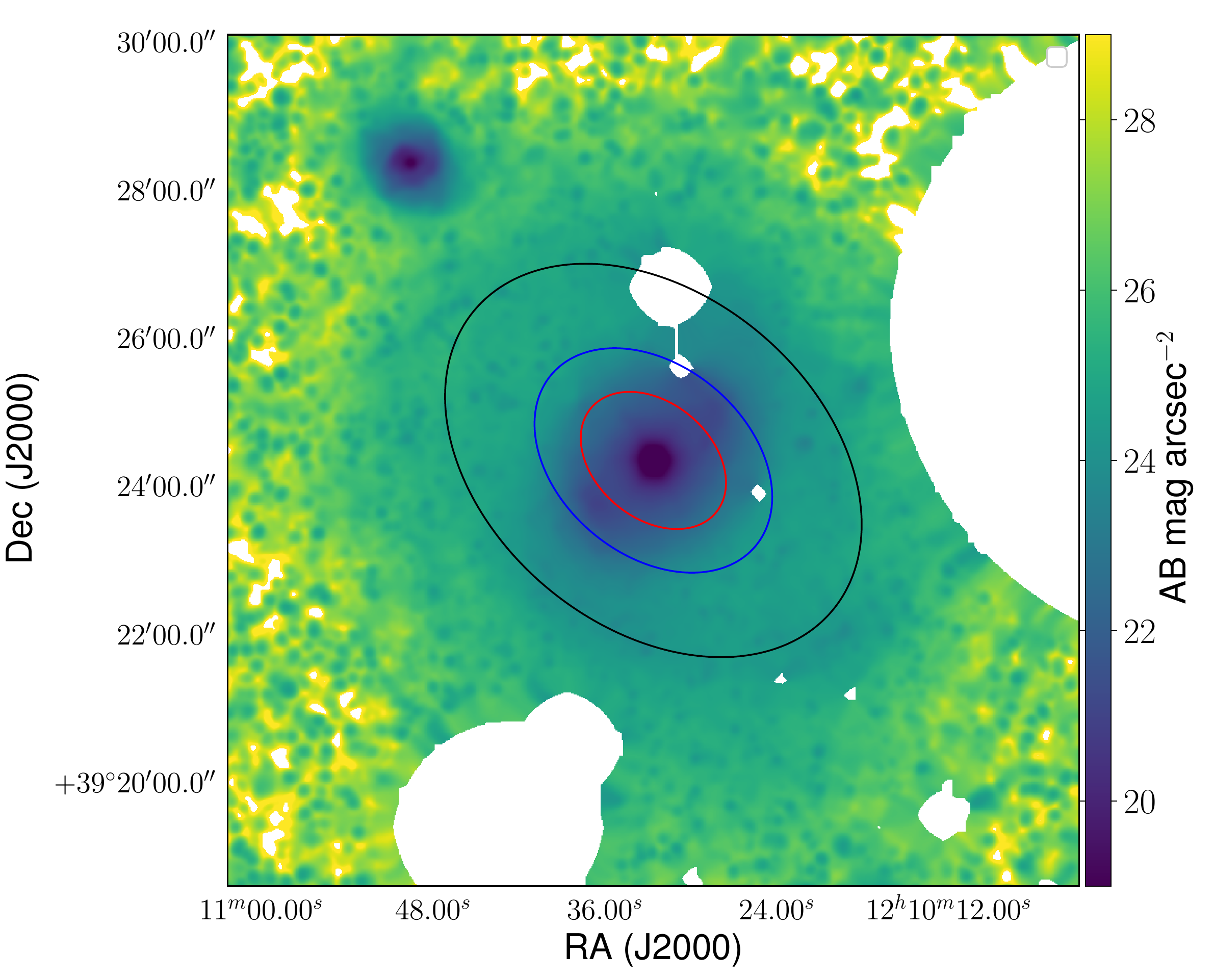} &
        \includegraphics[width=75mm]{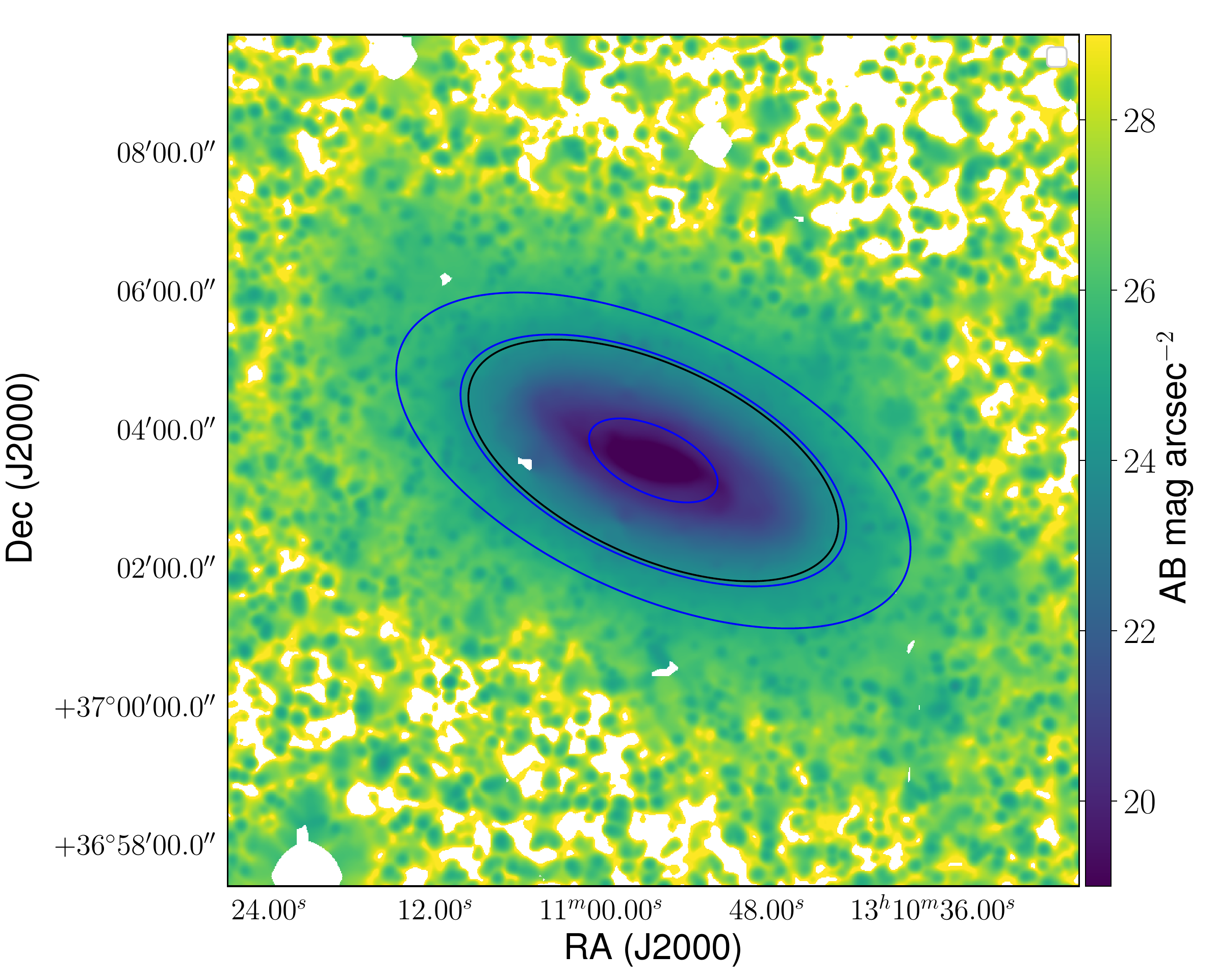} \\
        (c) NGC 4151 & (d) NGC 5005 \\[6pt]
    \end{tabular}
    \caption{A sampling of 3.6~\um\ mosaics. Black ellipses are $R_{25}$, red ellipses are Type-II breaks, and blue ellipses are Type-III breaks.}
    \label{fig:mosaics}
\end{figure*}

After masking, the mosaics are smoothed to reach the 3.6~\um\ Spitzer warm-mission limiting surface brightness of $\sim0.5$~kJy~sr$^{-1}$, (1~$\sigma$ per pixel; see \cite{Krick+11} for examples and discussion). The smoothing is accomplished using the python function ${\tt astropy.convolve}$ \citep{astropy+13} with a 2D-Gaussian kernel. The size of this kernel was determined by convolving every mosaic with a variety of filters of different sizes and measuring the standard deviation of the sky (using the same manually defined sky regions used in the measurement of the sky gradient) for the entire sample (see Figure~\ref{fig:smooth_test}). The 4 pixel ($3\arcsec$) wide kernel is the first to reach the Warm Mission's limiting surface brightness and is chosen for this study. We also note that the convolution algorithm uses the smoothing kernel to interpolate over missing values in the image. Normally this is not ideal as the interpolated data lack the noise characteristics of the original image, the structure beneath the interpolated-over source may not necessarily be smooth, and this interpolation may recover sources which have not been fully masked. However, missing data causes the surface brightness profile fitting (described below) to fail prematurely, and thus we use astropy's interpolative convolution method while noting the above issues. See Figure~\ref{fig:mosaics} for a suite of example mosaics.

\renewcommand{\thefigure}{\arabic{figure} (Cont.)}
\addtocounter{figure}{-1}

\begin{figure*}
    \centering
    \begin{tabular}{cc}
        \includegraphics[width=75mm]{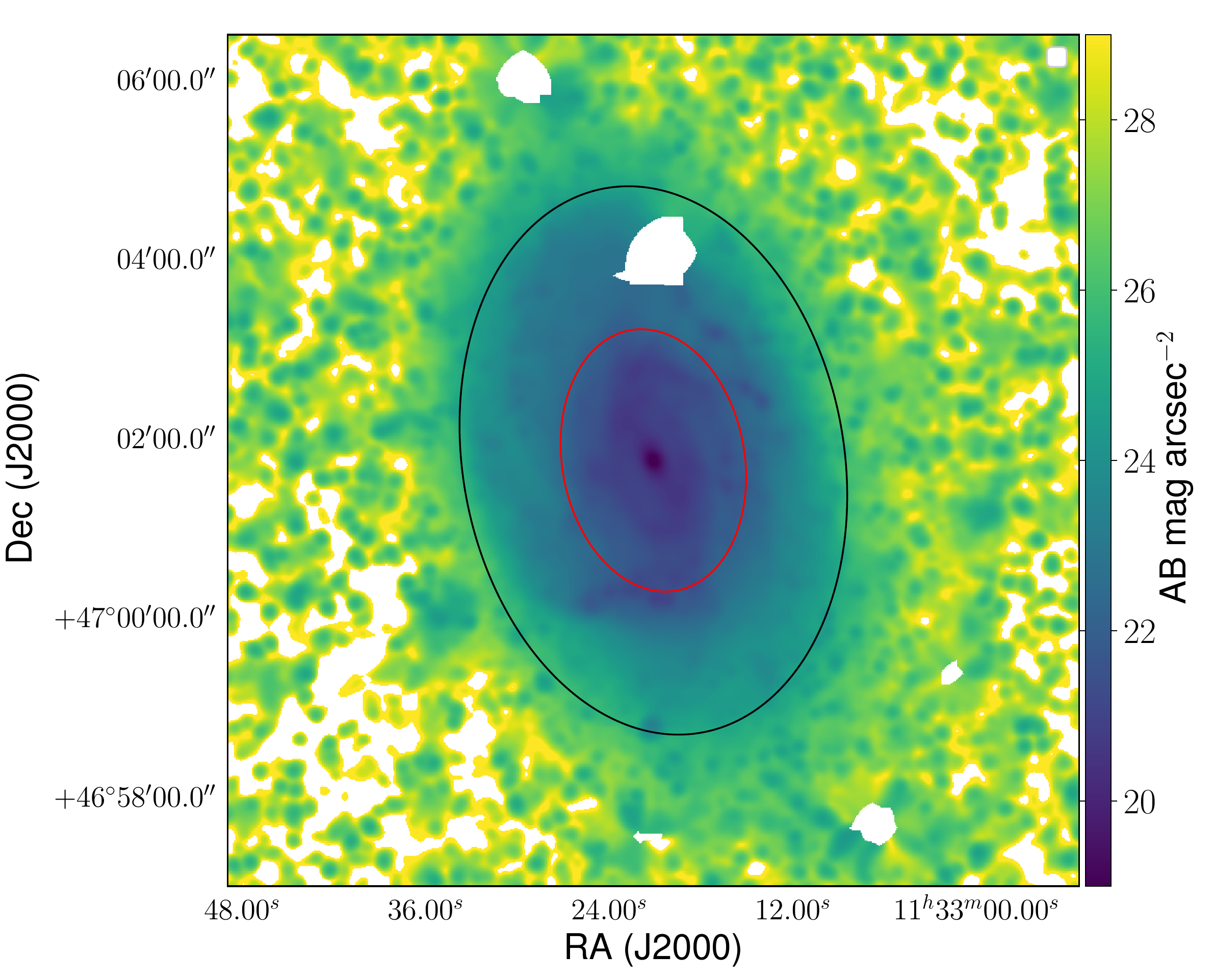} &
        \includegraphics[width=75mm]{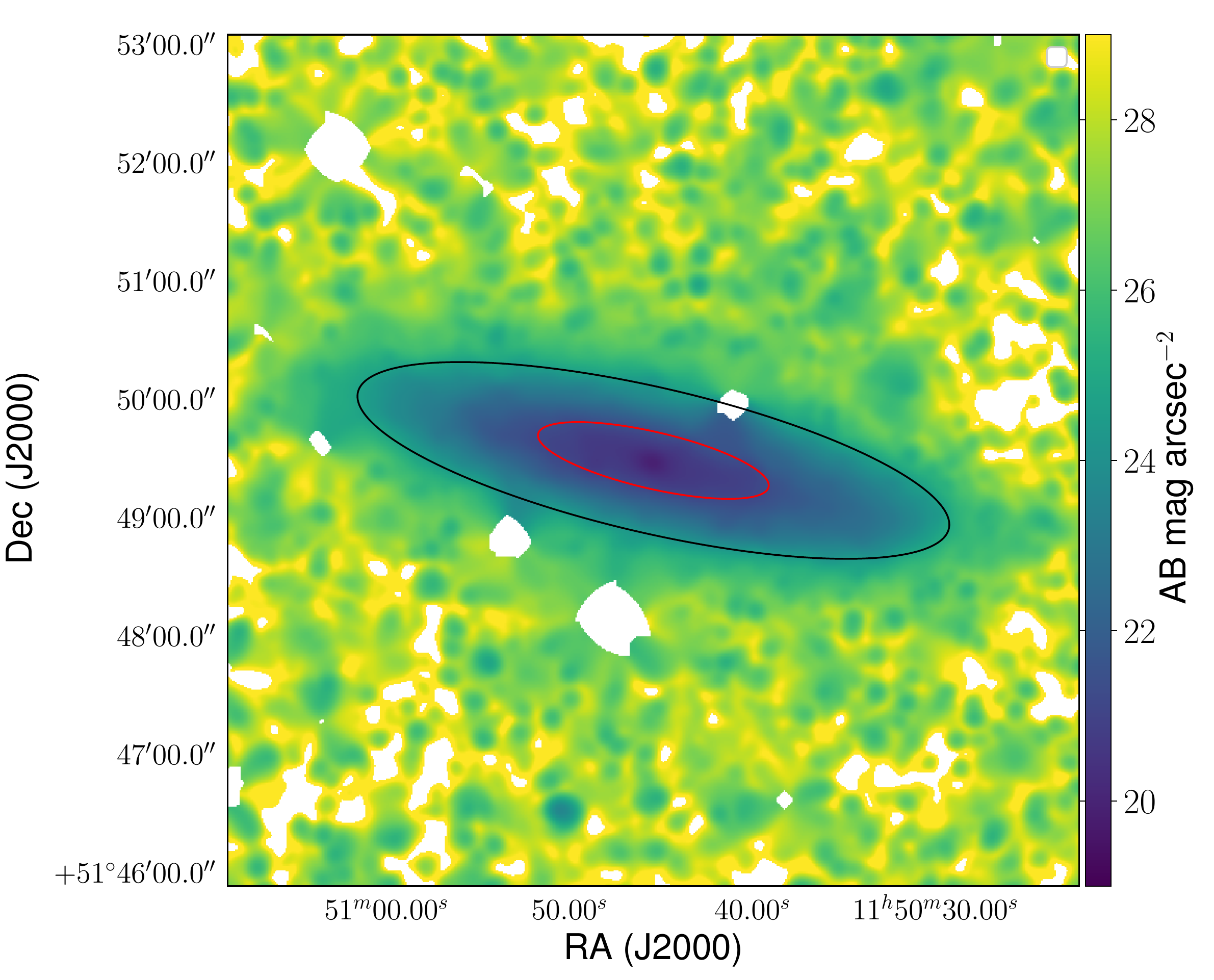} \\
        (e) NGC 3726 & (f) NGC 3917 \\[6pt]
        \includegraphics[width=75mm]{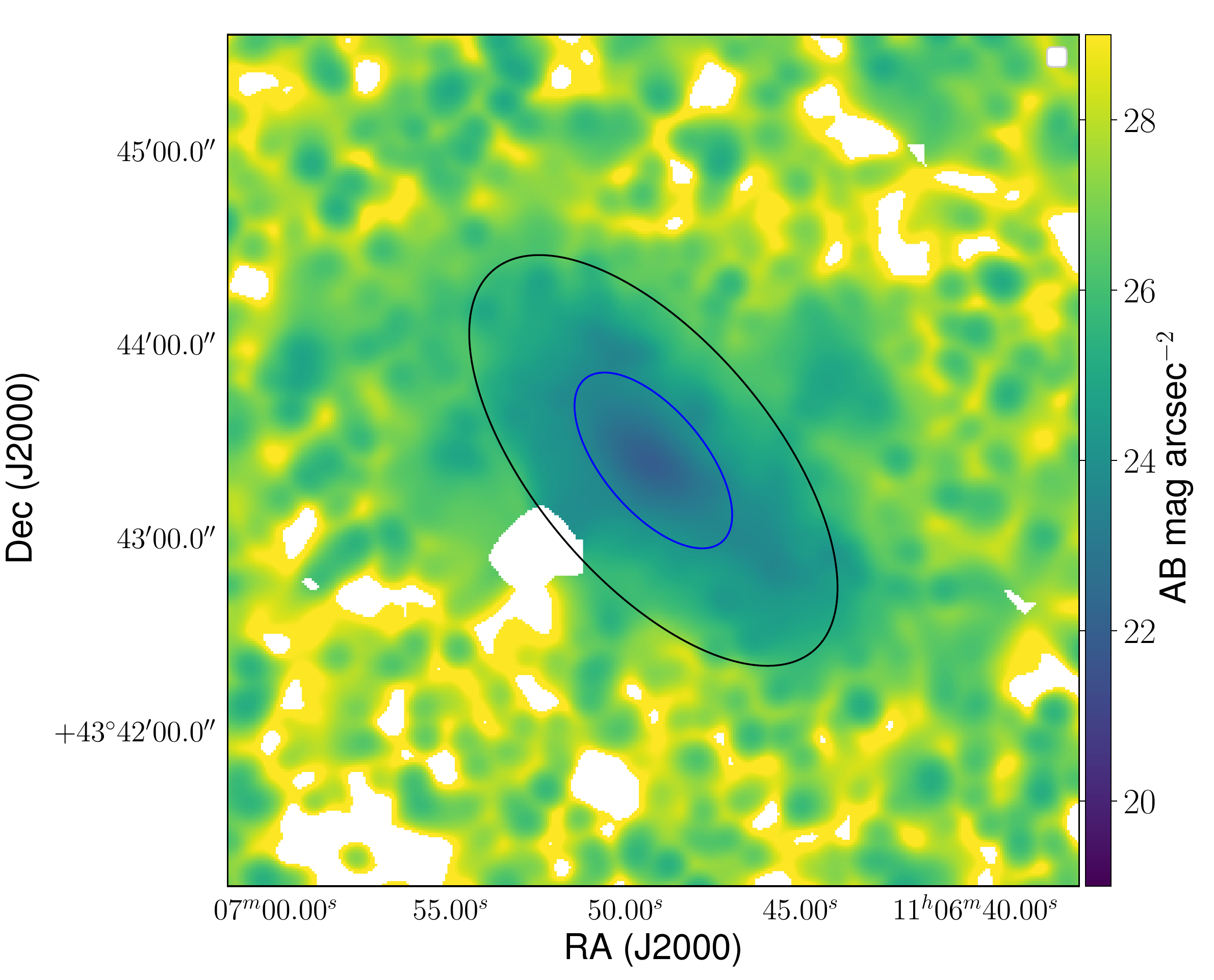} &
        \includegraphics[width=75mm]{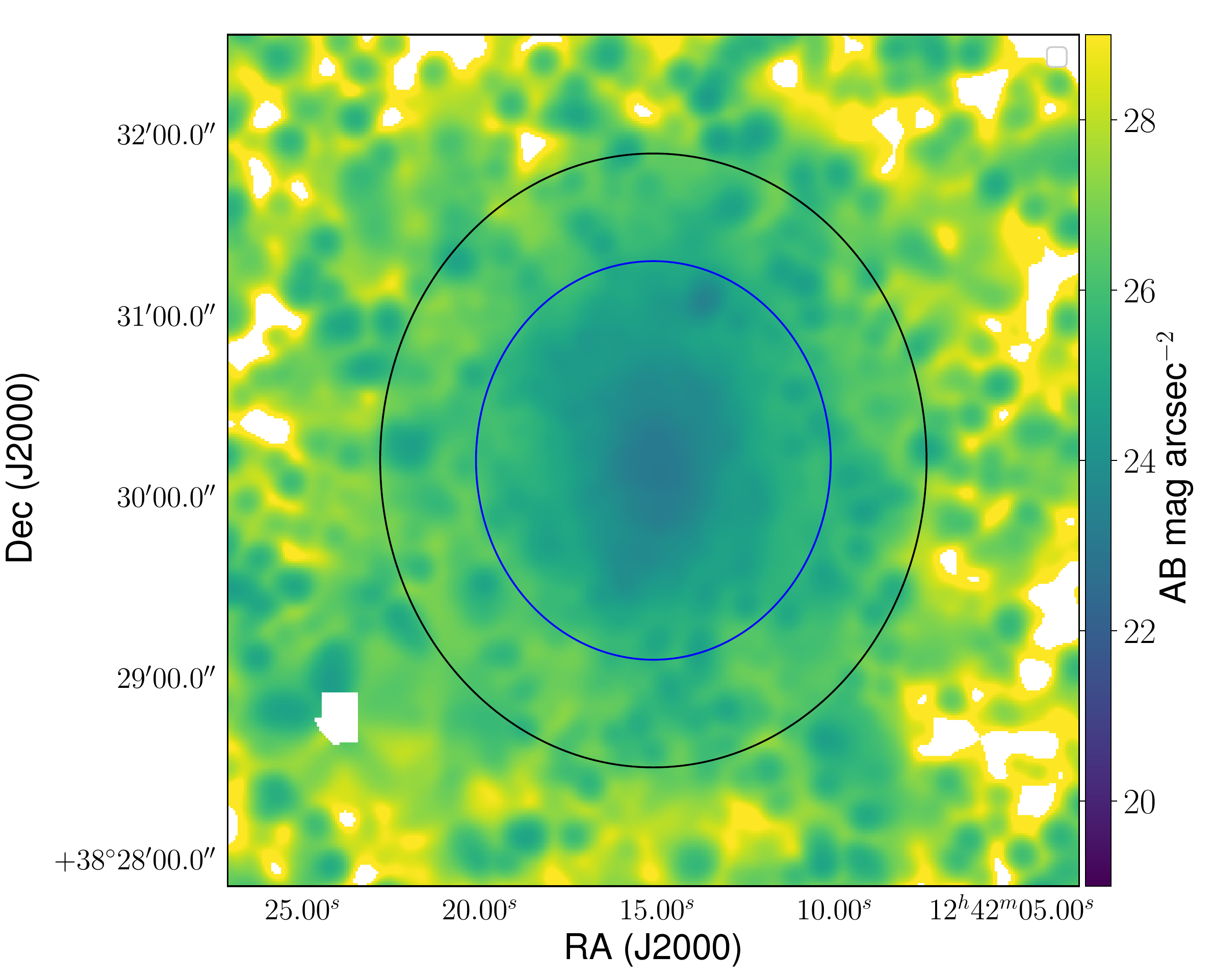} \\
        (g) UGC 06161 & (h) IC 3687 \\[6pt]
    \end{tabular}
    \caption{A sampling of 3.6~\um\ mosaics.}
\end{figure*}

\renewcommand{\thefigure}{\arabic{figure}}

\section{RESULTS}

\subsection{Surface brightness profiles}

\begin{figure*}
    \centering
    \begin{tabular}{cc}
        \includegraphics[width=90mm]{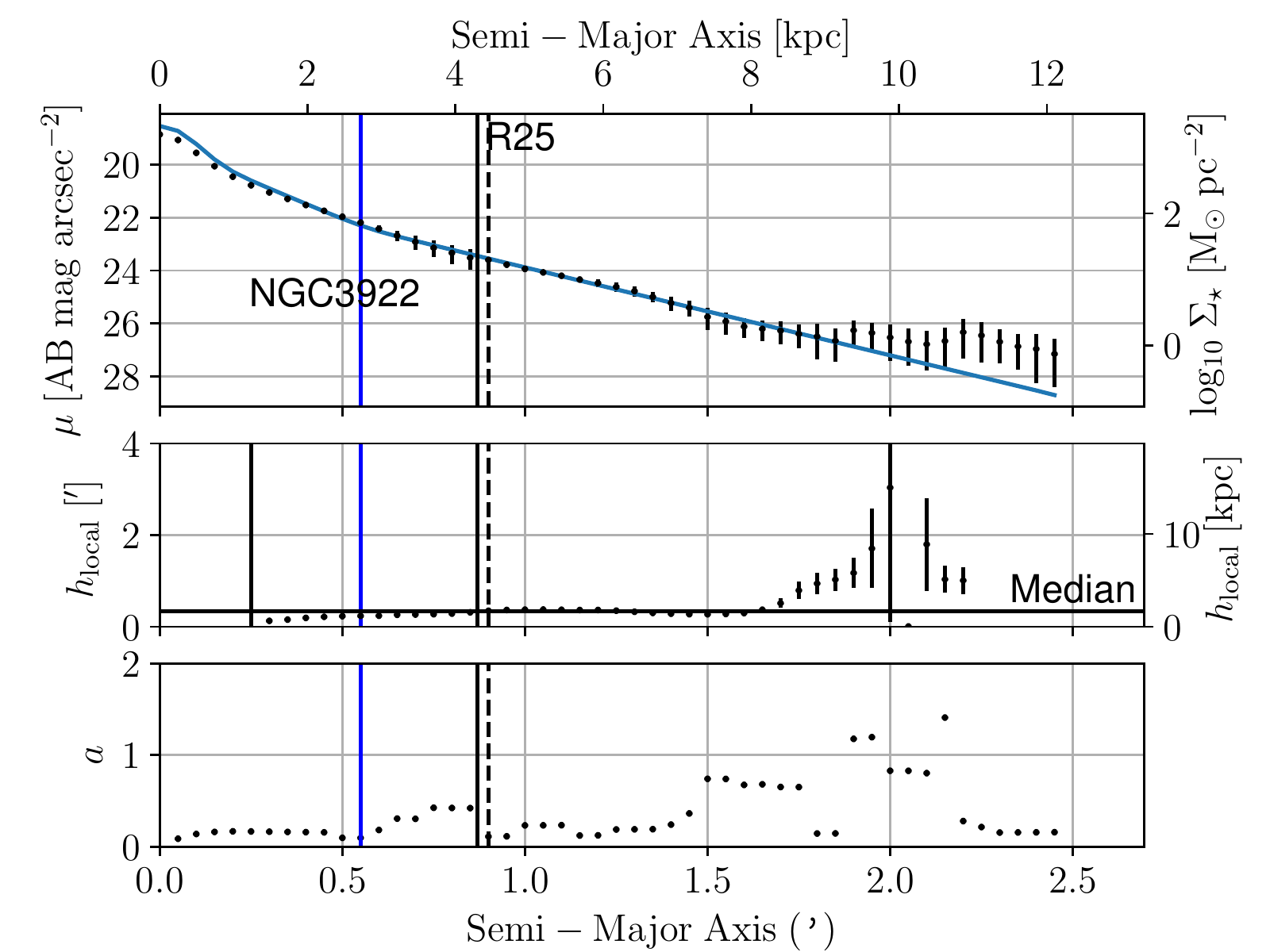} &
        \includegraphics[width=90mm]{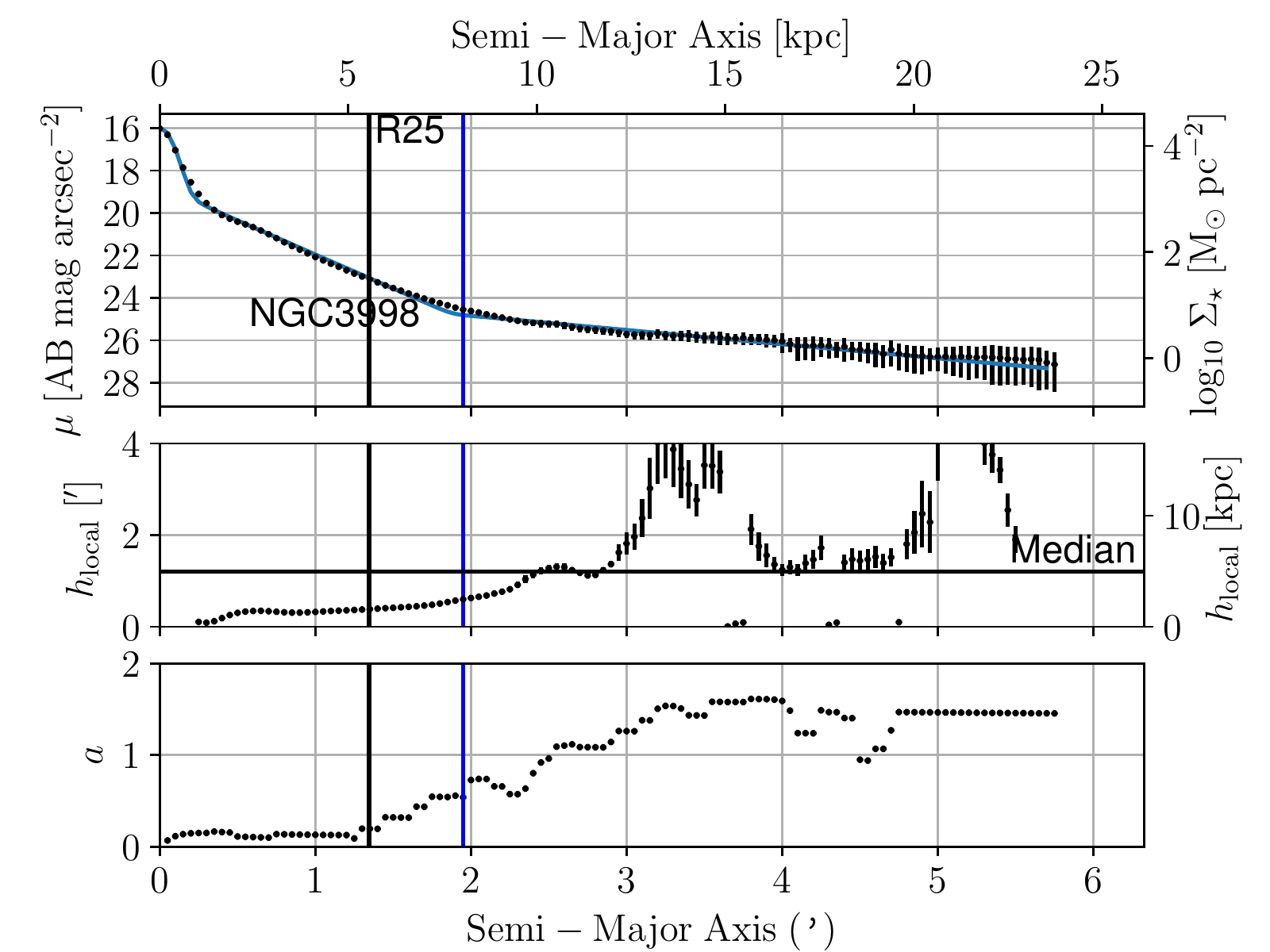} \\
        (a) NGC 3922 & (b) NGC 3998 \\[6pt]
        \includegraphics[width=90mm]{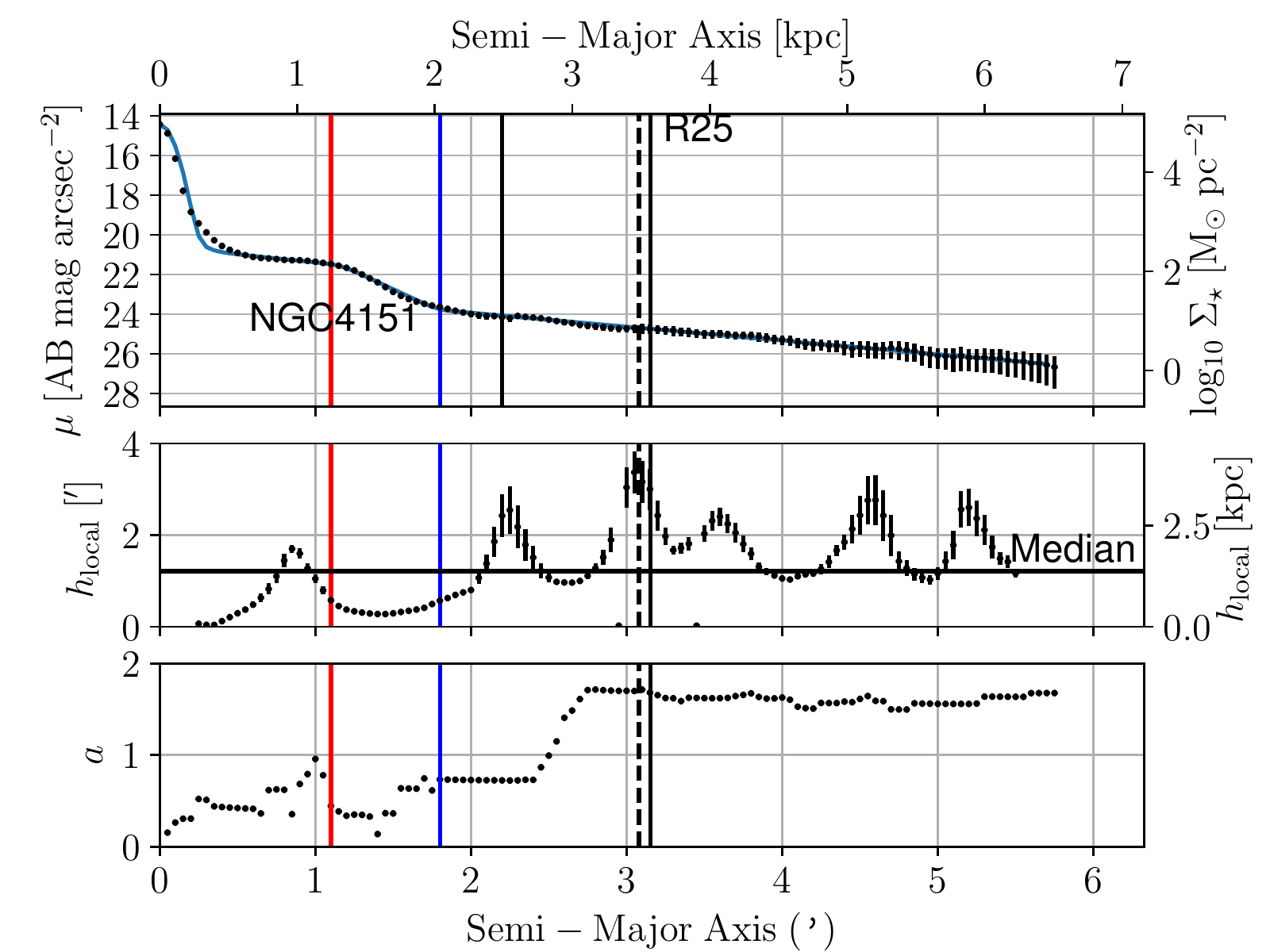} &
        \includegraphics[width=90mm]{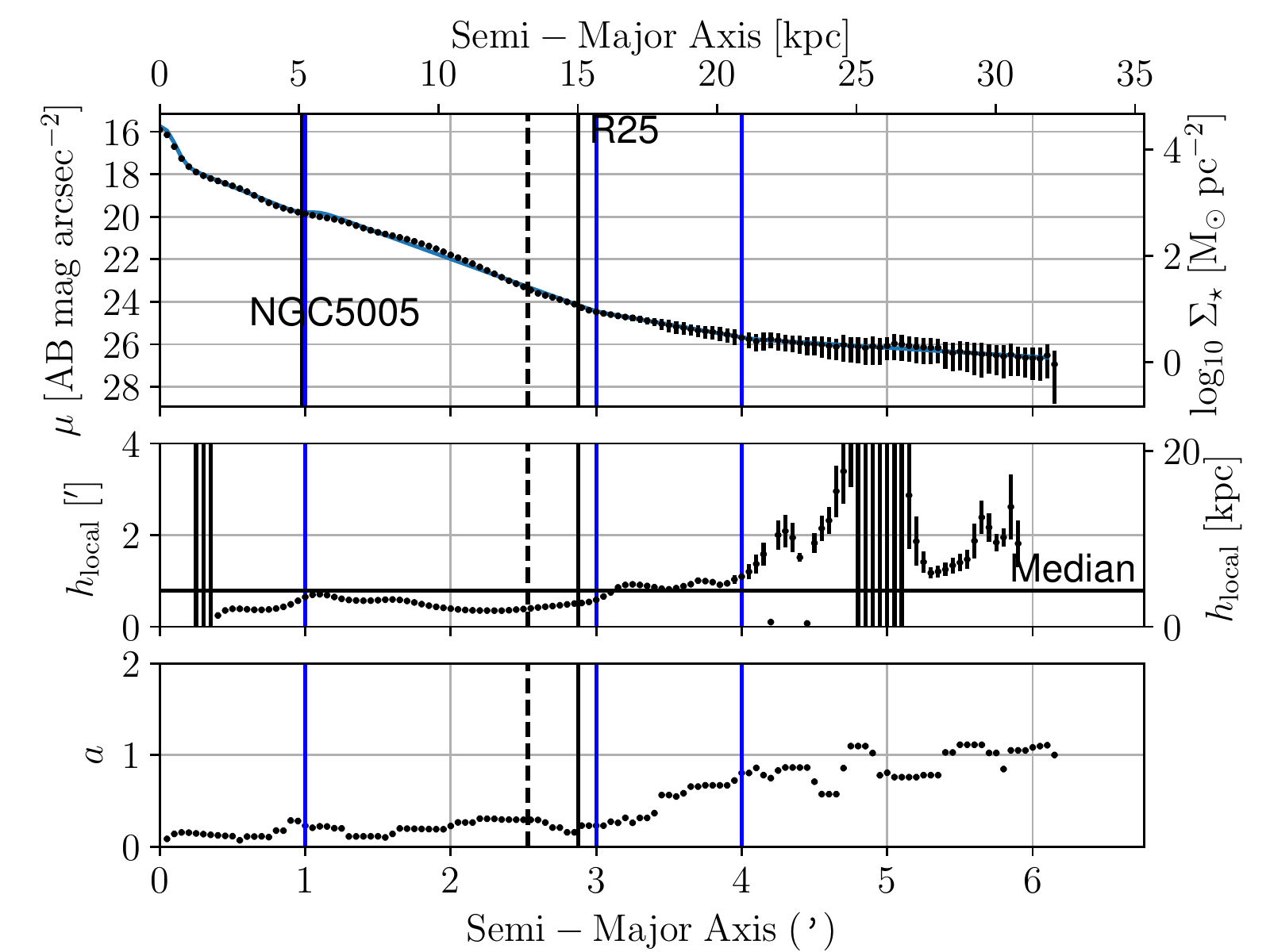} \\
        (c) NGC 4151 & (d) NGC 5005 \\[6pt]
    \end{tabular}
    \caption{Surface brightness profiles. $R_{25}$ is denoted as the solid and labeled line, the dashed line denotes the extent of the spiral structure, and the colored lines mark the locations of the breaks (blue for an up-bending break, red for a down-bending break). The blue line is the model fit convolved with the PSF. Also included is the local scale length, $h_{\rm local}$ (middle), and the asymmetry parameters $a$ (bottom).}
    \label{fig:profiles}
\end{figure*}

Once background and foreground sources have been removed from the mosaics and the mosaics have been smoothed, the IRAF task {\tt ELLIPSE} is run to generate surface brightness profiles. The ellipticity, central position, and position angle are left as free parameters to minimise uncertainties caused by differences of orientation and position between the outer and inner isophotes. The cause of these differences may be from interactions with nearby galaxies, the presence of distinct thick and thin discs or a bar, or extreme radial migration. The initial galaxy centers are defined by their RC3 values \citep{DeVaucouleurs+91} with the Spitzer-calibrated world coordinate system. In cases where ELLIPSE fitting fails to find a solution using the RC3 values the centers are defined manually. An iterative 4$\sigma$ rejection is used to minimise the effect of cosmic rays and unidentified point sources, and the surface brightness profile is sampled linearly with an annular width of 4~pixels ($3\arcsec$) per bin. Also, a pixel list file is generated to properly mask any missing data. Once the surface brightness is measured the profiles are converted into AB-based magnitudes with the calibration from \cite{Reach+05}. Also, the stellar mass density is calculated using a $M/L$ ratio of 0.5$\pm0.1$. This simple relation has been found to be consistent at 3.6~\um\ with a variety of techniques: optically calibrated stellar population synthesis models \citep{Oh+08, Eskew+12, McGaugh+14}, rotation curve decomposition \citep{Barnes+14}, stellar population synthesis calibrated with [3.6]-[4.5]~\um\ colors \cite{Meidt+14}, and with the Tully-Fisher relation \citep{McGaugh+15}. Examples which span the range of morphological types of EDGES is found in Figure~\ref{fig:profiles}, and surface brightness profiles for the entire sample are available within the supplementary online-only version of this paper.

\renewcommand{\thefigure}{\arabic{figure} (Cont.)}
\addtocounter{figure}{-1}

\begin{figure*}
    \centering
    \begin{tabular}{cc}
        \includegraphics[width=90mm]{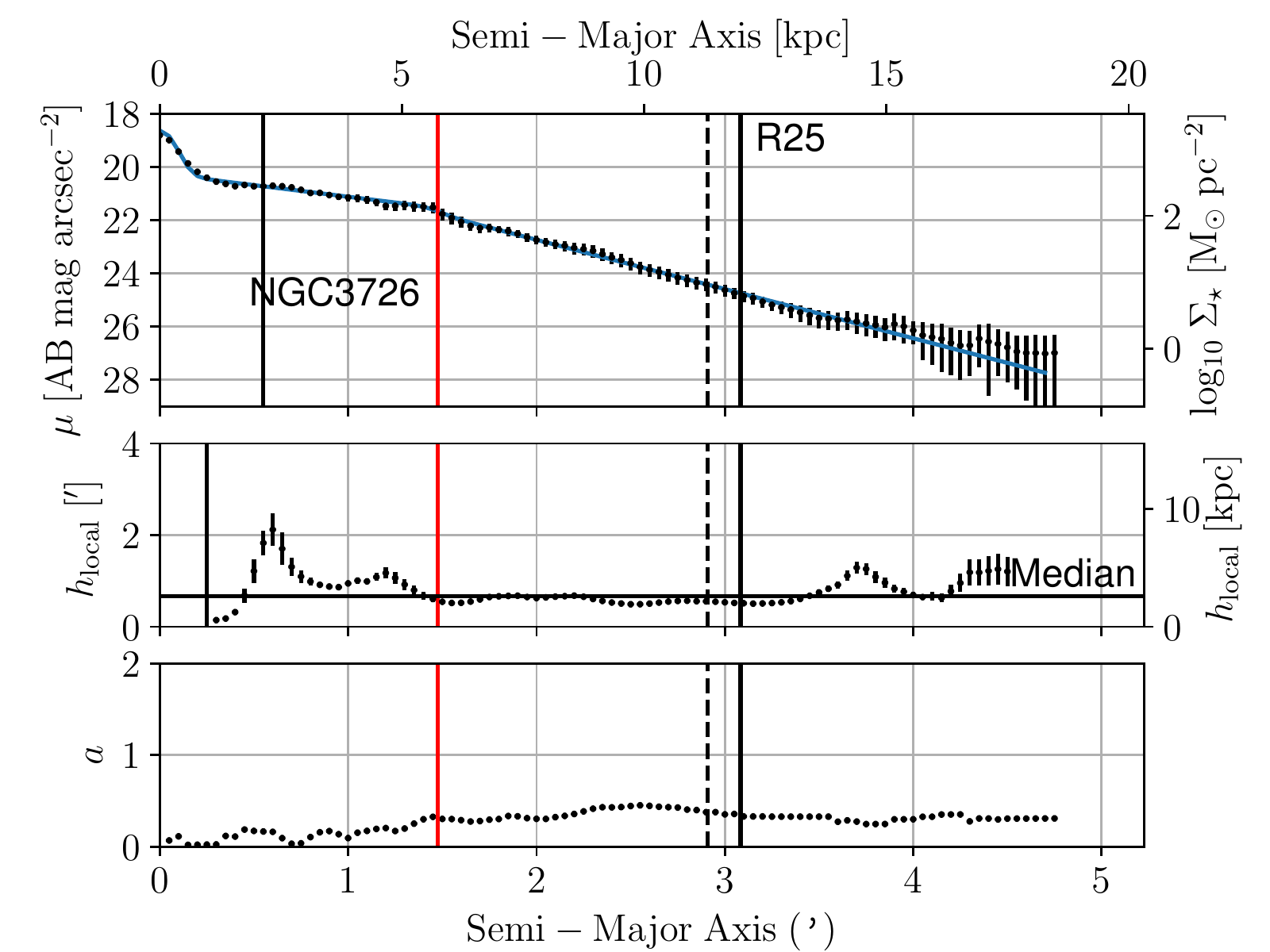} &
        \includegraphics[width=90mm]{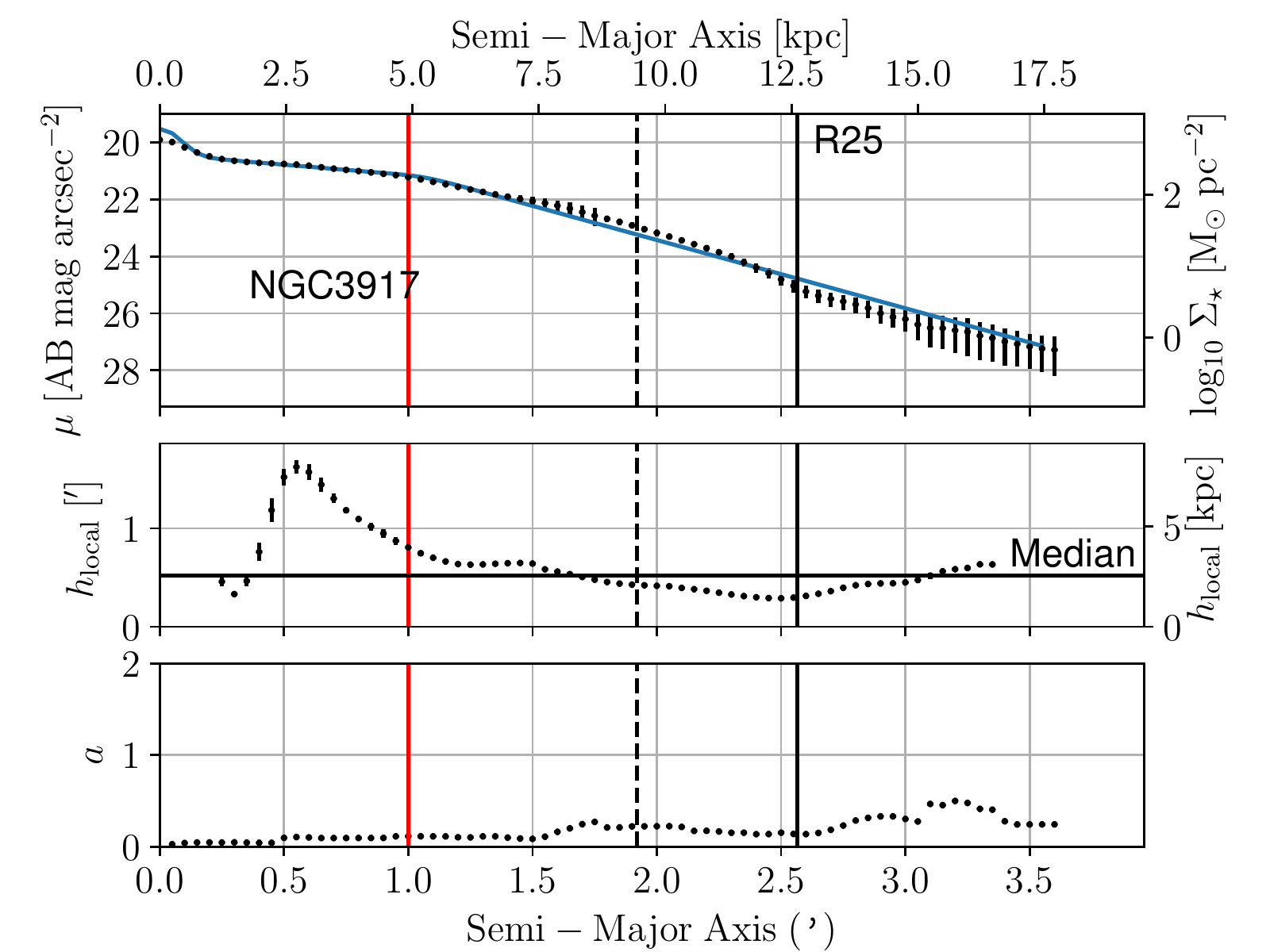} \\
        (e) NGC 3726 & (f) NGC 3917 \\[6pt]
        \includegraphics[width=90mm]{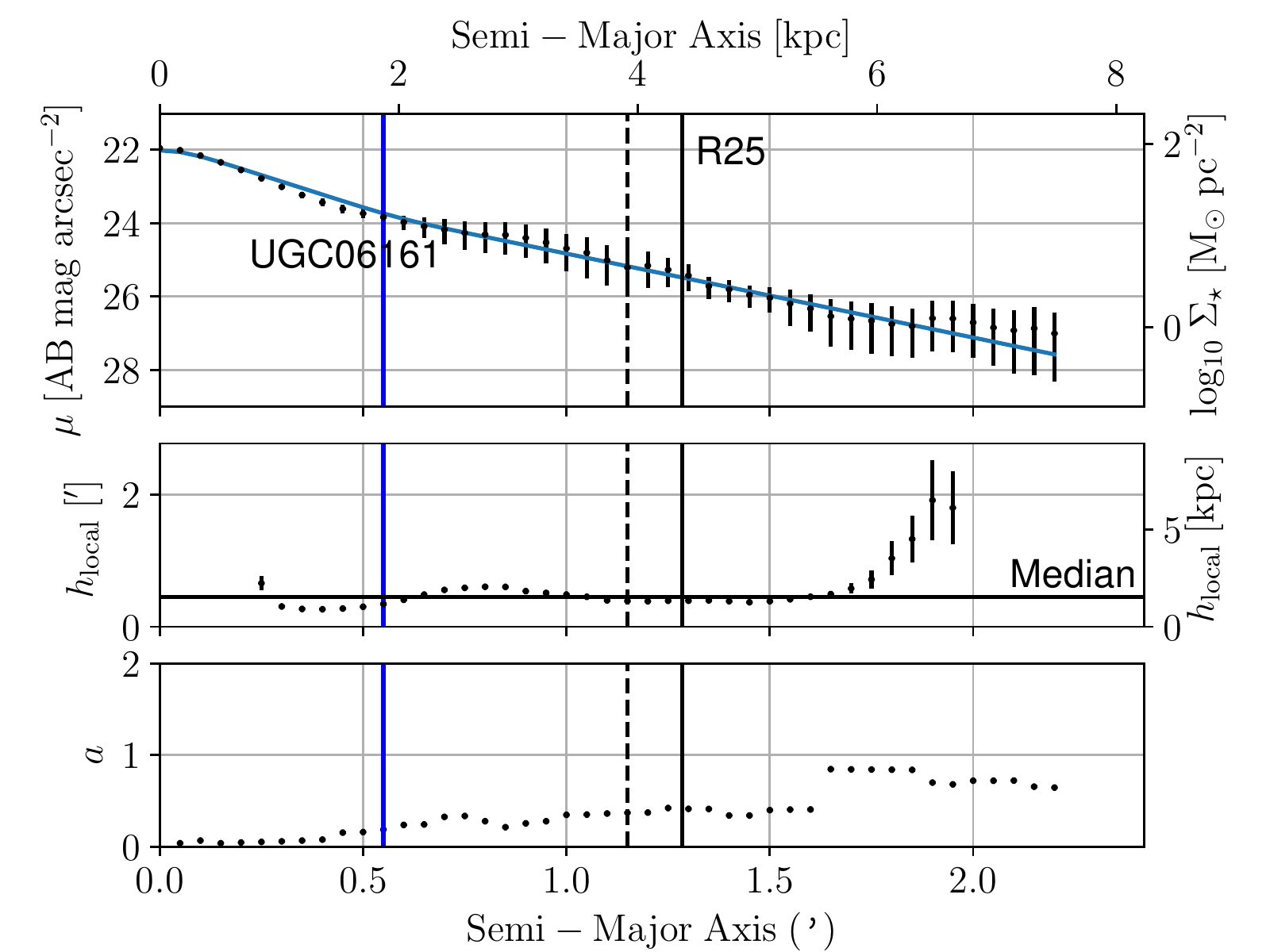} &
        \includegraphics[width=90mm]{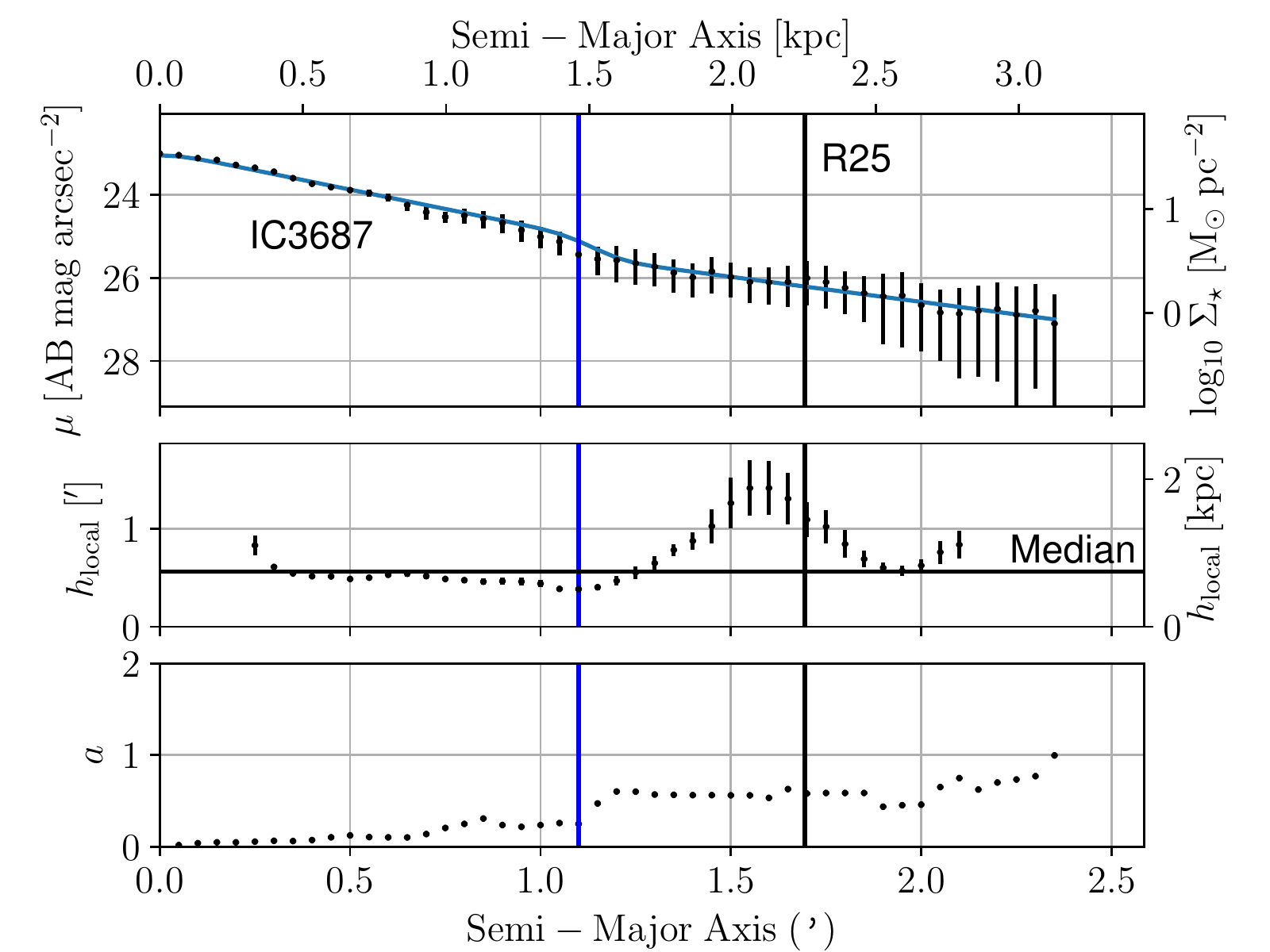} \\
        (g) UGC 06161 & (h) IC 3687 \\[6pt]
    \end{tabular}
    \caption{Surface brightness profiles.}
\end{figure*}

\renewcommand{\thefigure}{\arabic{figure}}

\subsection{Component fitting and modeling}

Variations within the surface brightness profiles are caused from a multitude of sources, including the PSF, transitions from one galactic component to another, (e.g. bulges, bars, discs, and stellar halos), spiral structure, asymmetries in a component's two-dimensional profile, and radial migration. To study these phenomena individually, we model the surface brightness profiles as a number of overlapping components. These models include a bulge component which is based upon the 3.6~\um\ PSF, and disc components based upon S{\'e}rsic disc profiles \cite{Sersic68}.

\subsubsection{The bulge model}

While traditionally bulges have been modeled as S{\'e}rsic profiles with $n>2$, we have found that the bulge profile is dominated by the PSF in the EDGES surface brightness profiles after smoothing the PSF to $3\arcsec$. To measure the bulge as a point source we have created a PSF specifically for EDGES with the aid of the Spitzer Warm Mission extended Point Response Function (PRF) from the NASA/IPAC Infrared Science Archive \footnote{http://irsa.ipac.caltech.edu/data/SPITZER/docs/irac/calibrationfiles/psfprf/}. The PRF is transformed to the EDGES platescale with the IDL task ${\tt CONGRID}$, then smoothed using a Gaussian with a 4~pixel standard deviation with the python function ${\tt astropy.convolve}$ \citep{astropy+13}, the same procedure used to smooth the mosaics. This result is then normalised to sum to unity, resulting in the EDGES PSF. We note that the PSF differs between mosaics since the set of images used to construct the mosaics were split into subsets taken on different days, resulting in subsets of images taken with different telescope orientation angles. The diffraction spikes do not overlap with different orientation angles, creating differences in the final PSF for each mosaic. However, with only a single spatial dimension, the surface brightness profile's PSF does not suffer from this orientation issue, and is the same for each surface brightness profile. The contribution from the central point source to the surface brightness profile is the following ratio:
\begin{equation}
	S_{\rm bulge} = \frac{S_0}{\rm PSF_{cen}},
\end{equation}
where $S_{\rm bulge}$ is the total surface brightness within the bulge, $S_0$ is the surface brightness at $R=0$ on the surface brightness profile, and ${\rm PSF_{cen}}$ is the value of the central pixel in the normalised PSF. We note that in many cases the galaxy in question will not have a significant contribution from a bulge or any other central point source. In these cases the modeled central surface brightness of the disc (described below) is compared to the measured central surface brightness. If they are within 2$\sigma$ of one another the contribution from the central point source is set to zero.

\subsubsection{disc models}
\label{sec:final_model}
To model the disc we assume S{\'e}rsic disc profiles (i.e. $n=1$), of the general form:
\begin{equation}
	S(R) = S_0 {\rm exp}[-(R/h)], \\
\end{equation}
where $S$ is the surface brightness at a semi-major axis, $R$, $S_0$ is the surface brightness of the disc at $R=0$, and $h$ is the scalelength. To include breaks in the model, where the slope of the surface brightness profile abruptly changes, we allow for multiple S{\'e}rsic disc profiles defined for specific radial regimes. Our final model takes the form:
\begin{equation}
\label{eq:final_model}
    S(R) = \begin{cases}
               S_{\text{bulge}}, 				& R=0, \\
               S_{0_1} {\rm exp}[-(R/h_1)], 	& 0 < R \leqslant R_{b_1}, \\
               S_{0_2} {\rm exp}[-(R/h_2)], 	& R_{b_1} < R \leqslant R_{b_2}, \\
			   &... \\
			   S_{0_n} {\rm exp}[-(R/h_n)], 	& R_{b_{n-1}} < R \leqslant {\rm 200~kpc}
           \end{cases}
\end{equation}
where the numerical subscripts denote with which S{\'e}rsic disc profile a parameter is associated, $n$ denotes the total number of these profiles, and $R_b$ is a break radius.

To locate the break radii a first pass is performed with the ``mark-the-disc'' method from \cite{Pohlen+Trujillo06}. The first step is to mark the extent of the disc-like section of the surface brightness profile where the S{\'e}rsic index is one. The inner mark denotes the extent of the bulge, or as is the case with the EDGES data, the extent of the contribution from the central point source. For galaxies with a prominent bulge this is determined by eye, and for galaxies without a prominent bulge (described above) the inner mark is set to zero. The outer mark is determined visually by examining the mosaics and drawing an aperture over the extent of each galaxy.

Once the disc has been marked, the local scalelength $h_{\rm local}$ profile is measured along the disc. This is done by calculating the slope of the surface brightness profile within eleven neighboring points with scipy's ${\tt curve\_fit}$, a Levenburg-Marquardt least squares fitting algorithm. The weighted mean of the local scalelength between the inner and outer marks is calculated and the points where the local scalelength crosses the weighted mean are used as the initial break radii. These values are then used to fit the initial parameters of the model (see above). For galaxies with no significant break in their 3.6~\um\ surface brightness profile many false breaks can be identified with this approach due to slight variations in the local scalelength caused by spiral structure, or from random scatter in $h_{\rm local}$ due to the uncertainties in the surface brightness profile. This method may also miss breaks in galaxies with multiple breaks, where a break may occur entirely above or below the weighted mean. For galaxies with either no breaks or multiple breaks the radii are determined by eye, then the model is fit (see Equation~\ref{eq:final_model}) and the breaks are adjusted iteratively until the model correctly reproduces the observed surface brightness profile. See Figure~\ref{fig:profiles} for example surface brightness profiles with the local scalelength profile, break locations, the fits to the surface brightness profile between each break, and the complete model.

\subsection{Extended PSF}
\label{sec:exten_psf}

A serious concern for any study of extended surface brightness profiles is the influence of the PSF on the extreme outskirts of the surface brightness profile. In some cases, the extended wings of PSFs cause a break to appear. Large ground-based reflectors in particular suffer from this problem \citep[e.g.][]{Sandin14, Trujillo+16}, and this is especially true for red optical observations with thinned CCDs, where even shallow images are affected \citep{Michard02}. The Dragonfly group foregoes large reflectors entirely and has achieved incredibly deep images ($\mu_B\sim$32~\SB) with a fully automated array of commercially manufactured refractors (48 lenses for an equivalent aperture of 99~cm at the time of this writing). These Canon-branded instruments were originally designed for sports or birding photography \citep{Abraham+14}, have an effective aperture of only $\sim$400~mm, and do not suffer from extended wings in their PSFs since they are refractors and because of a proprietary chemical coating on the lens designed to limit diffraction spikes from bright stadium lights. 

\begin{figure}
    \includegraphics[width=\columnwidth]{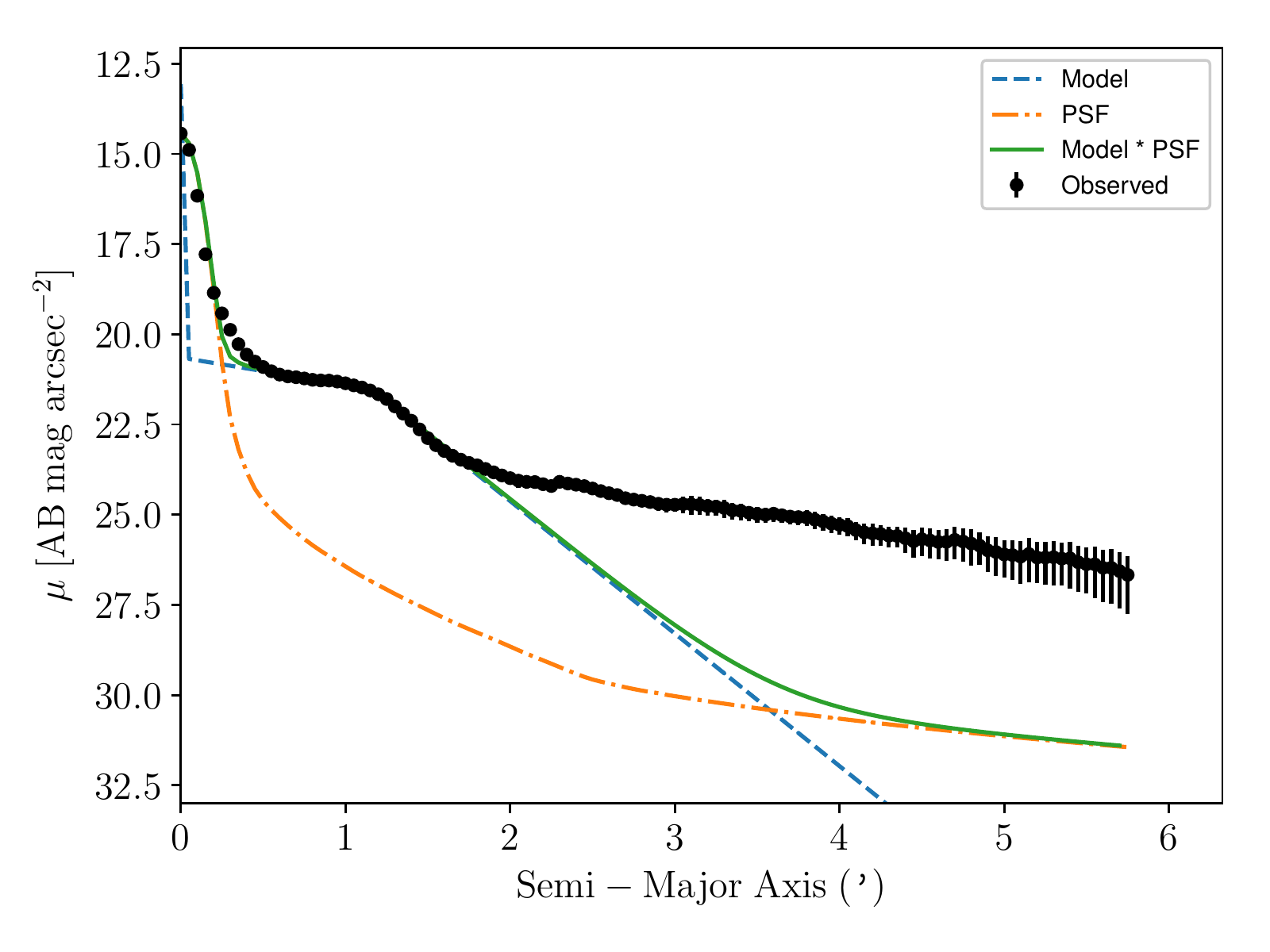}
    \caption{NGC~4151 modeled without the up-bending break at 1.8\arcmin. Observed data are black points with included uncertainties. The model is represented as a blue dashed line and contains only the bulge and first disc component. The PSF is scaled to appear on this plot and is the orange dashed-dotted line. The model convolved with the PSF is the green solid line. Note that a false up-bending break is produced but appears at 29~AB~mag~arcsec$^{-2}$, below the EDGES detection limit.}
    \label{fig:N4151PSF}
\end{figure}

We investigate the effect of the Spitzer 3.6~\um\ PSF on the EDGES data set by constructing a model with the method used in \cite{Sandin14, Sandin15}. This method is based upon the supposition that if a model bulge and disc are convolved with the PSF and an up-bending break is observed, then observed up-bending breaks may be due to the wings of the PSF and not from a distinct component within the galaxy. For our model the bulge and disc are based upon measurements from NGC~4151, an extreme case as it has the brightest 3.6~\um\ central pixel in the entire EDGES sample due to its status as a Seyfert galaxy. The bulge is modeled as a point source at $R=0$ with $S_0=1340$~MJy/sr, and a disc with S{\'e}rsic disc profile with $S_0=0.85$~MJy/sr and $h=1.44$\arcmin\ (see Equation~\ref{eq:final_model}). The inner section of the PSF is based upon the surface brightness profile of the model PSF used to measure the contribution from the bulge (see \S~\ref{sec:final_model}). The extent of the model PSF is 2\farcm2, but we extend it by fitting an inverse square law to the the model PSF beyond 1\arcmin\ (which most PSFs tend to follow in this radial regime). The PSF and model image are then convolved to produce Figure~\ref{fig:N4151PSF}. This convolution correctly reproduces the bulge and disc section of the galaxy but there is no evidence of a stellar halo caused by the wings of the PSF.  A false up-bending break is produced, but this break does not match the location and brightness of the observed up-bending break. This same procedure is used to test any up-bending break in the entire sample but we find no evidence of false up-bending breaks due to the extended PSF in EDGES data.

\subsection{Break classification}
\label{sec:break_class}

\begin{figure}
    \includegraphics[width=\columnwidth]{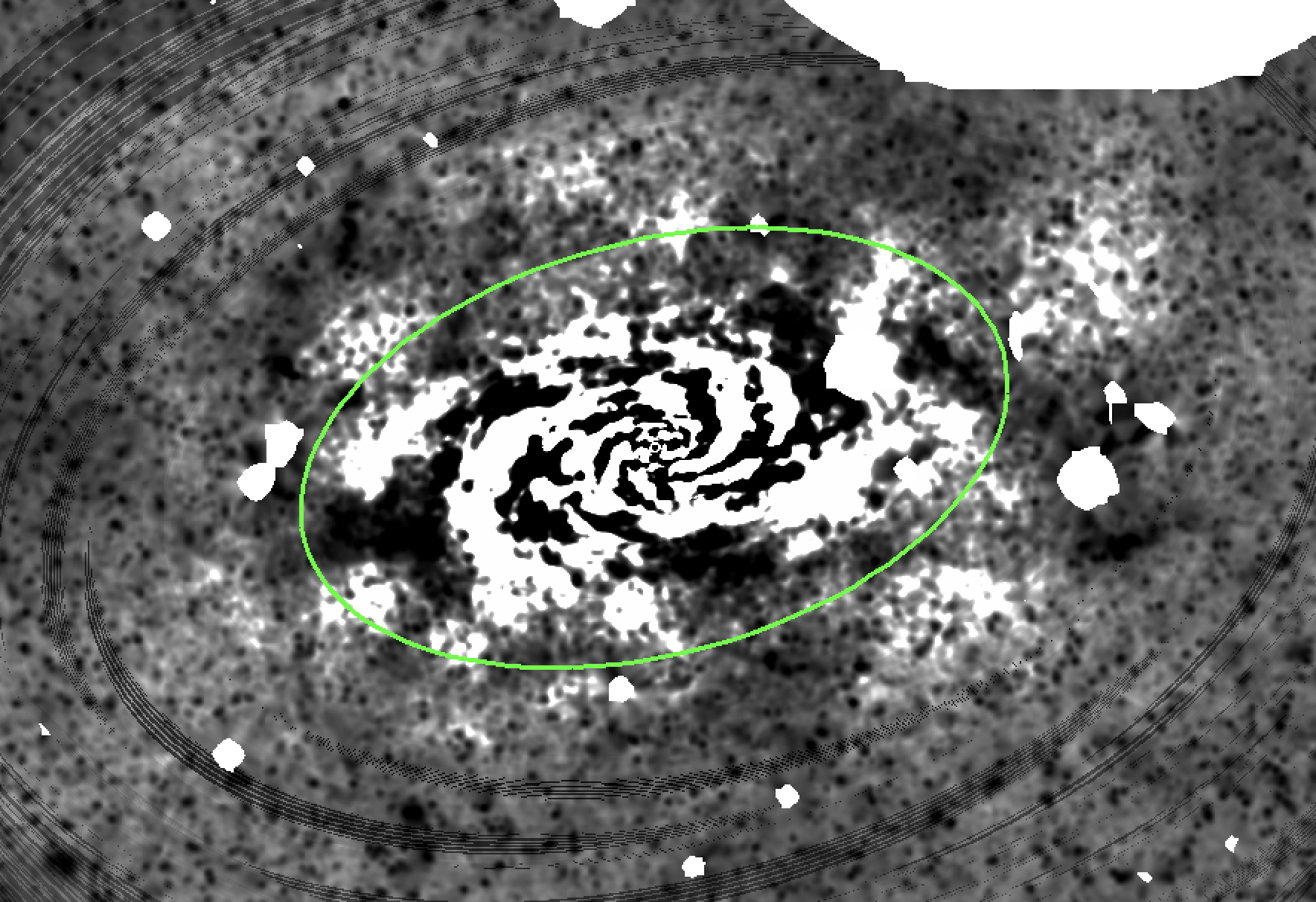}
    \caption{The subtracted {\tt bmodel} image of Messier~063, highlighting the grand-design spiral structure.}
    \label{fig:M63Sub}
\end{figure}

The discs of each galaxy in EDGES are decomposed into their individual components according to the breaks in the surface brightness profile (see above). This is accomplished by measuring the central surface brightness and scalelength of the profile between each break with Equation~\ref{eq:final_model} and the Levenburg-Marquardt least squares fitting algorithm ${\tt curve\_fit}$. Each break is then classified according to its type defined according to the following commonly used metric \citep{Erwin+08, Pohlen+Trujillo06, Gutierrez+11, Laine+14, Laine+16}, where {\bf Type-I} signifies no break, {\bf Type-II} signifies a down-bending break, or a decrease in scalelength, and {\bf Type-III} signifies an up-bending break, or an increase in scalelength. The Type-II breaks are subdivided into {\bf Type-II.OLR} (for Outer Linblad Resonance), and {\bf Type-II.CT} (for Classical Truncations). Outer Linblad Resonances cause down-bending breaks via interactions with the bar and occur at less than 2 times the bar radius; thus any Type-II break with a radius less than 2 times the bar radius (measured by examining the mosaics by eye) is classified as Type-II.OLR, and any other down-bending break is classified as Type-II.CT. Type-III breaks which occur within the spiral structure of the galaxy are classified as {\bf Type-III.S} (for spiral) and Type-III breaks which occur beyond the extent of the spiral structure are defined as {\bf Type-III.O} (for outer). The extent of the spiral structure was found by creating a difference image between the mosaic, and a smooth, homogeneous model created by the ${\tt IRAF}$ task ${\tt bmodel}$. This difference image removes the power-law profile from the images and leaves spiral structure. The extent of the structure is measured with a circular aperture. See Figure~\ref{fig:M63Sub} for an illustration of this process.

\subsection{Mass measurement}
\label{sec:mass_measure}

To examine the physical properties of the EDGES galaxies we measure the mass for each component within each galaxy. The total flux density between two points for the S{\'e}rsic disc components $(n=1)$ is:
\begin{equation}
	f(R_i < R < R_o)= 2 \pi S_0 h^2 
	\left\{ 
		\gamma \left( 2, \frac{R_o}{h} \right) 
		- \gamma \left( 2, \frac{R_i}{h} \right) 
	\right\}, \\
\end{equation}
where $f$ is the total flux density between the inner point, $R_i$, and the outer point, $R_o$, $S_0$ is the surface brightness at $R=0$, $h$ is the scalelength, and $\gamma$ is the incomplete gamma function. For components between breaks, $R_i$ and $R_o$ span the distance between the breaks, and we use $R_o=200$~kpc for the final component. To compute the mass we use:
\begin{equation}
	\label{eq:mass_derivation}
	M[M_{\odot}] = \Upsilon_{\star}^{3.6} \frac{f[Jy]}{c_m^{3.6}[Jy]} \times 10^{0.4(\mu + M_{\odot}^{3.6})},
\end{equation}
where $M$ is the mass, $\Upsilon_{\star}^{3.6}=0.5$ is the stellar mass-to-light ratio at 3.6~\um\ \citep{Oh+08, Eskew+12, Barnes+14, Meidt+14}, $c_m^{3.6}$ is the 3.6~\um\ calibration factor \citep{Reach+05}, $\mu$ is the distance modulus taken from the Extragalactic Distance Database \citep{Tully+09}, and $M_{\odot}^{3.6}=6.02$ is the absolute AB-based magnitude of the Sun at 3.6~\um\ \citep{Oh+08}. The masses of the individual galaxies, along with their respective model components, are found in Table~\ref{table:mass}.

\section{ANALYSIS AND DISCUSSION}

\subsection{Break statistics}

The entire sample contains 121 breaks; $6\%$ are Type-II.OLR, $31\%$ are Type-CT, $20\%$ are Type-III.S, and $43\%$ are Type-III.O. Seven galaxies have three breaks (8$\%$), twenty-nine galaxies have two breaks (32$\%$), forty-two galaxies have one break (46$\%$), and fourteen galaxies have no breaks (15$\%$). This is compared to only 7/90 (8$\%$) galaxies with two breaks in \cite{Pohlen+Trujillo06}, 5/47 (11$\%$) in \cite{Gutierrez+11}, and 4/328 (1$\%$) from \cite{Laine+14}. Note that these differences are sensitive to the wavelength of the surface brightness profiles under study \citep{Laine+16}, the method in which the breaks are determined, and the depth of the imaging. These details are further expanded upon below.

The mean break occurs at 6.6~kpc, the median at 5.2~kpc, and the furthest break occurs at 29.6~kpc. The distribution of break radii is found in Figure~\ref{fig:break_hist}. These break-radii are similar to a number of optical-based studies of disc-breaks \citep[i.e.][]{Pohlen+Trujillo06, Erwin+08, Gutierrez+11}, with a slightly higher fraction of Type-III breaks which occur further away. The differences from the optical studies \citep{Pohlen+Trujillo06, Erwin+08, Gutierrez+11} are likely due to differences between near-infrared and optical data (such as dust lanes and the greater effect of metallicity on optical data). In some cases these differences between optical and infrared data are enough to change the statistics of the break parameters, as seen in the near-infrared based S$^4$G and NIRS0S \citep{Laine+14, Laine+16}. Despite also being in the near-infrared, EDGES also differs from S$^4$G, with less Type-I galaxies and more Type-III galaxies. Galaxies labeled Type-III in EDGES would be labeled Type-I in S$^4$G when the break occurs beyond the surface brightness limit of S$^4$G (the unsmoothed noise level of EDGES is 2.5~kJy~sr$^{-1}$ versus 7.2~kJy~sr$^{-1}$ for S$^4$G \citep{Sheth+10}, a difference of 1.15~magnitudes). Of the five Type-I galaxies in \citep{Laine+14} which are also in EDGES, we classify four as Type-III and one as Type-II.

\begin{figure}
    \includegraphics[width=\columnwidth]{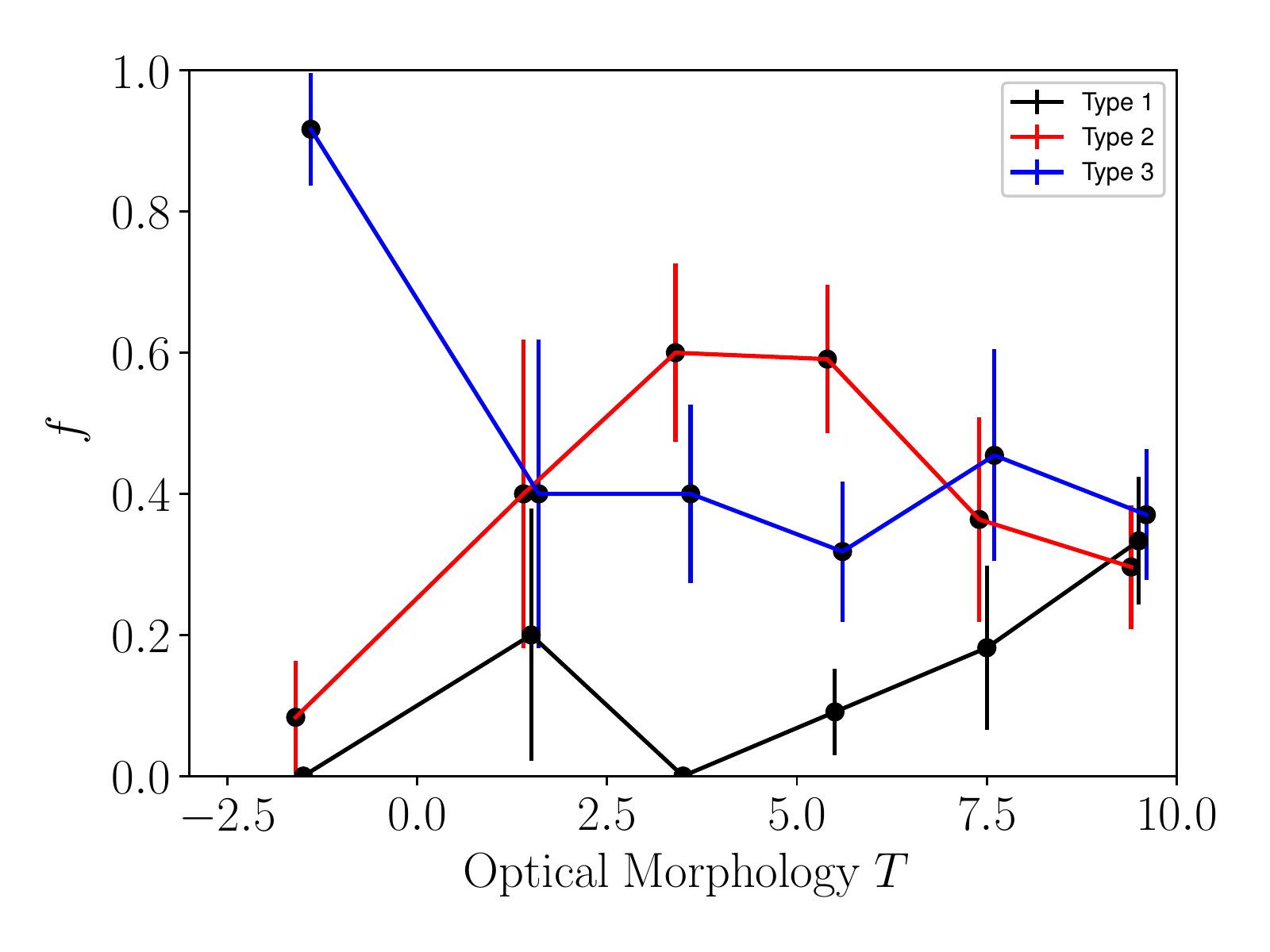}
    \caption{The percentage of galaxies with a certain T-type. Type-II and Type-III galaxies are slightly offset to emphasize their differences.}
    \label{fig:t_type}
\end{figure}

Figure~\ref{fig:t_type} shows break type by morphological $T$-type. Note that in this plot galaxies with multiple breaks are considered to be the type of their first break, which allows a comparison to studies which did not find many galaxies with multiple breaks. The bins are defined to mimic \cite{Pohlen+Trujillo06} with three bins added to represent ellipticals, Sa, and irregular/dwarf galaxies. The bins are thusly defined as: $T \leq 0.5$, $0.5 < T \leq 2.5$, $2.5 < T \leq 4.5$, $4.5 < T \leq 6.5$, $6.5 < T \leq 8.5$, $T > 8.5$ which roughly maps to ellipticals, Sa, Sb, Sc, Sd, and irregular/dwarfs respectively. EDGES has a relatively small sample of galaxies (92) and cannot fill every bin. With that caveat, our data fall within the uncertainties of Figure~5 in \cite{Laine+14}. In general, Type-I profiles are rare in all galaxies except irregulars and dwarfs, Type-II profiles dominate in disc galaxies, and Type-III profiles are common in all galaxies types.

\begin{figure}
    \includegraphics[width=\columnwidth]{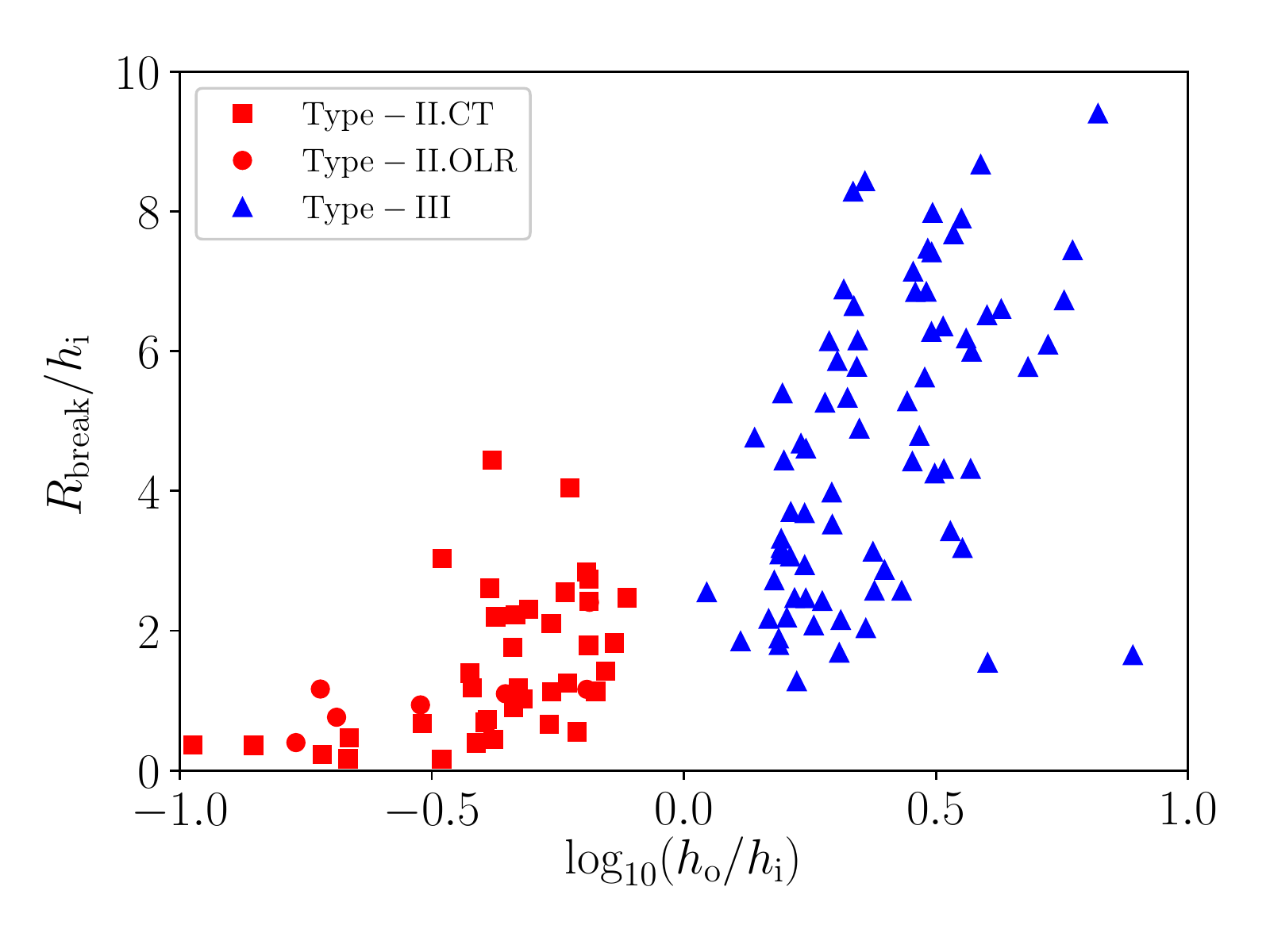}
    \caption{The relation between the ratio of the inner break radius to the inner scale-length, $R_{break}/h_i$ with the log of the ratio of the outer scale-length to the inner scale-length, log$_{10}(h_0/h_i)$ for each EDGES galaxy. The data are classified according to the break-type of the first break.}
    \label{fig:break}
\end{figure}

To further investigate the break statistics in EDGES, Figure~\ref{fig:break} presents the break radius, $R_{\rm break}$ normalized by the scalelength interior to the break $h_{\rm i}$, versus the log of the ratio between the interior scalelength and the outer scalelength, $h_{\rm o}$. The data-points are identified by their break type, either Type-II.CT, Type-II.OLR, Type-III.S, or Type-III.O as discussed in $\S~\ref{sec:break_class}$. The x-axis represents the strength and type of the break; large values for strong breaks, small values for weak breaks, and positive values for Type-III breaks, and negative values for Type-II breaks. The y-axis is analogous to the location of the break, normalised by the physical size of the galaxy (this normalisation is important because discs with high-valued scalelengths tend to be larger as they reach the surface brightness limit more gradually). These data are similar to the results in \cite{Laine+14} with a few key differences. Due to the greater depth of EDGES a larger number of Type-III breaks are detected farther out in the disc, and with a slightly larger difference between interior and outer scalelengths. This causes the overall break statistics to be more heavily weighted towards Type-III breaks (in $45\%$ of EDGES galaxies the first break is Type-III versus $38\%$ in \cite{Gutierrez+11} and $21\%$ in \cite{Laine+14}), and our results are a continuation of the general trend in Figure~\ref{fig:break}.  

\subsection{Mass analysis}

\begin{figure}
    \includegraphics[width=\columnwidth]{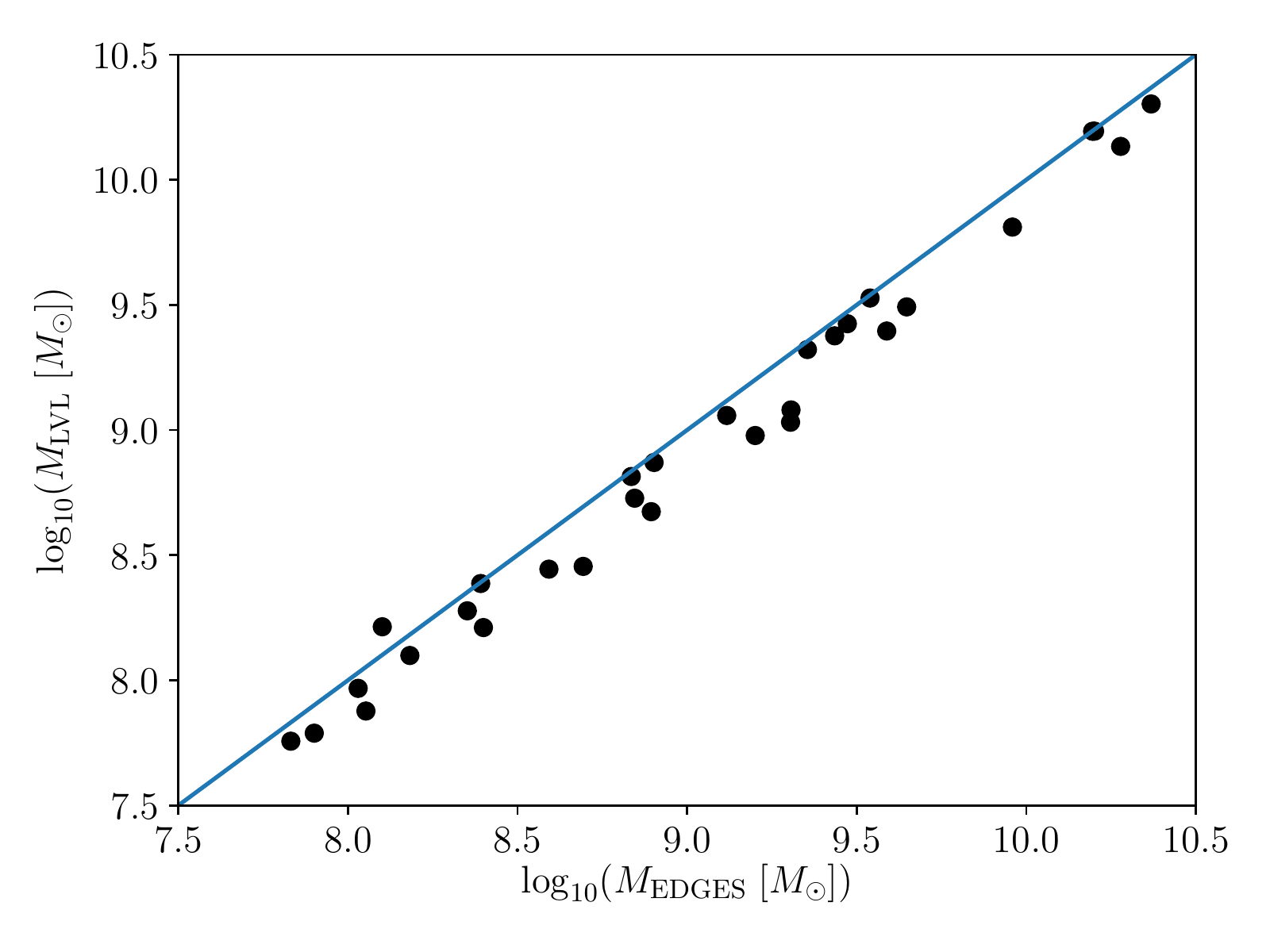}
    \caption{A comparison of LVL derived masses versus EDGES derived masses.}
    \label{fig:lvl_comparison}
\end{figure}

\begin{figure}
    \includegraphics[width=\columnwidth]{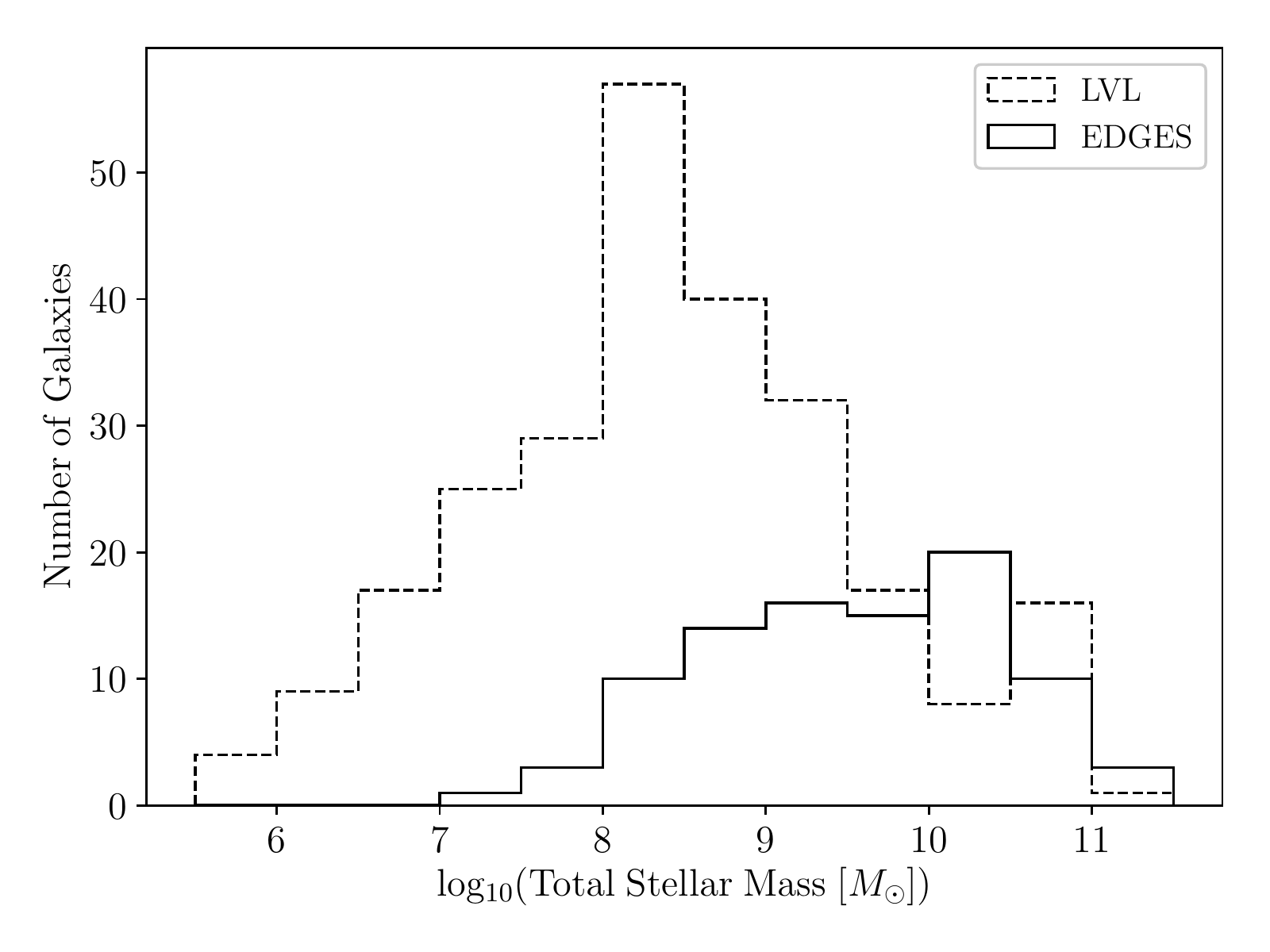}
    \caption{The distribution of total mass for every galaxy in EDGES (solid line) and in the LVL survey (dashed line).}
    \label{fig:total_mass_hist}
\end{figure}

The total stellar mass within each galaxy is found by summing the components of each model from Equation~\ref{eq:final_model}. To verify these results, we compare the total stellar mass for the galaxies which appear in EDGES and also in LVL \citep{Dale+09} in Figure~\ref{fig:lvl_comparison}. The LVL masses are found with Equation~\ref{eq:mass_derivation} along with the integrated flux from \citep{Dale+09}, and distances from the Extragalactic Distance Database \citep{Tully+09}. We find that, on average, the total stellar mass found in EDGES is larger than the total stellar mass in the LVL sample. This offset is due to the differences in integration techniques, and the depth of the data; EDGES integrates a model to 200~kpc, whereas in LVL the total flux was calculated using aperture photometry which encompasses the shallower LVL emission (which averages to 1.13~$R_{25}$). In addition, we have computed the stellar mass for the entirety of LVL to showcase the differences between EDGES and a volume-limited sample in Figure~\ref{fig:total_mass_hist}. In general, EDGES galaxies are more massive, and have a flatter distribution than the galaxies found within LVL. This is due to the sampling strategy (see \S~\ref{sec:sample}), and subsequent lack of dwarf-irregular galaxies in EDGES in comparison to LVL.
    
\begin{figure}
    \includegraphics[width=\columnwidth]{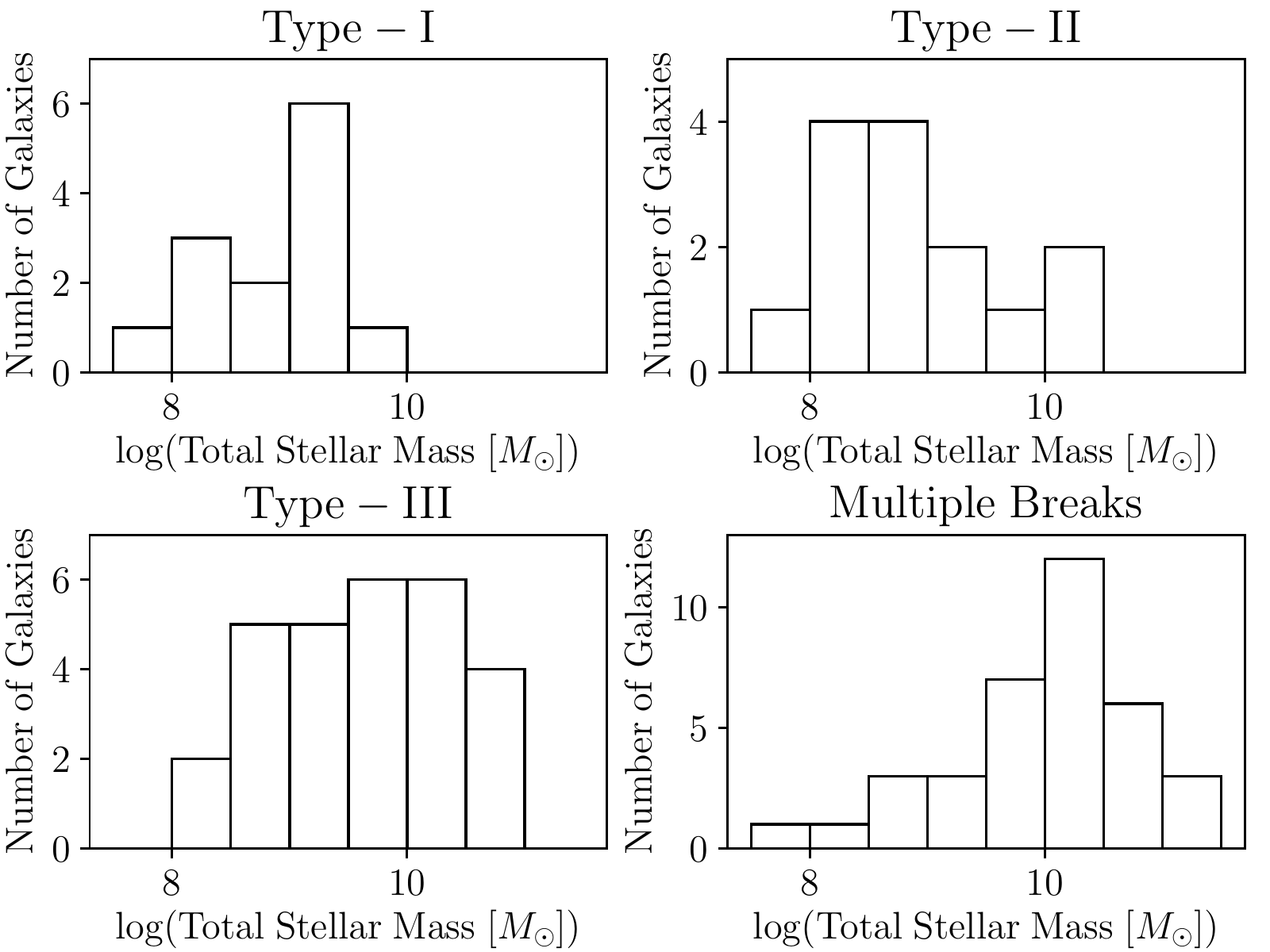}
    \caption{The distribution of total mass for the EDGES sample, subdivided by galaxies with a single Type-I break (top-left), a single Type-II break (top-right), a single Type-III break (bottom-left), and for galaxies with multiple breaks (bottom-right).}
    \label{fig:mass_type_hist}
\end{figure}

In Figure~\ref{fig:mass_type_hist} we plot the distribution of total stellar mass according to the galaxy break-type. In this case, galaxies with no breaks are considered Type-I, galaxies with a single down-bending break are Type-II, galaxies with a single up-bending break are Type-III, and galaxies with more than one break are classified as ``Multiple Breaks''. In general, Type-I galaxies have the least stellar mass, followed by Type-II galaxies, Type-III galaxies, and galaxies with multiple breaks have the greatest stellar mass.

\begin{figure}
    \includegraphics[width=\columnwidth]{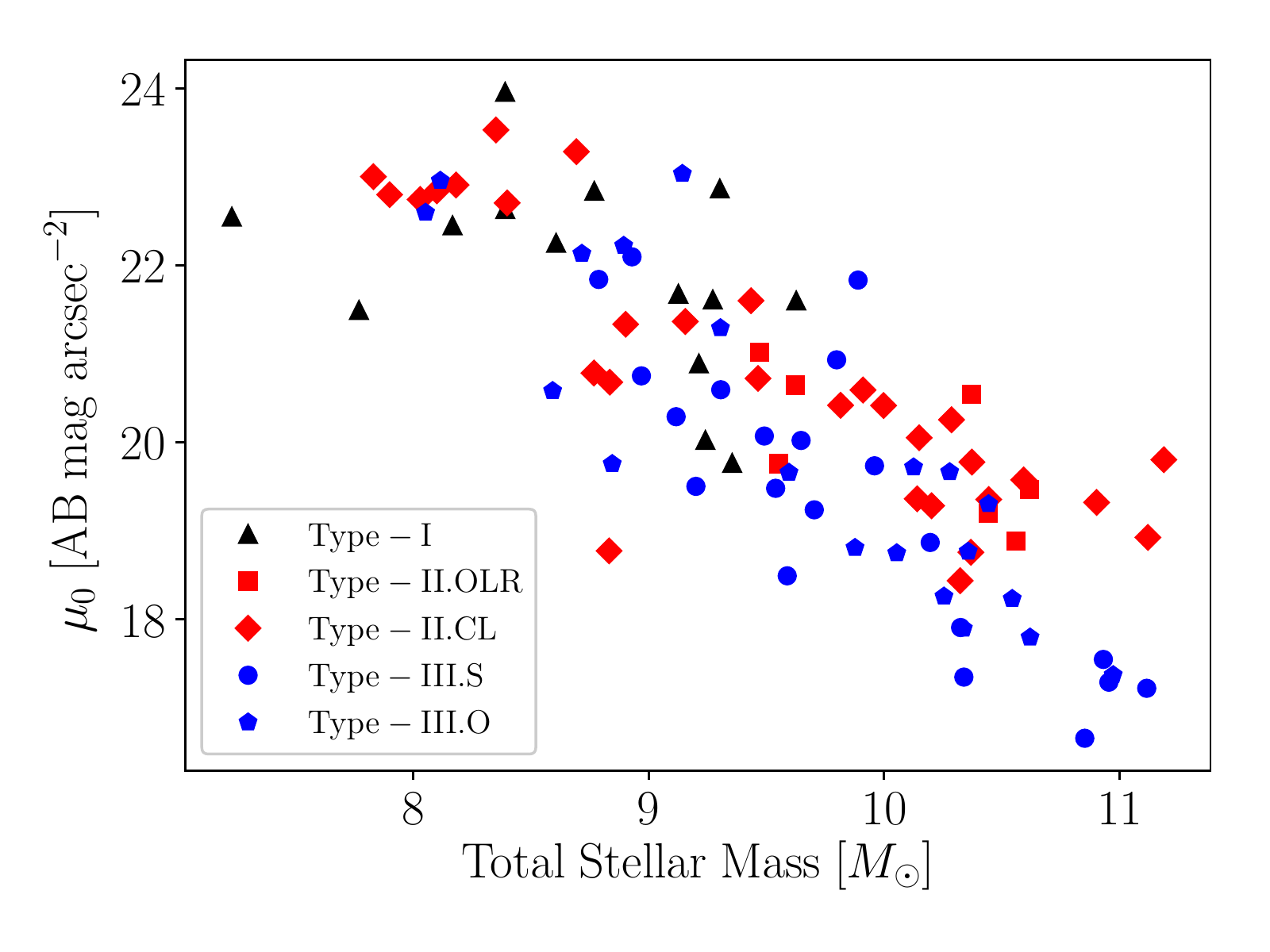}
    \caption{The dependence of the central surface brightness, $\mu_0$, of the first S{\'e}rsic disc on total stellar mass for each EDGES galaxy. The data are classified according to the break-type of the first break.}
    \label{fig:mu_mass}
\end{figure}

\begin{figure}
    \includegraphics[width=\columnwidth]{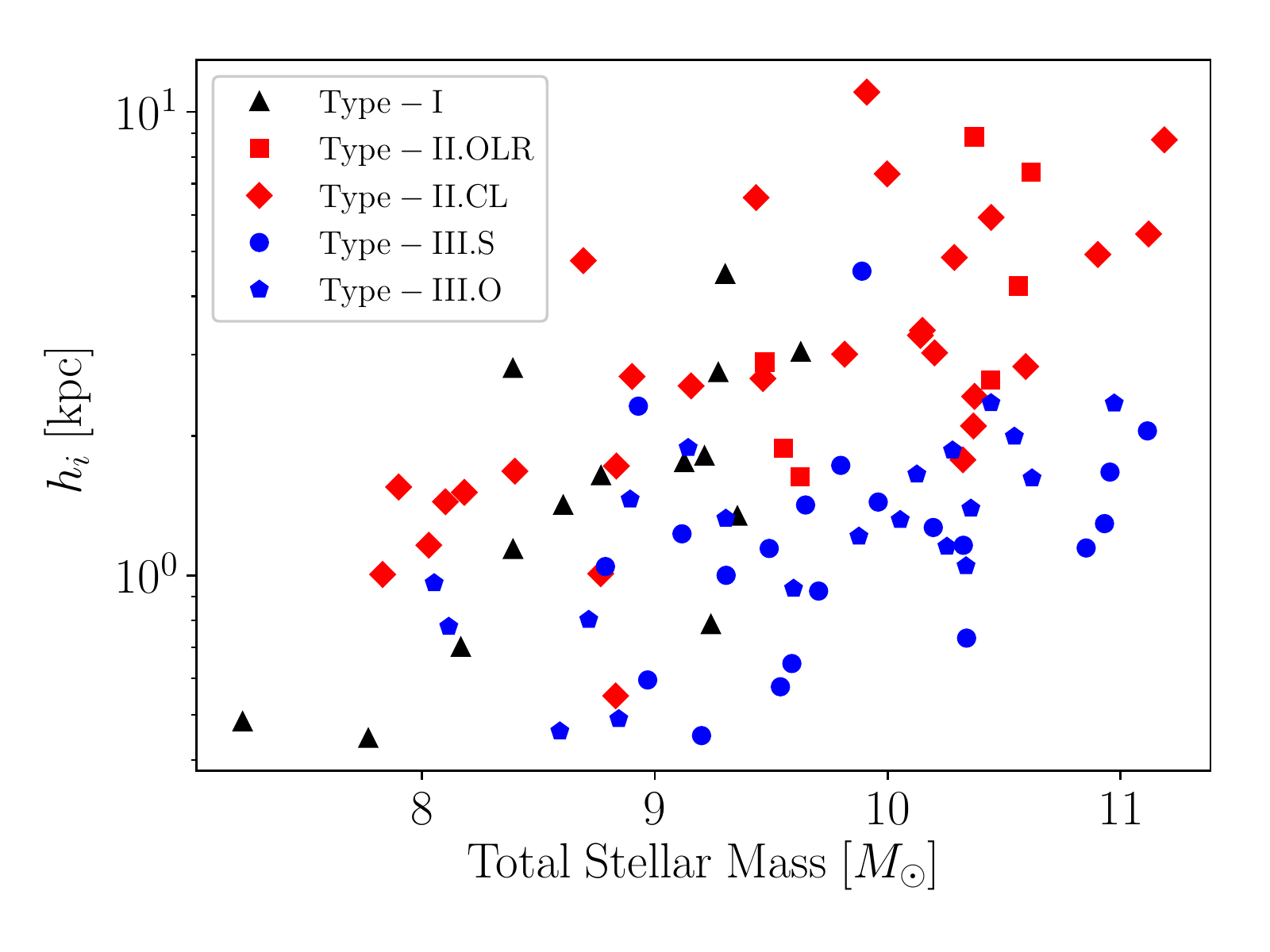}
    \caption{The dependence of the inner scale-length, $h_i$, on total stellar mass for each EDGES galaxy. The data are classified according to the break-type of the first break.}
    \label{fig:h_mass}
\end{figure}

These differences are coupled with the relation of galaxy break type with central surface brightness and scale-length of the disc components. In Figure~\ref{fig:mu_mass} we plot the relation between the central surface brightness of the inner-most disc and total stellar mass. EDGES galaxies with high stellar mass tend to have the brightest disc central surface brightness. In Figure~\ref{fig:h_mass} we plot the relation between the scale-length of the inner-most disc, and total stellar mass. This relation is relatively weak, but positive, where galaxies with high total stellar mass have longer scale-lengths. Both of these relations are found in the optically-based SDSS \citep{Gadotti+09}, as well as in the near-infrared based S$^4$G survey \citep{Laine+16}. In addition, we differentiate each point by the type of galaxy as defined above. Type-I galaxies tend to have low stellar mass, with dim cores, and short scale lengths. Type-II galaxies have slightly higher mass, brighter cores, and longer scale-lengths. Type-III galaxies have even higher mass, brighter cores, and longer scale-lengths. Galaxies with multiple breaks have the highest mass with the brightest cores and the longest scale-lengths.

\subsection{Stellar halos}
\label{sec:stellar_halos}

The Type-III, or up-bending, breaks in the EDGES sample may be caused by a disc which is distinct from the inner disc, a stellar halo, substructure in dwarf galaxies, the extended wings of the EDGES PSF, or by variance as the data approach the noise-floor. We address the extended wings possibility in \S~\ref{sec:exten_psf}, and while the noise-floor may cause up-bending breaks, these breaks are not statistically significant, and do not occur often in our analysis after the disc has been correctly marked (see \S~\ref{sec:final_model}). This leaves distinct discs, stellar halos, and substructure from dwarf galaxies. While distinct discs and substructure within dwarfs have implications for the dynamic processes which form and shape galaxies, we focus on stellar halos as they have a direct link to our broader, cosmological, understanding of the Universe. As discussed in the introduction, the ratio of the mass in the stellar halo to the total stellar mass of a galaxy follows a relation predicted by the latest cosmological models of galaxy evolution \citep{Purcell+07, Cooper+13, Rodriguez-Gomez+15}. The prohibitive depth required to detect stellar halos ($\mu$>28~AB~mag~arcsec$^{-2}$) makes verification of the results from simulations difficult. With the depth of EDGES, some percent of the Type-III breaks are likely associated with stellar halos. In this section we consider if a Type-III break is the result of a stellar halo, or if the break is associated with another cause.

A method to determine where discs end and stellar halos begin is with resolved stellar populations. Stellar halos are populated with stars born from multiple galaxies while disc stars are generally born within the parent galaxy, causing a myriad of differences between the stellar populations of disc and halo stars. The GHOSTS survey has used this strategy of differentiating between discs and halos with stellar populations to investigate stellar halos within their sample \citep{Bailin+11, Harmsen+17}. This method has also found success in detecting ultra-faint dwarfs \citep{Smercina+17}. Differing stellar populations are also reflected within radial optical color differences \citep{Chonis+11,Dale+16}. Lacking resolved stellar populations and optical colors, we compare our solely mass-based observations with predictions from simulations \citep{Cooper+13, Rodriguez-Gomez+15} and observations of aggregated stellar halos in SDSS \citep{DSouza+14}.

\begin{figure}
    \includegraphics[width=\columnwidth]{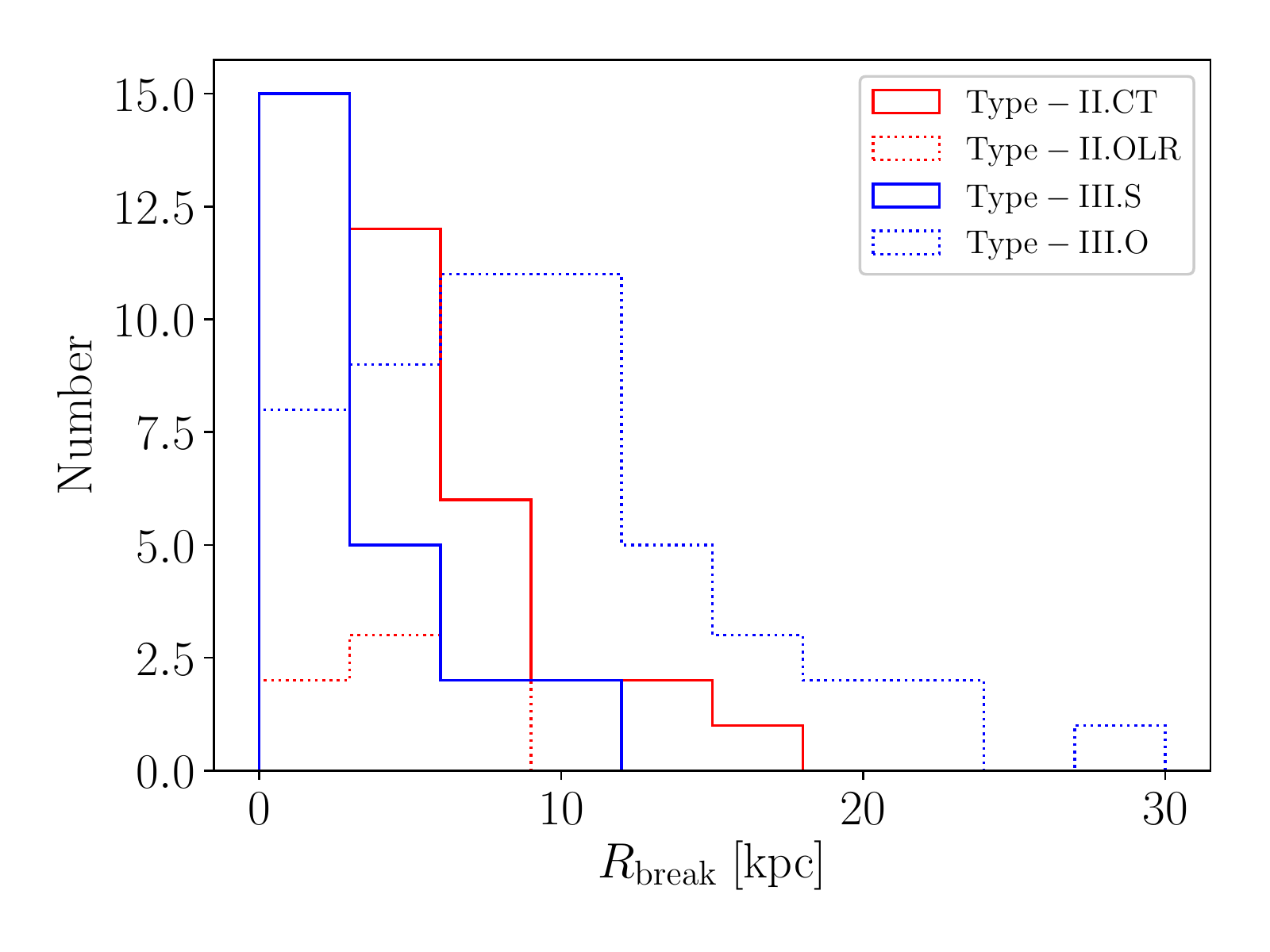}
    \caption{The distributions of break radii, $R_{\rm break}$. These distributions are grouped by break-type, with a red solid line representing Type-II.CT breaks, a red dashed line representing Type-II.OLR, a blue solid line representing Type-III.S, and a blue dashed line representing Type-III.O}
    \label{fig:break_hist}
\end{figure}

\begin{figure}
    \includegraphics[width=\columnwidth]{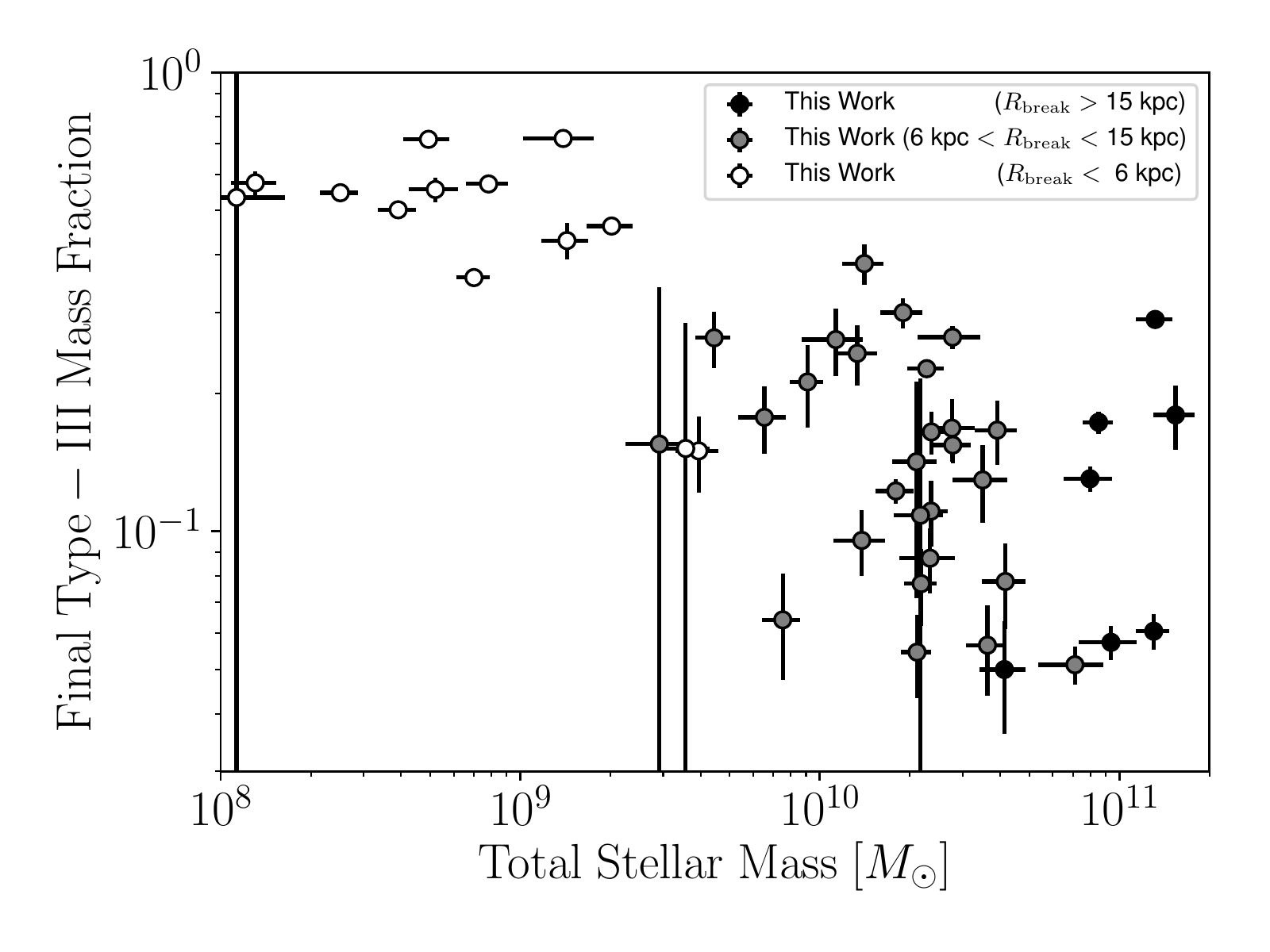}
    \caption{The dependence of stellar mass fraction of disc components which occur after a final Type-III break on total stellar mass. These data are subdivided according to the radius at which the Type-III break occurs, with filled circles being Type-III breaks found beyond 15~kpc, grey circles are found between 6 and 15~kpc, and white circles are found under 6~kpc.}
    \label{fig:type_III_frac}
\end{figure}

Type-III breaks associated with stellar halos are beyond the disc. Therefore, we only consider breaks beyond the observed spiral structure, Type-III.O. Also, for galaxies with multiple breaks, we only consider the final breaks (those which occur at the largest radii). In Figure~\ref{fig:type_III_frac} we plot component mass fraction (the mass integrated from the final Type-III.O component over total stellar mass) versus total stellar mass. These data are subdivided according to the distribution of Type-III break radii (Figure~\ref{fig:break_hist}). Type-III breaks with radii less than the peak of the distribution at $\sim6$~kpc are shown in white, galaxies above 6~kpc but below the tail at 15~kpc are shown in gray, and any break in the tail of the distribution, at 15~kpc, is shown as black. The white points are generally associated with the dwarf galaxies within the EDGES sample, with total stellar masses ranging from $\sim10^8-2\times10^9~M_{\odot}$. These breaks are most likely caused by substructures within these small, often disturbed, galaxies and are not necessarily caused by stellar halos. The twenty six grey points fall nicely within the locus of the simulations of \citep{Cooper+13}. While it is possible that these breaks are associated with stellar halos, they appear much closer to the center of their galaxies than simulations predict \citep{Purcell+07, Cooper+13, Rodriguez-Gomez+16}. A survey of the stellar populations found within these galaxies would determine if these are due to stellar halos or some other influence. The seven black points, the galaxies with final Type-III.O breaks beyond 15~kpc are retained as candidate galaxies with detected stellar halos. These galaxies are NGC~3675, NGC~3953, M~109, NGC~4013, NGC~4100, NGC~5005, and M~63. 

\begin{figure}
    \includegraphics[width=\columnwidth]{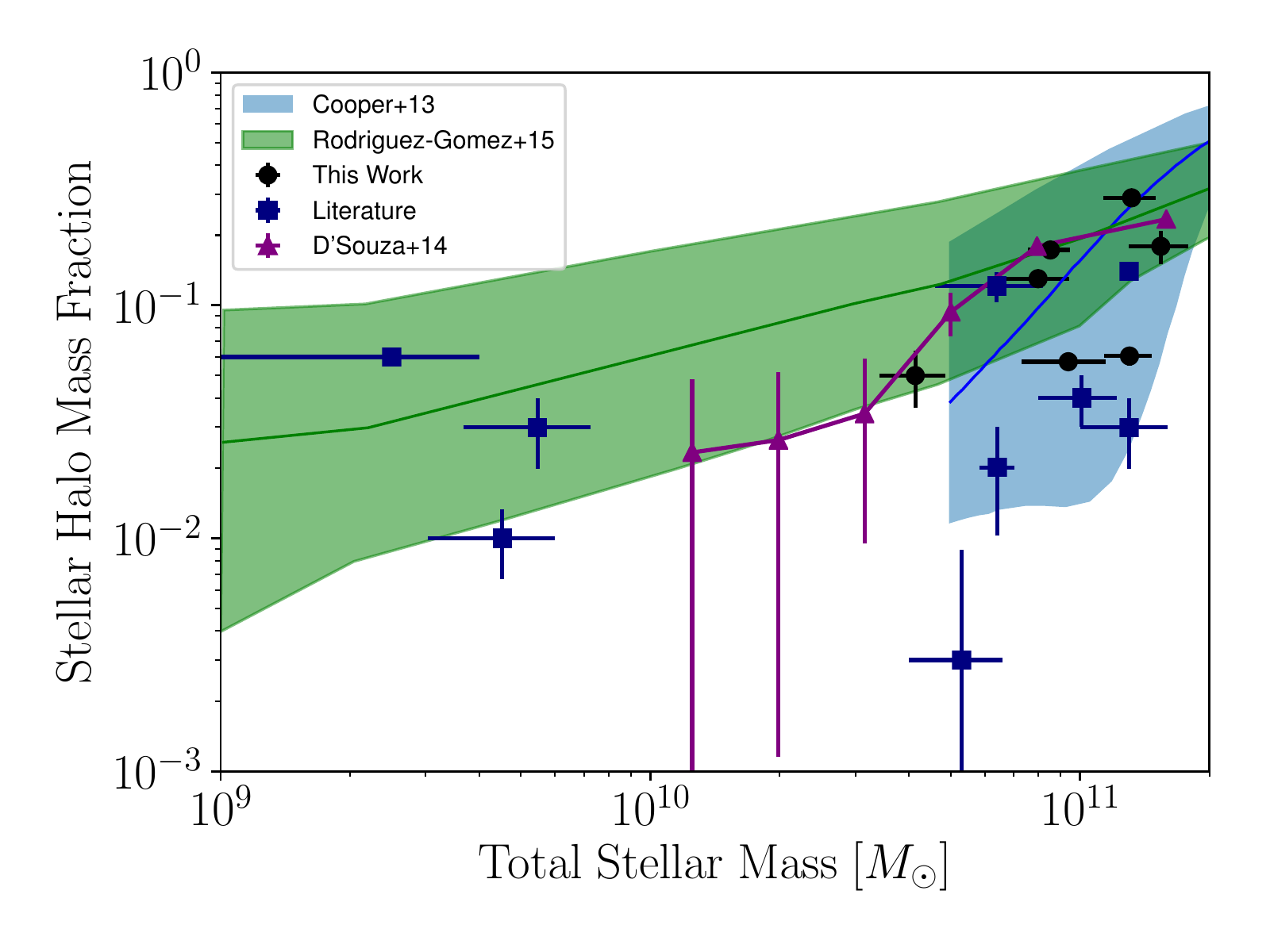}
    \caption{The dependence of stellar halo mass fraction on total stellar mass. The blue polygon represents the model from \protect\cite{Cooper+13}, where the middle line is the median, nd the lower and upper limits are a single standard deviation above and below the median. The green polygon is similarly defined and represents the model from \protect\cite{Rodriguez-Gomez+15}. The literature data-points (see \S~\ref{sec:stellar_halos}) are represented by blue squares. The aggregated stellar halo measurements from \protect\cite{DSouza+14} are purple triangles, and the 7 galaxies with stellar halo candidates from this work are represented by filled circles.
    }
    \label{fig:halo_frac}
\end{figure}

The ratio of the mass due to final Type-III.O breaks to the total stellar mass of the galaxy versus the total stellar mass is shown in Figure~\ref{fig:halo_frac}. We include the predictions from the simulations of \cite{Cooper+13} and \cite{Rodriguez-Gomez+15}, the average stellar halo in SDSS from \cite{DSouza+14}, and a sampling of observed stellar halos found within the literature: M~101 \citep{vanDokkum+14}, the Milky Way \citep{Carollo+10}, M~31 \citep{Courteau+11}, M~63 \citep{Staudaher+15}, NGC~0253 \citep{Bailin+11}, M~33 \citep{McConnachie+10}, NGC~2403 \citep{Barker+12}, NGC~3115 \citep{Peacock+15}, and UGC~00180 \citep{Trujillo+16}. In general, the literature points appear below the median values of the simulations. However, there is significant scatter amid the few literature points, and given the stochastic nature of merger events \citep{Bullock+05, DSouza+Bell18} a more robust sample is needed. Our seven candidate stellar halos are also included in this figure as black points. The EDGES data fall within a single standard deviation of the predictions from the models.

\begin{figure}
    \includegraphics[width=\columnwidth]{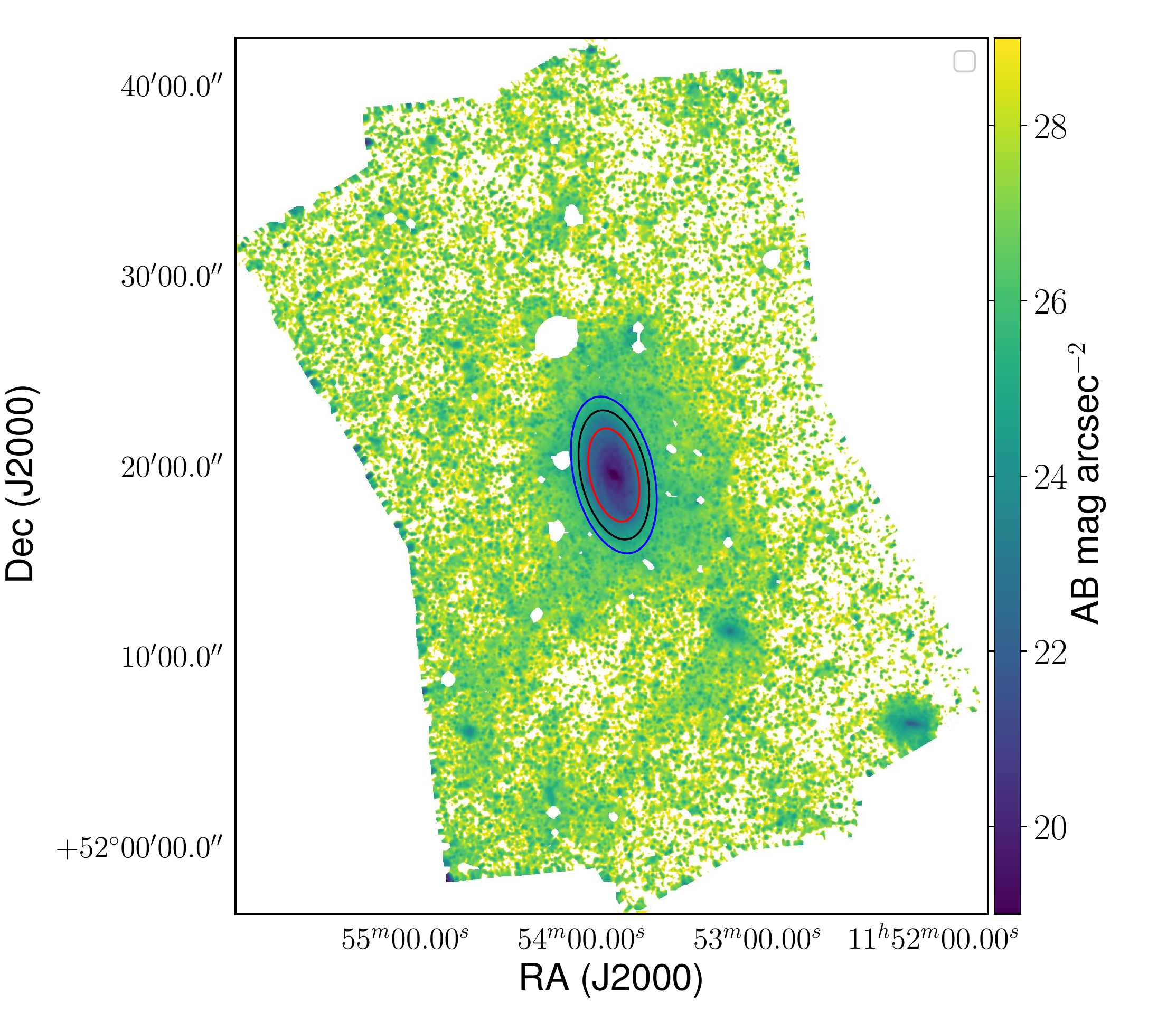}
    \caption{The full 3.6~\um\ mosaic of NGC~3953. Note the looping tidal stream south of the disc.}
    \label{fig:N3953 Stream}
\end{figure}

Another distinctive characteristic of stellar halos is highly structured filaments; tidal streams caused by the stochastic accretion of satellite galaxies \citep{Cooper+13, Rogriguez-Gomez+16} due to stochastic accretion of a handful of high mass systems \citep{Bullock+05}. We find similarly distinctive structures in the seven candidate galaxies, especially in M~63, NGC~4013, and NGC~3953, which either have known tidal streams in the cases of M~63 \citep{Chonis+11, Staudaher+15} and NGC~4013 \citep{Martinez-Delgado+09}, or their stellar streams are newly discovered in the EDGES data (for NGC~3953, see Figure~\ref{fig:N3953 Stream}). Barring new data, the Type-III.O breaks beyond 15~kpc within these seven galaxies are consistent with stellar halos.

\section{SUMMARY}

In this paper we describe the observational strategy for the EDGES survey, along with the data processing techniques used to create deep 3.6~\um\ mosaics for each of the 92 galaxies within the sample. The surface brightness profiles of these mosaics reach a depth of 28~AB~mag~arcsec$^{-2}$. We model these profiles as a combination of distinct S{\'e}rsic discs, along with a bulge component.

\begin{enumerate}
	\item We find many more galaxies with multiple breaks (where the slope of the profile increases or decreases) than previous, shallower, studies. 
	\item The type of the break depends upon the radius at which the break occurs, and also on whether the break occurred first, second, or third. In general, Type-II breaks occur closer to the cores of galaxies, and Type-III breaks occur further out. 
	\item Galaxies without breaks are preferentially lower mass than galaxies with breaks. In addition, galaxies with a single Type-II break are less massive than galaxies with a single Type-III break. Galaxies with multiple breaks tend to be the most massive galaxies in the EDGES sample.
	\item Type-III breaks are associated with stellar halos for at least seven galaxies in our sample, with the possibility of more. In general, the fraction of the mass within these halos agrees with simulations, but many galaxies in EDGES show no evidence for stellar halos despite the depth of the survey.
	\item A new tidal stream is discovered near NGC~3953.
\end{enumerate}

\section*{Acknowledgments}
This work is based [in part] on observations made with the Spitzer Space Telescope, which is operated by the Jet Propulsion Laboratory, California Institute of Technology under a contract with NASA. Support for this work was provided by NASA through an award issued by JPL/Caltech. This research has made use of the NASA/IPAC Extragalactic Database (NED) which is operated by the Jet Propulsion Laboratory, California Institute of Technology, under contract with the National Aeronautics and Space Administration. We acknowledge the usage of the HyperLeda database (http://leda.univ-lyon1.fr).
\clearpage
\onecolumn
\begin{deluxetable}{ lccccccccc }
\tablecolumns{ 10 }
\tablewidth{ 0pc }
\tablecaption{  }
\tablehead{
   \colhead{ Galaxy } & \colhead{ $\alpha_0$ } & \colhead{ $\delta_0$ } & \colhead{ $cz$ } & \colhead{ $m-M$ } & \colhead{ $m_B$ } & \colhead{ $d_{25}$ } & \colhead{ $b/a$ } & \colhead{ PA } & \colhead{ $T$ } \\
   \colhead{ } & \colhead{ J2000 } & \colhead{ J2000 } & \colhead{ (km~s$^{-1}$) } & \colhead{ (mag) } & \colhead{ (mag) } & \colhead{ (\arcsec) } & \colhead{  } & \colhead{ ($\degr$) } & \colhead{  }
}
\startdata
NGC0024        & 00:09:56.5 & $-$24:57:47 & 554 & 29.42 $\pm$ 0.09 & 12.0 $\pm$ 0.1 & 345 & 0.23 & 46 & 5.1 $\pm$ 0.4 \\
NGC0059        & 00:15:25.1 & $-$21:26:40 & 362 & 28.4 $\pm$ 0.1 & 13.1 $\pm$ 0.2 & 158 & 0.50 & 127 & -2.9 $\pm$ 0.7 \\
NGC0625        & 01:35:04.6 & $-$41:26:10 & 396 & 27.73 $\pm$ 0.09 & 11.8 $\pm$ 0.2 & 345 & 0.33 & 92 & 9 $\pm$ 1 \\
UGC05829       & 10:42:41.9 & $+$34:26:56 & 629 & 29.0 $\pm$ 0.2 & 14.0 $\pm$ 0.9 & 281 & 0.89 & \nodata & 9.8 $\pm$ 0.6 \\
NGC3344        & 10:43:31.2 & $+$24:55:20 & 581 & 30.0 $\pm$ 0.1 & 10.8 $\pm$ 0.2 & 425 & 0.91 & 150 & 4.0 $\pm$ 0.3 \\
NGC3486        & 11:00:23.9 & $+$28:58:30 & 682 & 30.7 $\pm$ 0.4 & 10.8 $\pm$ 0.2 & 425 & 0.74 & 80 & 5.2 $\pm$ 0.7 \\
UGC06112       & 11:02:35.3 & $+$16:44:05 & 1034 & 28.4 $\pm$ 0.1 & 13.9 $\pm$ 0.2 & 151 & 0.32 & 123 & 7.4 $\pm$ 0.8 \\
NGC3501        & 11:02:47.3 & $+$17:59:22 & 1131 & 28.4 $\pm$ 0.1 & 13.6 $\pm$ 0.2 & 233 & 0.13 & 27 & 5.8 $\pm$ 0.5 \\
NGC3507        & 11:03:25.4 & $+$18:08:08 & 980 & 31.9 $\pm$ 0.2 & \nodata & 203 & 0.85 & 110 & 3.1 $\pm$ 0.4 \\
UGC06161       & 11:06:49.2 & $+$43:43:24 & 757 & 30.3 $\pm$ 0.2 & 14.0 $\pm$ 0.3 & 154 & 0.48 & 40 & 7.9 $\pm$ 0.9 \\
UGC06399       & 11:23:23.2 & $+$50:53:34 & 792 & 31.6 $\pm$ 0.2 & 14.5 $\pm$ 0.3 & 165 & 0.28 & 142 & 9 $\pm$ 1 \\
NGC3675        & 11:26:08.6 & $+$43:35:09 & 770 & 31.5 $\pm$ 0.2 & 10.8 $\pm$ 0.1 & 353 & 0.53 & 178 & 3.0 $\pm$ 0.5 \\
NGC3718        & 11:32:34.9 & $+$53:04:05 & 994 & 28.8 $\pm$ 0.1 & 11.3 $\pm$ 0.2 & 488 & 0.49 & 15 & 1.1 $\pm$ 0.6 \\
NGC3726        & 11:33:21.1 & $+$47:01:45 & 866 & 30.6 $\pm$ 0.2 & 11.0 $\pm$ 0.2 & 370 & 0.69 & 10 & 5.1 $\pm$ 0.6 \\
NGC3729        & 11:33:49.3 & $+$53:07:32 & 1061 & 31.6 $\pm$ 0.3 & 12.3 $\pm$ 0.1 & 169 & 0.68 & 15 & 1.2 $\pm$ 0.7 \\
NGC3769        & 11:37:44.1 & $+$47:53:35 & 738 & 31.1 $\pm$ 0.4 & 12.3 $\pm$ 0.2 & 185 & 0.32 & 152 & 3.4 $\pm$ 0.9 \\
UGC06628       & 11:40:05.7 & $+$45:56:32 & 842 & 28.8 $\pm$ 0.1 & 13.9 $\pm$ 0.5 & 173 & 1.0  & 70 & 9 $\pm$ 1 \\
NGC3893        & 11:48:38.2 & $+$48:42:39 & 968 & 31.3 $\pm$ 0.3 & 10.8 $\pm$ 0.2 & 268 & 0.62 & 165 & 5.2 $\pm$ 0.8 \\
UGC06792       & 11:49:23.3 & $+$39:46:17 & 842 & 31.6 $\pm$ 0.2 & 14.6 $\pm$ 0.2 & 165 & 0.15 & 172 & 5.9 $\pm$ 0.5 \\
NGC3917        & 11:50:45.4 & $+$51:49:29 & 965 & 31.1 $\pm$ 0.2 & 12.4 $\pm$ 0.1 & 308 & 0.24 & 77 & 5.9 $\pm$ 0.5 \\
NGC3922        & 11:51:13.4 & $+$50:09:25 & 907 & 31.1 $\pm$ 0.4 & 13.8 $\pm$ 0.2 & 104 & 0.44 & 38 & -0.1 $\pm$ 0.4 \\
NGC3938        & 11:52:49.5 & $+$44:07:15 & 810 & 28.8 $\pm$ 0.1 & 10.9 $\pm$ 0.1 & 322 & 0.91 & 29 & 5.2 $\pm$ 0.5 \\
NGC3941        & 11:52:55.4 & $+$36:59:11 & 929 & 30.5 $\pm$ 0.1 & 11.3 $\pm$ 0.1 & 208 & 0.66 & 10 & -2.0 $\pm$ 0.4 \\
NGC3949        & 11:53:41.7 & $+$47:51:31 & 801 & 31.4 $\pm$ 0.2 & 11.5 $\pm$ 0.2 & 173 & 0.57 & 120 & 4.0 $\pm$ 0.1 \\
NGC3953        & 11:53:48.9 & $+$52:19:36 & 1053 & 31.4 $\pm$ 0.2 & 10.8 $\pm$ 0.1 & 415 & 0.50 & 13 & 4.0 $\pm$ 0.2 \\
UGC06900       & 11:55:39.7 & $+$31:31:07 & 590 & 28.8 $\pm$ 0.8 & 14.4 $\pm$ 0.5 & 125 & 0.62 & 115 & 9.7 $\pm$ 0.7 \\
NGC3972        & 11:55:45.1 & $+$55:19:15 & 853 & 31.6 $\pm$ 0.2 & 12.8 $\pm$ 0.2 & 233 & 0.28 & 120 & 4.0 $\pm$ 0.3 \\
UGC06917       & 11:56:26.5 & $+$50:25:43 & 911 & 31.4 $\pm$ 0.2 & 13.5 $\pm$ 0.3 & 213 & 0.57 & 130 & 9 $\pm$ 1 \\
UGC06930       & 11:57:17.4 & $+$49:16:59 & 778 & 28.8 $\pm$ 0.1 & 13.5 $\pm$ 0.5 & 262 & 0.63 & 47 & 6.6 $\pm$ 0.8 \\
MESSIER109     & 11:57:36.0 & $+$53:22:28 & 1049 & 32.1 $\pm$ 0.2 & 10.5 $\pm$ 0.1 & 455 & 0.62 & 68 & 4.0 $\pm$ 0.2 \\
NGC3998        & 11:57:56.1 & $+$55:27:13 & 1041 & 30.8 $\pm$ 0.1 & 11.4 $\pm$ 0.1 & 162 & 0.83 & 140 & -2.1 $\pm$ 0.5 \\
UGC06955       & 11:58:29.8 & $+$38:04:35 & 906 & 31.4 $\pm$ 0.2 & 14.2 $\pm$ 0.8 & 301 & 0.51 & 70 & 9.8 $\pm$ 0.6 \\
NGC4013        & 11:58:31.4 & $+$43:56:48 & 832 & 31.6 $\pm$ 0.2 & 12.3 $\pm$ 0.2 & 315 & 0.20 & 66 & 3.1 $\pm$ 0.5 \\
IC0749         & 11:58:34.0 & $+$42:44:03 & 808 & 28.8 $\pm$ 0.1 & 13.0 $\pm$ 0.2 & 141 & 0.81 & 150 & 5.9 $\pm$ 0.5 \\
NGC4010        & 11:58:37.9 & $+$47:15:41 & 902 & 31.4 $\pm$ 0.2 & 13.0 $\pm$ 0.2 & 256 & 0.19 & 66 & 6.8 $\pm$ 0.9 \\
IC0750         & 11:58:52.2 & $+$42:43:21 & 701 & 31.9 $\pm$ 0.4 & 12.7 $\pm$ 0.1 & 158 & 0.45 & 43 & 2.2 $\pm$ 0.7 \\
UGC06983       & 11:59:09.3 & $+$52:42:27 & 1083 & 31.6 $\pm$ 0.3 & 13.8 $\pm$ 0.5 & 208 & 0.69 & 85 & 5.9 $\pm$ 0.5 \\
NGC4026        & 11:59:25.2 & $+$50:57:42 & 931 & 30.8 $\pm$ 0.1 & 11.6 $\pm$ 0.1 & 315 & 0.24 & 178 & -1.8 $\pm$ 0.9 \\
NGC4051        & 12:03:09.6 & $+$44:31:53 & 701 & 30.2 $\pm$ 0.2 & 11.3 $\pm$ 0.2 & 315 & 0.74 & 135 & 4.0 $\pm$ 0.1 \\
NGC4068        & 12:04:00.8 & $+$52:35:18 & 210 & 28.2 $\pm$ 0.1 & 13.0 $\pm$ 0.2 & 199 & 0.53 & 30 & 9.9 $\pm$ 0.3 \\
NGC4085        & 12:05:22.7 & $+$50:21:11 & 746 & 31.6 $\pm$ 0.2 & 12.9 $\pm$ 0.1 & 169 & 0.28 & 78 & 5.2 $\pm$ 0.5 \\
NGC4088        & 12:05:34.2 & $+$50:32:21 & 757 & 30.8 $\pm$ 0.2 & 11.1 $\pm$ 0.1 & 345 & 0.39 & 43 & 4.7 $\pm$ 0.9 \\
UGC07089       & 12:05:57.7 & $+$43:08:36 & 770 & 30.3 $\pm$ 0.2 & 14.4 $\pm$ 0.3 & 194 & 0.20 & 36 & 7.9 $\pm$ 0.8 \\
NGC4096        & 12:06:01.1 & $+$47:28:42 & 566 & 30.5 $\pm$ 0.2 & 11.6 $\pm$ 0.1 & 396 & 0.27 & 20 & 5.3 $\pm$ 0.6 \\
NGC4100        & 12:06:08.5 & $+$49:34:58 & 1075 & 31.5 $\pm$ 0.2 & 11.6 $\pm$ 0.1 & 322 & 0.33 & 167 & 4.1 $\pm$ 0.6 \\
NGC4102        & 12:06:23.0 & $+$52:42:40 & 847 & 31.6 $\pm$ 0.2 & 11.9 $\pm$ 0.1 & 181 & 0.57 & 38 & 3.1 $\pm$ 0.5 \\
NGC4111        & 12:07:03.1 & $+$43:03:57 & 808 & 29.9 $\pm$ 0.1 & 11.6 $\pm$ 0.1 & 274 & 0.21 & 150 & -1 $\pm$ 1 \\
UGC07125       & 12:08:42.3 & $+$36:48:10 & 1072 & 32.7 $\pm$ 0.4 & 14.2 $\pm$ 0.3 & 281 & 0.16 & 85 & 8.8 $\pm$ 0.7 \\
NGC4138        & 12:09:29.8 & $+$43:41:07 & 889 & 31.0 $\pm$ 0.1 & 12.1 $\pm$ 0.1 & 154 & 0.66 & 150 & 0 $\pm$ 1 \\
NGC4143        & 12:09:36.1 & $+$42:32:03 & 959 & 31.0 $\pm$ 0.1 & 12.0 $\pm$ 0.1 & 138 & 0.63 & 144 & -2.0 $\pm$ 0.6 \\
NGC4144        & 12:09:58.6 & $+$46:27:26 & 265 & 29.17 $\pm$ 0.09 & 12.1 $\pm$ 0.1 & 362 & 0.22 & 104 & 5.9 $\pm$ 0.5 \\
NGC4151        & 12:10:32.6 & $+$39:24:21 & 996 & 27.9 $\pm$ 0.4 & 10.9 $\pm$ 0.1 & 379 & 0.71 & 50 & 2.0 $\pm$ 0.5 \\
NGC4183        & 12:13:16.9 & $+$43:41:55 & 931 & 31.1 $\pm$ 0.2 & 12.8 $\pm$ 0.1 & 315 & 0.15 & 166 & 5.8 $\pm$ 0.5 \\
NGC4203        & 12:15:05.1 & $+$33:11:50 & 1087 & 30.9 $\pm$ 0.1 & 11.6 $\pm$ 0.1 & 203 & 0.93 & 10 & -2.7 $\pm$ 0.7 \\
NGC4214        & 12:15:39.2 & $+$36:19:37 & 291 & 27.34 $\pm$ 0.08 & 10.2 $\pm$ 0.2 & 511 & 0.78 & 128 & 9.8 $\pm$ 0.5 \\
NGC4220        & 12:16:11.7 & $+$47:53:00 & 915 & 31.5 $\pm$ 0.2 & 12.2 $\pm$ 0.1 & 233 & 0.35 & 141 & -0.3 $\pm$ 0.9 \\
UGC07301       & 12:16:42.1 & $+$46:04:44 & 690 & 32.0 $\pm$ 0.2 & 15.5 $\pm$ 0.2 & 109 & 0.13 & 82 & 6.5 $\pm$ 0.9 \\
NGC4242        & 12:17:30.2 & $+$45:37:09 & 506 & 28.7 $\pm$ 0.2 & 11.6 $\pm$ 0.2 & 301 & 0.76 & 25 & 7.9 $\pm$ 0.5 \\
UGC07408       & 12:21:15.0 & $+$45:48:41 & 462 & 29.3 $\pm$ 0.1 & 14.3 $\pm$ 0.3 & 158 & 0.46 & 100 & 10 $\pm$ 1 \\
NGC4369        & 12:24:36.2 & $+$39:22:59 & 1046 & 29.8 $\pm$ 0.2 & 12.2 $\pm$ 0.1 & 125 & 0.98 & 175 & 1.0 $\pm$ 0.9 \\
NGC4389        & 12:25:35.1 & $+$45:41:05 & 719 & 29.9 $\pm$ 0.2 & 12.7 $\pm$ 0.1 & 158 & 0.51 & 105 & 4.1 $\pm$ 0.5 \\
UGC07577       & 12:27:40.9 & $+$43:29:44 & 195 & 27.07 $\pm$ 0.08 & 13.9 $\pm$ 0.6 & 256 & 0.56 & 130 & 9.8 $\pm$ 0.9 \\
UGC07608       & 12:28:44.2 & $+$43:13:27 & 539 & 30.2 $\pm$ 0.1 & \nodata & 203 & 0.98 & \nodata & 9.9 $\pm$ 0.5 \\
NGC4460        & 12:28:45.6 & $+$44:51:51 & 490 & 29.8 $\pm$ 0.1 & 12.3 $\pm$ 0.1 & 239 & 0.29 & 40 & 0 $\pm$ 2 \\
UGC07639       & 12:29:53.4 & $+$47:31:52 & 382 & 30.2 $\pm$ 0.1 & 14.3 $\pm$ 0.4 & 138 & 0.71 & 153 & 9.9 $\pm$ 0.4 \\
NGC4485        & 12:30:31.1 & $+$41:42:04 & 493 & 29.7 $\pm$ 0.1 & 12.5 $\pm$ 0.2 & 138 & 0.71 & 15 & 10 $\pm$ 1 \\ 
NGC4490        & 12:30:36.2 & $+$41:38:38 & 566 & 30.2 $\pm$ 0.1 & 10.3 $\pm$ 0.2 & 379 & 0.49 & 125 & 7.0 $\pm$ 0.2 \\
UGC07699       & 12:32:48.0 & $+$37:37:18 & 496 & 31.3 $\pm$ 0.4 & 13.2 $\pm$ 0.2 & 228 & 0.27 & 32 & 6.1 $\pm$ 0.6 \\
UGC07774       & 12:36:22.7 & $+$40:00:19 & 527 & 32.2 $\pm$ 0.2 & 14.8 $\pm$ 0.3 & 218 & 0.12 & 102 & 6.3 $\pm$ 0.8 \\
NGC4618        & 12:41:32.8 & $+$41:09:03 & 545 & 30.2 $\pm$ 0.1 & 11.3 $\pm$ 0.2 & 250 & 0.81 & 25 & 9 $\pm$ 1 \\
IC3687         & 12:42:15.1 & $+$38:30:12 & 354 & 28.3 $\pm$ 0.1 & 14.5 $\pm$ 0.8 & 203 & 0.89 & \nodata & 9.9 $\pm$ 0.3 \\
NGC4707        & 12:48:22.9 & $+$51:09:53 & 468 & 29.1 $\pm$ 0.1 & 14.5 $\pm$ 0.5 & 134 & 0.93 & 25 & 8.8 $\pm$ 0.5 \\
NGC4861        & 12:59:02.3 & $+$34:51:34 & 836 & 29.4 $\pm$ 0.4 & 12.5 $\pm$ 0.2 & 239 & 0.37 & 15 & 8.9 $\pm$ 0.6 \\
IC4182         & 13:05:49.5 & $+$37:36:18 & 321 & 28.22 $\pm$ 0.06 & 13.0 $\pm$ 0.7 & 362 & 0.91 & \nodata & 8.8 $\pm$ 0.5 \\
NGC5005        & 13:10:56.2 & $+$37:03:33 & 947 & 31.27 $\pm$ 0.08 & 10.9 $\pm$ 0.2 & 345 & 0.48 & 65 & 4.0 $\pm$ 0.2 \\
IC4213         & 13:12:11.2 & $+$35:40:11 & 816 & 31.4 $\pm$ 0.2 & 14.0 $\pm$ 0.2 & 151 & 0.19 & 174 & 5.8 $\pm$ 0.5 \\
NGC5023        & 13:12:12.6 & $+$44:02:28 & 407 & 29.13 $\pm$ 0.09 & 13.0 $\pm$ 0.2 & 362 & 0.13 & 28 & 5.9 $\pm$ 0.4 \\
UGC08303       & 13:13:17.6 & $+$36:13:03 & 945 & 28.7 $\pm$ 0.1 & 14.1 $\pm$ 0.4 & 134 & 0.85 & 155 & 9.9 $\pm$ 0.3 \\
NGC5033        & 13:13:27.5 & $+$36:35:38 & 876 & 31.4 $\pm$ 0.2 & 10.6 $\pm$ 0.2 & 643 & 0.47 & 170 & 5.1 $\pm$ 0.8 \\
UGC08320       & 13:14:28.0 & $+$45:55:09 & 192 & 28.1 $\pm$ 0.1 & 13.5 $\pm$ 0.3 & 218 & 0.38 & 150 & 9.9 $\pm$ 0.4 \\
MESSIER063     & 13:15:49.3 & $+$42:01:45 & 484 & 29.78 $\pm$ 0.09 & 9.6 $\pm$ 0.1 & 755 & 0.57 & 105 & 4.0 $\pm$ 0.2 \\
NGC5229        & 13:34:02.8 & $+$47:54:56 & 364 & 30.5 $\pm$ 0.2 & 14.3 $\pm$ 0.2 & 199 & 0.17 & 167 & 6.7 $\pm$ 0.7 \\
NGC5273        & 13:42:08.3 & $+$35:39:15 & 1065 & 31.1 $\pm$ 0.1 & 12.4 $\pm$ 0.1 & 165 & 0.91 & 10 & -1.9 $\pm$ 0.4 \\
UGC08839       & 13:55:25.0 & $+$17:47:42 & 958 & 31.7 $\pm$ 0.4 & 13.6 $\pm$ 0.9 & 239 & 0.68 & 120 & 9.9 $\pm$ 0.5 \\
NGC5523        & 14:14:52.3 & $+$25:19:03 & 1040 & 31.6 $\pm$ 0.2 & 12.7 $\pm$ 0.2 & 274 & 0.28 & 99 & 5.8 $\pm$ 0.5 \\
NGC5608        & 14:23:17.9 & $+$41:46:33 & 664 & 31.2 $\pm$ 0.4 & 13.9 $\pm$ 0.3 & 158 & 0.51 & 95 & 10 $\pm$ 1 \\
ESO290-G028    & 22:57:09.0 & $-$42:48:16 & 931 & 27.1 $\pm$ 0.4 & 14.3 $\pm$ 0.2 & 208 & 0.14 & 91 & 8.0 $\pm$ 0.5 \\
NGC7713        & 23:36:15.0 & $-$37:56:17 & 692 & 29.9 $\pm$ 0.2 & 11.7 $\pm$ 0.2 & 268 & 0.41 & 168 & 6.7 $\pm$ 0.8 \\
UGCA442        & 23:43:45.6 & $-$31:57:24 & 267 & 28.2 $\pm$ 0.1 & 13.4 $\pm$ 0.2 & 213 & 0.18 & 48 & 9 $\pm$ 1 \\
ESO348-G009    & 23:49:23.5 & $-$37:46:19 & 648 & 30.0 $\pm$ 0.5 & \nodata & 144 & 0.41 & 83 & 10.0 $\pm$ 0.4 \\
ESO149-G003    & 23:52:02.8 & $-$52:34:40 & 590 & 29.2 $\pm$ 0.1 & 15.0 $\pm$ 0.2 & 131 & 0.18 & 148 & 9.7 $\pm$ 0.8 \\
NGC7793        & 23:57:49.8 & $-$32:35:28 & 227 & 27.77 $\pm$ 0.07 & 9.6 $\pm$ 0.2 & 560 & 0.68 & 98 & 7.4 $\pm$ 0.6 \\
\enddata
\tablecomments{ $\alpha_0$, $\delta_0$, $cz$, $m_B$, $d_{25}$, $d_{25}$, and PA are from NED, $m-M$ is from the Extragalactic Distance Database \citep{Tully+09}, $T$ is from HyperLeda\footnote{http://leda.univ-lyon1.fr/} \citep{Makarov+14}. }
\label{table:general}
\end{deluxetable}
\twocolumn
\clearpage
\onecolumn
\begin{deluxetable}{ lrrrrrr }
\tablecolumns{ 7 }
\tablewidth{ 0pc }
\tablecaption{  }
\tablehead{
   \colhead{ Galaxy } & \colhead{ Total Mass } & \colhead{ Bulge Mass } & \colhead{ Disk$_0$ Mass } & \colhead{ Disk$_1$ Mass } & \colhead{ Disk$_2$ Mass } & \colhead{ Disk$_3$ Mass } \\
   \colhead{  } & \colhead{ $M_{\odot}$ } & \colhead{ $M_{\odot}$ } & \colhead{ $M_{\odot}$ } & \colhead{ $M_{\odot}$ } & \colhead{ $M_{\odot}$ } & \colhead{ $M_{\odot}$ }
}
\startdata
NGC0024 & 2.3$\pm$0.4e+9 & \nodata & 2.2$\pm$0.4e+9 & \nodata & \nodata & \nodata \\
NGC0059 & 3.9$\pm$0.5e+8 & 3.1$\pm$0.7e+7 & 1.6$\pm$0.4e+8 & 2.0$\pm$0.4e+8 & \nodata & \nodata \\
NGC0625 & 7.0$\pm$0.9e+8 & 1.2$\pm$0.3e+7 & 2.7$\pm$0.6e+8 & 1.6$\pm$0.3e+8 & 2.5$\pm$0.5e+8 & \nodata \\
UGC05829 & 2.2$\pm$0.4e+8 & \nodata & 6$\pm$1.0e+7 & 1.6$\pm$0.4e+8 & \nodata & \nodata \\
NGC3344 & 1.9$\pm$0.3e+10 & 8$\pm$2.0e+8 & 1.3$\pm$0.3e+10 & 6$\pm$1.0e+9 & \nodata & \nodata \\
NGC3486 & 1.6$\pm$0.3e+10 & 1.3$\pm$0.4e+9 & 1.0$\pm$0.3e+10 & 4$\pm$1.0e+9 & 6$\pm$2.0e+8 & \nodata \\
UGC06112 & 6$\pm$1.0e+7 & \nodata & 5$\pm$1.0e+7 & \nodata & \nodata & \nodata \\
NGC3501 & 7$\pm$1.0e+8 & \nodata & 6$\pm$1.0e+8 & 3.3$\pm$0.8e+7 & \nodata & \nodata \\
NGC3507 & 3.9$\pm$0.6e+10 & 2.6$\pm$0.6e+9 & 2.1$\pm$0.5e+10 & 9$\pm$2.0e+9 & 6$\pm$2.0e+9 & \nodata \\
UGC06161 & 6$\pm$1.0e+8 & \nodata & 1.8$\pm$0.4e+8 & 4$\pm$1.0e+8 & \nodata & \nodata \\
UGC06399 & 1.9$\pm$0.4e+9 & \nodata & 1.8$\pm$0.4e+9 & \nodata & \nodata & \nodata \\
NGC3675 & 1.3$\pm$0.2e+11 & 7$\pm$2.0e+9 & 5$\pm$1.0e+10 & 4$\pm$1.0e+10 & 3.8$\pm$1.0e+10 & \nodata \\
NGC3718 & 3.1$\pm$0.4e+9 & 4.0$\pm$0.8e+8 & 1.1$\pm$0.2e+9 & 1.6$\pm$0.3e+9 & \nodata & \nodata \\
NGC3726 & 1.9$\pm$0.3e+10 & 5$\pm$1.0e+8 & 1.1$\pm$0.3e+10 & 7$\pm$2.0e+9 & \nodata & \nodata \\
NGC3729 & 2.8$\pm$0.5e+10 & 2.2$\pm$0.7e+9 & 9$\pm$3.0e+9 & 1.2$\pm$0.4e+10 & 5$\pm$2.0e+9 & \nodata \\
NGC3769 & 1.1$\pm$0.3e+10 & \nodata & 8$\pm$2.0e+9 & 3$\pm$1.0e+9 & \nodata & \nodata \\
UGC06628 & 5$\pm$1.0e+8 & 6$\pm$1.0e+6 & 2.3$\pm$0.5e+8 & 2.9$\pm$0.9e+8 & \nodata & \nodata \\
NGC3893 & 2.8$\pm$0.6e+10 & 2.4$\pm$0.8e+9 & 1.8$\pm$0.6e+10 & 7$\pm$3.0e+9 & \nodata & \nodata \\
UGC06792 & 1.6$\pm$0.3e+9 & \nodata & 1.5$\pm$0.3e+9 & \nodata & \nodata & \nodata \\
NGC3917 & 1$\pm$2.0e+10 & 3.2$\pm$0.8e+8 & 4$\pm$1.0e+9 & 5$\pm$1.0e+9 & \nodata & \nodata \\
NGC3922 & 5$\pm$1.0e+9 & 8$\pm$3.0e+8 & 1.8$\pm$0.6e+9 & 2.4$\pm$0.9e+9 & \nodata & \nodata \\
NGC3938 & 3.9$\pm$0.7e+9 & 1.3$\pm$0.3e+8 & 3.2$\pm$0.7e+9 & 6$\pm$2.0e+8 & \nodata & \nodata \\
NGC3941 & 2.2$\pm$0.3e+10 & 4.2$\pm$1.0e+9 & 6$\pm$1.0e+9 & 1$\pm$2.0e+10 & 1.7$\pm$0.6e+9 & \nodata \\
NGC3949 & 2.8$\pm$0.4e+10 & 1.6$\pm$0.4e+9 & 1.0$\pm$0.2e+10 & 1.2$\pm$0.3e+10 & 4$\pm$1.0e+9 & \nodata \\
NGC3953 & 8$\pm$1.0e+10 & 4$\pm$1.0e+9 & 6$\pm$1.0e+10 & 8$\pm$2.0e+9 & 1.0$\pm$0.3e+10 & \nodata \\
UGC06900 & 1.1$\pm$0.4e+8 & \nodata & 6$\pm$3.0e+7 & 4$\pm$2.0e+7 & \nodata & \nodata \\
NGC3972 & 1.4$\pm$0.2e+10 & 6$\pm$1.0e+8 & 6$\pm$1.0e+9 & 2.4$\pm$0.5e+9 & 5$\pm$2.0e+9 & \nodata \\
UGC06917 & 4$\pm$1.0e+9 & 1$\pm$3.0e+8 & 4$\pm$1.0e+9 & \nodata & \nodata & \nodata \\
UGC06930 & 9$\pm$1.0e+8 & \nodata & 3.8$\pm$0.9e+8 & 5$\pm$1.0e+8 & \nodata & \nodata \\
MESSIER109 & 1.5$\pm$0.2e+11 & 7$\pm$2.0e+9 & 8$\pm$2.0e+10 & 4.1$\pm$1.0e+10 & 2.8$\pm$0.8e+10 & \nodata \\
NGC3998 & 2.3$\pm$0.3e+10 & 8$\pm$2.0e+9 & 1$\pm$2.0e+10 & 5$\pm$1.0e+9 & \nodata & \nodata \\
UGC06955 & 2.0$\pm$0.5e+9 & \nodata & 2.0$\pm$0.5e+9 & \nodata & \nodata & \nodata \\
NGC4013 & 9$\pm$2.0e+10 & \nodata & 8$\pm$2.0e+10 & 5$\pm$1.0e+9 & \nodata & \nodata \\
IC0749 & 5.9$\pm$0.9e+8 & 2.3$\pm$0.5e+7 & 3.5$\pm$0.8e+8 & 2.1$\pm$0.5e+8 & \nodata & \nodata \\
NGC4010 & 1.4$\pm$0.3e+10 & \nodata & 1.1$\pm$0.3e+10 & 9$\pm$2.0e+8 & 1.3$\pm$0.5e+9 & \nodata \\
IC0750 & 7$\pm$2.0e+10 & \nodata & 5$\pm$2.0e+10 & 4$\pm$1.0e+9 & 4$\pm$1.0e+9 & \nodata \\
UGC06983 & 6$\pm$2.0e+9 & 1.5$\pm$0.6e+8 & 2.0$\pm$0.7e+9 & 4$\pm$1.0e+9 & \nodata & \nodata \\
NGC4026 & 2.1$\pm$0.2e+10 & 5$\pm$1.0e+9 & 5$\pm$1.0e+9 & 8$\pm$2.0e+9 & 2.6$\pm$0.6e+9 & 1.2$\pm$0.4e+9 \\
NGC4051 & 2.4$\pm$0.3e+10 & 4$\pm$1.0e+9 & 6$\pm$2.0e+9 & 1.0$\pm$0.2e+10 & 2.6$\pm$0.8e+9 & \nodata \\
NGC4068 & 1.3$\pm$0.2e+8 & \nodata & 2.8$\pm$0.6e+7 & 1$\pm$2.0e+8 & \nodata & \nodata \\
NGC4085 & 2.1$\pm$0.3e+10 & 2.2$\pm$0.5e+9 & 1.1$\pm$0.3e+10 & 5$\pm$1.0e+9 & 3$\pm$2.0e+9 & \nodata \\
NGC4088 & 3.6$\pm$0.6e+10 & 1.9$\pm$0.5e+9 & 2.1$\pm$0.5e+10 & 1.1$\pm$0.3e+10 & 2.0$\pm$0.8e+9 & \nodata \\
UGC07089 & 1.4$\pm$0.2e+9 & \nodata & 2.6$\pm$0.6e+8 & 5$\pm$1.0e+8 & 6$\pm$2.0e+8 & \nodata \\
NGC4096 & 1.6$\pm$0.3e+10 & 7$\pm$2.0e+8 & 1.2$\pm$0.3e+10 & 2.9$\pm$0.7e+9 & \nodata & \nodata \\
NGC4100 & 4.1$\pm$0.7e+10 & 3.8$\pm$1.0e+9 & 2.5$\pm$0.6e+10 & 1.1$\pm$0.3e+10 & 2.1$\pm$0.9e+9 & \nodata \\
NGC4102 & 4.2$\pm$0.6e+10 & 2.0$\pm$0.5e+10 & 1.8$\pm$0.4e+10 & 3$\pm$1.0e+9 & \nodata & \nodata \\
NGC4111 & 8$\pm$1.0e+9 & 3.4$\pm$0.8e+9 & 3.6$\pm$0.8e+9 & 5$\pm$2.0e+8 & \nodata & \nodata \\
UGC07125 & 8$\pm$2.0e+9 & \nodata & 2.7$\pm$0.9e+9 & 5$\pm$2.0e+9 & \nodata & \nodata \\
NGC4138 & 1.8$\pm$0.2e+10 & 3.7$\pm$0.8e+9 & 1.1$\pm$0.2e+10 & 1.4$\pm$0.3e+9 & 2.2$\pm$0.5e+9 & \nodata \\
NGC4143 & 2.2$\pm$0.4e+10 & 6$\pm$1.0e+9 & 1.3$\pm$0.3e+10 & 2$\pm$2.0e+9 & \nodata & \nodata \\
NGC4144 & 1.3$\pm$0.2e+9 & \nodata & 8$\pm$1.0e+8 & 3.8$\pm$0.7e+8 & 1.5$\pm$0.3e+8 & \nodata \\
NGC4151 & 4.2$\pm$0.9e+9 & 2.6$\pm$0.9e+9 & 5$\pm$2.0e+8 & 4$\pm$1.0e+8 & 8$\pm$3.0e+8 & \nodata \\
NGC4183 & 7$\pm$1.0e+9 & 3.4$\pm$0.8e+8 & 4$\pm$1.0e+9 & 9$\pm$2.0e+8 & 1.2$\pm$0.4e+9 & \nodata \\
NGC4203 & 2.4$\pm$0.3e+10 & 6$\pm$1.0e+9 & 7$\pm$2.0e+9 & 7$\pm$1.0e+9 & 4$\pm$1.0e+9 & \nodata \\
NGC4214 & 1.6$\pm$0.2e+9 & \nodata & 9$\pm$2.0e+8 & 6$\pm$1.0e+8 & \nodata & \nodata \\
NGC4220 & 3.5$\pm$0.6e+10 & 3.6$\pm$0.8e+9 & 2.7$\pm$0.6e+10 & 5$\pm$1.0e+9 & \nodata & \nodata \\
UGC07301 & 9$\pm$1.0e+8 & \nodata & 2.1$\pm$0.6e+8 & 2.3$\pm$0.6e+8 & 3.2$\pm$0.8e+8 & \nodata \\
NGC4242 & 2.0$\pm$0.3e+9 & \nodata & 1.1$\pm$0.2e+9 & 9$\pm$2.0e+8 & \nodata & \nodata \\
UGC07408 & 2.5$\pm$0.5e+8 & \nodata & 2.4$\pm$0.5e+8 & \nodata & \nodata & \nodata \\
NGC4369 & 1.7$\pm$0.3e+9 & 5$\pm$1.0e+8 & 1.2$\pm$0.3e+9 & \nodata & \nodata & \nodata \\
NGC4389 & 3.0$\pm$0.6e+9 & 2.2$\pm$0.6e+8 & 2.0$\pm$0.5e+9 & 6$\pm$2.0e+8 & 2.2$\pm$0.6e+8 & \nodata \\
UGC07577 & 6.8$\pm$1.0e+7 & \nodata & 3.6$\pm$0.7e+7 & 3.2$\pm$0.7e+7 & \nodata & \nodata \\
UGC07608 & 5$\pm$1.0e+8 & \nodata & 3.5$\pm$0.8e+7 & 1$\pm$2.0e+8 & 3.5$\pm$1.0e+8 & \nodata \\
NGC4460 & 3.9$\pm$0.6e+9 & \nodata & 1.2$\pm$0.3e+9 & 2.4$\pm$0.5e+9 & \nodata & \nodata \\
UGC07639 & 8$\pm$1.0e+8 & \nodata & 3.2$\pm$0.7e+8 & 4$\pm$1.0e+8 & \nodata & \nodata \\
NGC4485 & 2.0$\pm$0.4e+9 & \nodata & 9$\pm$2.0e+8 & 1.0$\pm$0.4e+9 & \nodata & \nodata \\
NGC4490 & 2.3$\pm$0.5e+10 & \nodata & 2.1$\pm$0.5e+10 & 4$\pm$1.0e+8 & 2.0$\pm$0.7e+9 & \nodata \\
UGC07699 & 3.0$\pm$0.8e+9 & \nodata & 2.2$\pm$0.8e+9 & 7$\pm$2.0e+8 & \nodata & \nodata \\
UGC07774 & 2.7$\pm$0.5e+9 & \nodata & 1.3$\pm$0.3e+9 & 1.3$\pm$0.3e+9 & \nodata & \nodata \\
NGC4618 & 9$\pm$1.0e+9 & 1.5$\pm$0.3e+8 & 3.2$\pm$0.7e+9 & 2.1$\pm$0.5e+9 & 1.8$\pm$0.4e+9 & 1.9$\pm$0.7e+9 \\
IC3687 & 1.3$\pm$0.2e+8 & \nodata & 5$\pm$1.0e+7 & 8$\pm$2.0e+7 & \nodata & \nodata \\
NGC4707 & 2.5$\pm$0.4e+8 & \nodata & 3.4$\pm$0.7e+7 & 8$\pm$2.0e+7 & 1.4$\pm$0.3e+8 & \nodata \\
NGC4861 & 4$\pm$1.0e+8 & \nodata & 4$\pm$1.0e+8 & \nodata & \nodata & \nodata \\
IC4182 & 6$\pm$1.0e+8 & \nodata & 6$\pm$1.0e+8 & \nodata & \nodata & \nodata \\
NGC5005 & 1.3$\pm$0.2e+11 & 1.4$\pm$0.3e+10 & 6$\pm$1.0e+10 & 4.8$\pm$1.0e+10 & 1.7$\pm$0.4e+9 & 8$\pm$2.0e+9 \\
IC4213 & 2.9$\pm$0.7e+9 & \nodata & 1.9$\pm$0.4e+9 & 4.1$\pm$0.9e+8 & 5$\pm$5.0e+8 & \nodata \\
NGC5023 & 6.8$\pm$1.0e+8 & \nodata & 2.9$\pm$0.6e+8 & 3.7$\pm$0.8e+8 & \nodata & \nodata \\
UGC08303 & 1.5$\pm$0.3e+8 & \nodata & 1.4$\pm$0.3e+8 & \nodata & \nodata & \nodata \\
NGC5033 & 9$\pm$2.0e+10 & 9$\pm$2.0e+9 & 4$\pm$1.0e+10 & 4$\pm$1.0e+10 & \nodata & \nodata \\
UGC08320 & 1.5$\pm$0.3e+8 & \nodata & 3.1$\pm$0.7e+7 & 1.2$\pm$0.3e+8 & \nodata & \nodata \\
MESSIER063 & 8.5$\pm$1.0e+10 & 2.6$\pm$0.5e+9 & 1.7$\pm$0.4e+10 & 3.9$\pm$0.8e+10 & 1.2$\pm$0.2e+10 & 1.5$\pm$0.3e+10 \\
NGC5229 & 8$\pm$1.0e+8 & \nodata & 3.2$\pm$0.7e+8 & 4.4$\pm$1.0e+8 & \nodata & \nodata \\
NGC5273 & 1.3$\pm$0.2e+10 & 2.0$\pm$0.5e+9 & 8$\pm$2.0e+9 & 3.3$\pm$0.9e+9 & \nodata & \nodata \\
UGC08839 & 1.4$\pm$0.4e+9 & \nodata & 4$\pm$1.0e+8 & 1$\pm$4.0e+9 & \nodata & \nodata \\
NGC5523 & 8$\pm$1.0e+9 & 7$\pm$2.0e+8 & 2.9$\pm$0.8e+9 & 2.2$\pm$0.6e+9 & 1.6$\pm$0.4e+9 & 7$\pm$2.0e+8 \\
NGC5608 & 1.3$\pm$0.4e+9 & \nodata & 1.3$\pm$0.4e+9 & \nodata & \nodata & \nodata \\
ESO290-G028 & 1.7$\pm$0.5e+7 & \nodata & 1.6$\pm$0.5e+7 & \nodata & \nodata & \nodata \\
NGC7713 & 4.4$\pm$0.6e+9 & \nodata & 1.6$\pm$0.4e+9 & 1$\pm$2.0e+9 & 6$\pm$2.0e+8 & 1.2$\pm$0.3e+9 \\
UGCA442 & 1.1$\pm$0.5e+8 & \nodata & 5$\pm$1.0e+7 & 6$\pm$5.0e+7 & \nodata & \nodata \\
ESO348-G009 & 2.5$\pm$1.0e+8 & \nodata & 2.4$\pm$1.0e+8 & \nodata & \nodata & \nodata \\
ESO149-G003 & 8$\pm$2.0e+7 & \nodata & 2.5$\pm$0.5e+6 & 2.5$\pm$0.6e+7 & 5$\pm$1.0e+7 & \nodata \\
NGC7793 & 3.5$\pm$0.4e+9 & 3.6$\pm$0.7e+7 & 5$\pm$1.0e+8 & 1.2$\pm$0.2e+9 & 1.7$\pm$0.3e+9 & \nodata \\
\enddata
\tablecomments{Mass measurements as described in \S~\ref{sec:mass_measure}.}
\label{table:mass}
\end{deluxetable}
\twocolumn
\clearpage
\bibliographystyle{mnras}
\bibliography{refs}

\newcommand{\noop}[1]{}
\begin{thebibliography}{}
\makeatletter
\relax
\def\mn@urlcharsother{\let\do\@makeother \do\$\do\&\do\#\do\^\do\_\do\%\do\~}
\def\mn@doi{\begingroup\mn@urlcharsother \@ifnextchar [ {\mn@doi@}
  {\mn@doi@[]}}
\def\mn@doi@[#1]#2{\def\@tempa{#1}\ifx\@tempa\@empty \href
  {http://dx.doi.org/#2} {doi:#2}\else \href {http://dx.doi.org/#2} {#1}\fi
  \endgroup}
\def\mn@eprint#1#2{\mn@eprint@#1:#2::\@nil}
\def\mn@eprint@arXiv#1{\href {http://arxiv.org/abs/#1} {{\tt arXiv:#1}}}
\def\mn@eprint@dblp#1{\href {http://dblp.uni-trier.de/rec/bibtex/#1.xml}
  {dblp:#1}}
\def\mn@eprint@#1:#2:#3:#4\@nil{\def\@tempa {#1}\def\@tempb {#2}\def\@tempc
  {#3}\ifx \@tempc \@empty \let \@tempc \@tempb \let \@tempb \@tempa \fi \ifx
  \@tempb \@empty \def\@tempb {arXiv}\fi \@ifundefined
  {mn@eprint@\@tempb}{\@tempb:\@tempc}{\expandafter \expandafter \csname
  mn@eprint@\@tempb\endcsname \expandafter{\@tempc}}}

\bibitem[\protect\citeauthoryear{{Abraham} \& {van Dokkum}}{{Abraham} \& {van
  Dokkum}}{2014}]{Abraham+14}
{Abraham} R.~G.,  {van Dokkum} P.~G.,  2014, \mn@doi [\pasp] {10.1086/674875},
  \href {http://adsabs.harvard.edu/abs/2014PASP..126...55A} {126, 55}

\bibitem[\protect\citeauthoryear{{Aihara} et~al.,}{{Aihara}
  et~al.}{2018}]{Aihara+17}
{Aihara} H.,  et~al., 2018, \mn@doi [\pasj] {10.1093/pasj/psx066}, \href
  {http://adsabs.harvard.edu/abs/2018PASJ...70S...4A} {70, S4}

\bibitem[\protect\citeauthoryear{{Astropy Collaboration} et~al.,}{{Astropy
  Collaboration} et~al.}{2013}]{astropy+13}
{Astropy Collaboration} et~al., 2013, \mn@doi [\aap]
  {10.1051/0004-6361/201322068}, \href
  {http://adsabs.harvard.edu/abs/2013A%26A...558A..33A} {558, A33}

\bibitem[\protect\citeauthoryear{{Bailin}, {Bell}, {Chappell}, {Radburn-Smith}
  \& {de Jong}}{{Bailin} et~al.}{2011}]{Bailin+11}
{Bailin} J.,  {Bell} E.~F.,  {Chappell} S.~N.,  {Radburn-Smith} D.~J.,   {de
  Jong} R.~S.,  2011, \mn@doi [\apj] {10.1088/0004-637X/736/1/24}, \href
  {http://adsabs.harvard.edu/abs/2011ApJ...736...24B} {736, 24}

\bibitem[\protect\citeauthoryear{{Barker}, {Ferguson}, {Irwin}, {Arimoto}  \&
  {Jablonka}}{{Barker} et~al.}{2012}]{Barker+12}
{Barker} M.~K.,  {Ferguson} A.~M.~N.,  {Irwin} M.~J.,  {Arimoto} N.,
  {Jablonka} P.,  2012, \mn@doi [\mnras] {10.1111/j.1365-2966.2011.19814.x},
  \href {http://adsabs.harvard.edu/abs/2012MNRAS.419.1489B} {419, 1489}

\bibitem[\protect\citeauthoryear{{Barnes}, {van Zee}, {Dale}, {Staudaher},
  {Bullock}, {Calzetti}, {Chandar}  \& {Dalcanton}}{{Barnes}
  et~al.}{2014}]{Barnes+14}
{Barnes} K.~L.,  {van Zee} L.,  {Dale} D.~A.,  {Staudaher} S.,  {Bullock}
  J.~S.,  {Calzetti} D.,  {Chandar} R.,   {Dalcanton} J.~J.,  2014, \mn@doi
  [\apj] {10.1088/0004-637X/789/2/126}, \href
  {http://adsabs.harvard.edu/abs/2014ApJ...789..126B} {789, 126}

\bibitem[\protect\citeauthoryear{{Begum}, {Chengalur}, {Kennicutt},
  {Karachentsev}  \& {Lee}}{{Begum} et~al.}{2008}]{Begum+08}
{Begum} A.,  {Chengalur} J.~N.,  {Kennicutt} R.~C.,  {Karachentsev} I.~D.,
  {Lee} J.~C.,  2008, \mn@doi [\mnras] {10.1111/j.1365-2966.2007.12592.x},
  \href {http://adsabs.harvard.edu/abs/2008MNRAS.383..809B} {383, 809}

\bibitem[\protect\citeauthoryear{{Belokurov} et~al.,}{{Belokurov}
  et~al.}{2006}]{Belokurov+06}
{Belokurov} V.,  et~al., 2006, \mn@doi [\apjl] {10.1086/504797}, \href
  {http://adsabs.harvard.edu/abs/2006ApJ...642L.137B} {642, L137}

\bibitem[\protect\citeauthoryear{{Bertin} \& {Arnouts}}{{Bertin} \&
  {Arnouts}}{1996}]{Bertin+96}
{Bertin} E.,  {Arnouts} S.,  1996, \mn@doi [\aaps] {10.1051/aas:1996164}, \href
  {http://adsabs.harvard.edu/abs/1996A%26AS..117..393B} {117, 393}

\bibitem[\protect\citeauthoryear{{Bullock} \& {Johnston}}{{Bullock} \&
  {Johnston}}{2005}]{Bullock+05}
{Bullock} J.~S.,  {Johnston} K.~V.,  2005, \mn@doi [\apj] {10.1086/497422},
  \href {http://adsabs.harvard.edu/abs/2005ApJ...635..931B} {635, 931}

\bibitem[\protect\citeauthoryear{{Bullock}, {Kravtsov}  \&
  {Weinberg}}{{Bullock} et~al.}{2000}]{Bullock+00}
{Bullock} J.~S.,  {Kravtsov} A.~V.,   {Weinberg} D.~H.,  2000, \mn@doi [\apj]
  {10.1086/309279}, \href {http://adsabs.harvard.edu/abs/2000ApJ...539..517B}
  {539, 517}

\bibitem[\protect\citeauthoryear{{Bullock}, {Stewart}, {Kaplinghat}, {Tollerud}
   \& {Wolf}}{{Bullock} et~al.}{2010}]{Bullock+10}
{Bullock} J.~S.,  {Stewart} K.~R.,  {Kaplinghat} M.,  {Tollerud} E.~J.,
  {Wolf} J.,  2010, \mn@doi [\apj] {10.1088/0004-637X/717/2/1043}, \href
  {http://adsabs.harvard.edu/abs/2010ApJ...717.1043B} {717, 1043}

\bibitem[\protect\citeauthoryear{{Carlin} et~al.,}{{Carlin}
  et~al.}{2016}]{Carlin+16}
{Carlin} J.~L.,  et~al., 2016, \mn@doi [\apjl] {10.3847/2041-8205/828/1/L5},
  \href {http://adsabs.harvard.edu/abs/2016ApJ...828L...5C} {828, L5}

\bibitem[\protect\citeauthoryear{{Carollo} et~al.,}{{Carollo}
  et~al.}{2010}]{Carollo+10}
{Carollo} D.,  et~al., 2010, \mn@doi [\apj] {10.1088/0004-637X/712/1/692},
  \href {http://adsabs.harvard.edu/abs/2010ApJ...712..692C} {712, 692}

\bibitem[\protect\citeauthoryear{{Chonis}, {Mart{\'{\i}}nez-Delgado}, {Gabany},
  {Majewski}, {Hill}, {Gralak}  \& {Trujillo}}{{Chonis}
  et~al.}{2011}]{Chonis+11}
{Chonis} T.~S.,  {Mart{\'{\i}}nez-Delgado} D.,  {Gabany} R.~J.,  {Majewski}
  S.~R.,  {Hill} G.~J.,  {Gralak} R.,   {Trujillo} I.,  2011, \mn@doi [\aj]
  {10.1088/0004-6256/142/5/166}, \href
  {http://adsabs.harvard.edu/abs/2011AJ....142..166C} {142, 166}

\bibitem[\protect\citeauthoryear{{Christlein} \& {Zaritsky}}{{Christlein} \&
  {Zaritsky}}{2008}]{Christlein+Zaritsky08}
{Christlein} D.,  {Zaritsky} D.,  2008, \mn@doi [\apj] {10.1086/587468}, \href
  {http://adsabs.harvard.edu/abs/2008ApJ...680.1053C} {680, 1053}

\bibitem[\protect\citeauthoryear{{Cooper}, {D'Souza}, {Kauffmann}, {Wang},
  {Boylan-Kolchin}, {Guo}, {Frenk}  \& {White}}{{Cooper}
  et~al.}{2013}]{Cooper+13}
{Cooper} A.~P.,  {D'Souza} R.,  {Kauffmann} G.,  {Wang} J.,  {Boylan-Kolchin}
  M.,  {Guo} Q.,  {Frenk} C.~S.,   {White} S.~D.~M.,  2013, \mn@doi [\mnras]
  {10.1093/mnras/stt1245}, \href
  {http://adsabs.harvard.edu/abs/2013MNRAS.434.3348C} {434, 3348}

\bibitem[\protect\citeauthoryear{{Courteau}, {Widrow}, {McDonald},
  {Guhathakurta}, {Gilbert}, {Zhu}, {Beaton}  \& {Majewski}}{{Courteau}
  et~al.}{2011}]{Courteau+11}
{Courteau} S.,  {Widrow} L.~M.,  {McDonald} M.,  {Guhathakurta} P.,  {Gilbert}
  K.~M.,  {Zhu} Y.,  {Beaton} R.~L.,   {Majewski} S.~R.,  2011, \mn@doi [\apj]
  {10.1088/0004-637X/739/1/20}, \href
  {http://adsabs.harvard.edu/abs/2011ApJ...739...20C} {739, 20}

\bibitem[\protect\citeauthoryear{{D'Souza} \& {Bell}}{{D'Souza} \&
  {Bell}}{2018}]{DSouza+Bell18}
{D'Souza} R.,  {Bell} E.~F.,  2018, \mn@doi [\mnras] {10.1093/mnras/stx3081},
  \href {http://adsabs.harvard.edu/abs/2018MNRAS.474.5300D} {474, 5300}

\bibitem[\protect\citeauthoryear{{D'Souza}, {Kauffman}, {Wang}  \&
  {Vegetti}}{{D'Souza} et~al.}{2014}]{DSouza+14}
{D'Souza} R.,  {Kauffman} G.,  {Wang} J.,   {Vegetti} S.,  2014, \mn@doi
  [\mnras] {10.1093/mnras/stu1194}, \href
  {http://adsabs.harvard.edu/abs/2014MNRAS.443.1433D} {443, 1433}

\bibitem[\protect\citeauthoryear{{Dale} et~al.,}{{Dale} et~al.}{2009}]{Dale+09}
{Dale} D.~A.,  et~al., 2009, \mn@doi [\apj] {10.1088/0004-637X/703/1/517},
  \href {http://adsabs.harvard.edu/abs/2009ApJ...703..517D} {703, 517}

\bibitem[\protect\citeauthoryear{{Dale} et~al.,}{{Dale} et~al.}{2016}]{Dale+16}
{Dale} D.~A.,  et~al., 2016, \mn@doi [\aj] {10.3847/0004-6256/151/1/4}, \href
  {http://adsabs.harvard.edu/abs/2016AJ....151....4D} {151, 4}

\bibitem[\protect\citeauthoryear{{Elias}, {Sales}, {Creasey}, {Cooper},
  {Bullock}, {Rich}  \& {Hernquist}}{{Elias} et~al.}{2018}]{Elias+18}
{Elias} L.~M.,  {Sales} L.~V.,  {Creasey} P.,  {Cooper} M.~C.,  {Bullock}
  J.~S.,  {Rich} R.~M.,   {Hernquist} L.,  2018, \mn@doi [\mnras]
  {10.1093/mnras/sty1718}, \href
  {http://adsabs.harvard.edu/abs/2018MNRAS.479.4004E} {479, 4004}

\bibitem[\protect\citeauthoryear{{Erwin}, {Pohlen}  \& {Beckman}}{{Erwin}
  et~al.}{2008}]{Erwin+08}
{Erwin} P.,  {Pohlen} M.,   {Beckman} J.~E.,  2008, \mn@doi [\aj]
  {10.1088/0004-6256/135/1/20}, \href
  {http://adsabs.harvard.edu/abs/2008AJ....135...20E} {135, 20}

\bibitem[\protect\citeauthoryear{{Eskew}, {Zaritsky}  \& {Meidt}}{{Eskew}
  et~al.}{2012}]{Eskew+12}
{Eskew} M.,  {Zaritsky} D.,   {Meidt} S.,  2012, \mn@doi [\aj]
  {10.1088/0004-6256/143/6/139}, \href
  {http://adsabs.harvard.edu/abs/2012AJ....143..139E} {143, 139}

\bibitem[\protect\citeauthoryear{{Gadotti}}{{Gadotti}}{2009}]{Gadotti+09}
{Gadotti} D.~A.,  2009, \mn@doi [\mnras] {10.1111/j.1365-2966.2008.14257.x},
  \href {http://adsabs.harvard.edu/abs/2009MNRAS.393.1531G} {393, 1531}

\bibitem[\protect\citeauthoryear{{Guo} et~al.,}{{Guo} et~al.}{2011}]{Guo+11}
{Guo} Q.,  et~al., 2011, \mn@doi [\mnras] {10.1111/j.1365-2966.2010.18114.x},
  \href {http://adsabs.harvard.edu/abs/2011MNRAS.413..101G} {413, 101}

\bibitem[\protect\citeauthoryear{{Guo}, {Cooper}, {Frenk}, {Helly}  \&
  {Hellwing}}{{Guo} et~al.}{2015}]{Guo+15}
{Guo} Q.,  {Cooper} A.~P.,  {Frenk} C.,  {Helly} J.,   {Hellwing} W.~A.,  2015,
  \mn@doi [\mnras] {10.1093/mnras/stv1938}, \href
  {http://adsabs.harvard.edu/abs/2015MNRAS.454..550G} {454, 550}

\bibitem[\protect\citeauthoryear{{Guti{\'e}rrez}, {Erwin}, {Aladro}  \&
  {Beckman}}{{Guti{\'e}rrez} et~al.}{2011}]{Gutierrez+11}
{Guti{\'e}rrez} L.,  {Erwin} P.,  {Aladro} R.,   {Beckman} J.~E.,  2011,
  \mn@doi [\aj] {10.1088/0004-6256/142/5/145}, \href
  {http://adsabs.harvard.edu/abs/2011AJ....142..145G} {142, 145}

\bibitem[\protect\citeauthoryear{{Harmsen}, {Monachesi}, {Bell}, {de Jong},
  {Bailin}, {Radburn-Smith}  \& {Holwerda}}{{Harmsen}
  et~al.}{2017}]{Harmsen+17}
{Harmsen} B.,  {Monachesi} A.,  {Bell} E.~F.,  {de Jong} R.~S.,  {Bailin} J.,
  {Radburn-Smith} D.~J.,   {Holwerda} B.~W.,  2017, \mn@doi [\mnras]
  {10.1093/mnras/stw2992}, \href
  {http://adsabs.harvard.edu/abs/2017MNRAS.466.1491H} {466, 1491}

\bibitem[\protect\citeauthoryear{{Huang}, {Leauthaud}, {Greene}, {Bundy},
  {Lin}, {Tanaka}, {Miyazaki}  \& {Komiyama}}{{Huang} et~al.}{2018}]{Huang+17}
{Huang} S.,  {Leauthaud} A.,  {Greene} J.~E.,  {Bundy} K.,  {Lin} Y.-T.,
  {Tanaka} M.,  {Miyazaki} S.,   {Komiyama} Y.,  2018, \mn@doi [\mnras]
  {10.1093/mnras/stx3200}, \href
  {http://adsabs.harvard.edu/abs/2018MNRAS.475.3348H} {475, 3348}

\bibitem[\protect\citeauthoryear{{Irwin}, {Ferguson}, {Huxor}, {Tanvir},
  {Ibata}  \& {Lewis}}{{Irwin} et~al.}{2008}]{Irwin+08}
{Irwin} M.~J.,  {Ferguson} A.~M.~N.,  {Huxor} A.~P.,  {Tanvir} N.~R.,  {Ibata}
  R.~A.,   {Lewis} G.~F.,  2008, \mn@doi [\apjl] {10.1086/587100}, \href
  {http://adsabs.harvard.edu/abs/2008ApJ...676L..17I} {676, L17}

\bibitem[\protect\citeauthoryear{{Javanmardi} et~al.,}{{Javanmardi}
  et~al.}{2016}]{Javanmardi+16}
{Javanmardi} B.,  et~al., 2016, \mn@doi [\aap] {10.1051/0004-6361/201527745},
  \href {http://adsabs.harvard.edu/abs/2016A%26A...588A..89J} {588, A89}

\bibitem[\protect\citeauthoryear{{Johnston}, {Bullock}, {Sharma}, {Font},
  {Robertson}  \& {Leitner}}{{Johnston} et~al.}{2008}]{Johnston+08}
{Johnston} K.~V.,  {Bullock} J.~S.,  {Sharma} S.,  {Font} A.,  {Robertson}
  B.~E.,   {Leitner} S.~N.,  2008, \mn@doi [\apj] {10.1086/592228}, \href
  {http://adsabs.harvard.edu/abs/2008ApJ...689..936J} {689, 936}

\bibitem[\protect\citeauthoryear{{Kauffmann}, {White}  \&
  {Guiderdoni}}{{Kauffmann} et~al.}{1993}]{Kauffmann+93}
{Kauffmann} G.,  {White} S.~D.~M.,   {Guiderdoni} B.,  1993, \mn@doi [\mnras]
  {10.1093/mnras/264.1.201}, \href
  {http://adsabs.harvard.edu/abs/1993MNRAS.264..201K} {264, 201}

\bibitem[\protect\citeauthoryear{{Kennicutt} Jr. et~al.,}{{Kennicutt}
  et~al.}{2003}]{Kennicutt+03}
{Kennicutt} Jr. R.~C.,  et~al., 2003, \mn@doi [\pasp] {10.1086/376941}, \href
  {http://adsabs.harvard.edu/abs/2003PASP..115..928K} {115, 928}

\bibitem[\protect\citeauthoryear{{Klypin}, {Kravtsov}, {Valenzuela}  \&
  {Prada}}{{Klypin} et~al.}{1999}]{Klypin+99}
{Klypin} A.,  {Kravtsov} A.~V.,  {Valenzuela} O.,   {Prada} F.,  1999, \mn@doi
  [\apj] {10.1086/307643}, \href
  {http://adsabs.harvard.edu/abs/1999ApJ...522...82K} {522, 82}

\bibitem[\protect\citeauthoryear{{Koposov}, {Belokurov}, {Torrealba}  \&
  {Evans}}{{Koposov} et~al.}{2015}]{Koposov+15}
{Koposov} S.~E.,  {Belokurov} V.,  {Torrealba} G.,   {Evans} N.~W.,  2015,
  \mn@doi [\apj] {10.1088/0004-637X/805/2/130}, \href
  {http://adsabs.harvard.edu/abs/2015ApJ...805..130K} {805, 130}

\bibitem[\protect\citeauthoryear{{Kravtsov}, {Gnedin}  \& {Klypin}}{{Kravtsov}
  et~al.}{2004}]{Kravtsov+04}
{Kravtsov} A.~V.,  {Gnedin} O.~Y.,   {Klypin} A.~A.,  2004, \mn@doi [\apj]
  {10.1086/421322}, \href {http://adsabs.harvard.edu/abs/2004ApJ...609..482K}
  {609, 482}

\bibitem[\protect\citeauthoryear{{Krick}, {Bridge}, {Desai}, {Mihos}, {Murphy},
  {Rudick}, {Surace}  \& {Neill}}{{Krick} et~al.}{2011}]{Krick+11}
{Krick} J.~E.,  {Bridge} C.,  {Desai} V.,  {Mihos} J.~C.,  {Murphy} E.,
  {Rudick} C.,  {Surace} J.,   {Neill} J.,  2011, \mn@doi [\apj]
  {10.1088/0004-637X/735/2/76}, \href
  {http://adsabs.harvard.edu/abs/2011ApJ...735...76K} {735, 76}

\bibitem[\protect\citeauthoryear{{Laine} et~al.,}{{Laine}
  et~al.}{2014}]{Laine+14}
{Laine} J.,  et~al., 2014, \mn@doi [\mnras] {10.1093/mnras/stu628}, \href
  {http://adsabs.harvard.edu/abs/2014MNRAS.441.1992L} {441, 1992}

\bibitem[\protect\citeauthoryear{{Laine}, {Laurikainen}  \& {Salo}}{{Laine}
  et~al.}{2016}]{Laine+16}
{Laine} J.,  {Laurikainen} E.,   {Salo} H.,  2016, \mn@doi [\aap]
  {10.1051/0004-6361/201628397}, \href
  {http://adsabs.harvard.edu/abs/2016A%26A...596A..25L} {596, A25}

\bibitem[\protect\citeauthoryear{{Makarov}, {Prugniel}, {Terekhova}, {Courtois}
   \& {Vauglin}}{{Makarov} et~al.}{2014}]{Makarov+14}
{Makarov} D.,  {Prugniel} P.,  {Terekhova} N.,  {Courtois} H.,   {Vauglin} I.,
  2014, \mn@doi [\aap] {10.1051/0004-6361/201423496}, \href
  {http://adsabs.harvard.edu/abs/2014A%26A...570A..13M} {570, A13}

\bibitem[\protect\citeauthoryear{{Martin}, {Ibata}, {Chapman}, {Irwin}  \&
  {Lewis}}{{Martin} et~al.}{2007}]{Martin+07}
{Martin} N.~F.,  {Ibata} R.~A.,  {Chapman} S.~C.,  {Irwin} M.,   {Lewis} G.~F.,
   2007, \mn@doi [\mnras] {10.1111/j.1365-2966.2007.12055.x}, \href
  {http://adsabs.harvard.edu/abs/2007MNRAS.380..281M} {380, 281}

\bibitem[\protect\citeauthoryear{{Mart{\'{\i}}nez-Delgado}, {Pohlen}, {Gabany},
  {Majewski}, {Pe{\~n}arrubia}  \& {Palma}}{{Mart{\'{\i}}nez-Delgado}
  et~al.}{2009}]{Martinez-Delgado+09}
{Mart{\'{\i}}nez-Delgado} D.,  {Pohlen} M.,  {Gabany} R.~J.,  {Majewski} S.~R.,
   {Pe{\~n}arrubia} J.,   {Palma} C.,  2009, \mn@doi [\apj]
  {10.1088/0004-637X/692/2/955}, \href
  {http://adsabs.harvard.edu/abs/2009ApJ...692..955M} {692, 955}

\bibitem[\protect\citeauthoryear{{McConnachie}, {Ferguson}, {Irwin},
  {Dubinski}, {Widrow}, {Dotter}, {Ibata}  \& {Lewis}}{{McConnachie}
  et~al.}{2010}]{McConnachie+10}
{McConnachie} A.~W.,  {Ferguson} A.~M.~N.,  {Irwin} M.~J.,  {Dubinski} J.,
  {Widrow} L.~M.,  {Dotter} A.,  {Ibata} R.,   {Lewis} G.~F.,  2010, \mn@doi
  [\apj] {10.1088/0004-637X/723/2/1038}, \href
  {http://adsabs.harvard.edu/abs/2010ApJ...723.1038M} {723, 1038}

\bibitem[\protect\citeauthoryear{{McGaugh} \& {Schombert}}{{McGaugh} \&
  {Schombert}}{2014}]{McGaugh+14}
{McGaugh} S.~S.,  {Schombert} J.~M.,  2014, \mn@doi [\aj]
  {10.1088/0004-6256/148/5/77}, \href
  {http://adsabs.harvard.edu/abs/2014AJ....148...77M} {148, 77}

\bibitem[\protect\citeauthoryear{{McGaugh} \& {Schombert}}{{McGaugh} \&
  {Schombert}}{2015}]{McGaugh+15}
{McGaugh} S.~S.,  {Schombert} J.~M.,  2015, \mn@doi [\apj]
  {10.1088/0004-637X/802/1/18}, \href
  {http://adsabs.harvard.edu/abs/2015ApJ...802...18M} {802, 18}

\bibitem[\protect\citeauthoryear{{Meidt} et~al.,}{{Meidt}
  et~al.}{2014}]{Meidt+14}
{Meidt} S.~E.,  et~al., 2014, \mn@doi [\apj] {10.1088/0004-637X/788/2/144},
  \href {http://adsabs.harvard.edu/abs/2014ApJ...788..144M} {788, 144}

\bibitem[\protect\citeauthoryear{{Merritt}, {van Dokkum}, {Abraham}  \&
  {Zhang}}{{Merritt} et~al.}{2016}]{Merritt+16}
{Merritt} A.,  {van Dokkum} P.,  {Abraham} R.,   {Zhang} J.,  2016, \mn@doi
  [\apj] {10.3847/0004-637X/830/2/62}, \href
  {http://adsabs.harvard.edu/abs/2016ApJ...830...62M} {830, 62}

\bibitem[\protect\citeauthoryear{{Michard}}{{Michard}}{2002}]{Michard02}
{Michard} R.,  2002, \mn@doi [\aap] {10.1051/0004-6361:20011813}, \href
  {http://adsabs.harvard.edu/abs/2002A%26A...384..763M} {384, 763}

\bibitem[\protect\citeauthoryear{{Moore}, {Lake}, {Quinn}  \& {Stadel}}{{Moore}
  et~al.}{1999}]{Moore+99}
{Moore} B.,  {Lake} G.,  {Quinn} T.,   {Stadel} J.,  1999, \mn@doi [\mnras]
  {10.1046/j.1365-8711.1999.02345.x}, \href
  {http://adsabs.harvard.edu/abs/1999MNRAS.304..465M} {304, 465}

\bibitem[\protect\citeauthoryear{{Mu{\~n}oz} et~al.,}{{Mu{\~n}oz}
  et~al.}{2015}]{Munoz+15}
{Mu{\~n}oz} R.~P.,  et~al., 2015, \mn@doi [\apjl]
  {10.1088/2041-8205/813/1/L15}, \href
  {http://adsabs.harvard.edu/abs/2015ApJ...813L..15M} {813, L15}

\bibitem[\protect\citeauthoryear{{Navarro}, {Frenk}  \& {White}}{{Navarro}
  et~al.}{1996}]{Navarro+96}
{Navarro} J.~F.,  {Frenk} C.~S.,   {White} S.~D.~M.,  1996, \mn@doi [\apj]
  {10.1086/177173}, \href {http://adsabs.harvard.edu/abs/1996ApJ...462..563N}
  {462, 563}

\bibitem[\protect\citeauthoryear{{Oh}, {de Blok}, {Walter}, {Brinks}  \&
  {Kennicutt}}{{Oh} et~al.}{2008}]{Oh+08}
{Oh} S.-H.,  {de Blok} W.~J.~G.,  {Walter} F.,  {Brinks} E.,   {Kennicutt} Jr.
  R.~C.,  2008, \mn@doi [\aj] {10.1088/0004-6256/136/6/2761}, \href
  {http://adsabs.harvard.edu/abs/2008AJ....136.2761O} {136, 2761}

\bibitem[\protect\citeauthoryear{{Peacock}, {Strader}, {Romanowsky}  \&
  {Brodie}}{{Peacock} et~al.}{2015}]{Peacock+15}
{Peacock} M.~B.,  {Strader} J.,  {Romanowsky} A.~J.,   {Brodie} J.~P.,  2015,
  \mn@doi [\apj] {10.1088/0004-637X/800/1/13}, \href
  {http://adsabs.harvard.edu/abs/2015ApJ...800...13P} {800, 13}

\bibitem[\protect\citeauthoryear{{Pohlen} \& {Trujillo}}{{Pohlen} \&
  {Trujillo}}{2006}]{Pohlen+Trujillo06}
{Pohlen} M.,  {Trujillo} I.,  2006, \mn@doi [\aap]
  {10.1051/0004-6361:20064883}, \href
  {http://adsabs.harvard.edu/abs/2006A%26A...454..759P} {454, 759}

\bibitem[\protect\citeauthoryear{{Purcell}, {Bullock}  \& {Zentner}}{{Purcell}
  et~al.}{2007}]{Purcell+07}
{Purcell} C.~W.,  {Bullock} J.~S.,   {Zentner} A.~R.,  2007, \mn@doi [\apj]
  {10.1086/519787}, \href {http://adsabs.harvard.edu/abs/2007ApJ...666...20P}
  {666, 20}

\bibitem[\protect\citeauthoryear{{Purcell}, {Bullock}, {Tollerud}, {Rocha}  \&
  {Chakrabarti}}{{Purcell} et~al.}{2011}]{Purcell+11}
{Purcell} C.~W.,  {Bullock} J.~S.,  {Tollerud} E.~J.,  {Rocha} M.,
  {Chakrabarti} S.,  2011, \mn@doi [\nat] {10.1038/nature10417}, \href
  {http://adsabs.harvard.edu/abs/2011Natur.477..301P} {477, 301}

\bibitem[\protect\citeauthoryear{{Reach} et~al.,}{{Reach}
  et~al.}{2005}]{Reach+05}
{Reach} W.~T.,  et~al., 2005, \mn@doi [\pasp] {10.1086/432670}, \href
  {http://adsabs.harvard.edu/abs/2005PASP..117..978R} {117, 978}

\bibitem[\protect\citeauthoryear{{Regan} et~al.,}{{Regan}
  et~al.}{2006}]{Regan+06}
{Regan} M.~W.,  et~al., 2006, \mn@doi [\apj] {10.1086/505382}, \href
  {http://adsabs.harvard.edu/abs/2006ApJ...652.1112R} {652, 1112}

\bibitem[\protect\citeauthoryear{{Rhode}, {Zepf}, {Kundu}  \& {Larner}}{{Rhode}
  et~al.}{2007}]{Rhode+07}
{Rhode} K.~L.,  {Zepf} S.~E.,  {Kundu} A.,   {Larner} A.~N.,  2007, \mn@doi
  [\aj] {10.1086/521397}, \href
  {http://adsabs.harvard.edu/abs/2007AJ....134.1403R} {134, 1403}

\bibitem[\protect\citeauthoryear{{Rodriguez-Gomez} et~al.,}{{Rodriguez-Gomez}
  et~al.}{2015}]{Rodriguez-Gomez+15}
{Rodriguez-Gomez} V.,  et~al., 2015, \mn@doi [\mnras] {10.1093/mnras/stv264},
  \href {http://adsabs.harvard.edu/abs/2015MNRAS.449...49R} {449, 49}

\bibitem[\protect\citeauthoryear{{Rodriguez-Gomez} et~al.,}{{Rodriguez-Gomez}
  et~al.}{2016}]{Rodriguez-Gomez+16}
{Rodriguez-Gomez} V.,  et~al., 2016, \mn@doi [\mnras] {10.1093/mnras/stw456},
  \href {http://adsabs.harvard.edu/abs/2016MNRAS.458.2371R} {458, 2371}

\bibitem[\protect\citeauthoryear{{Sandin}}{{Sandin}}{2014}]{Sandin14}
{Sandin} C.,  2014, \mn@doi [\aap] {10.1051/0004-6361/201423429}, \href
  {http://adsabs.harvard.edu/abs/2014A%26A...567A..97S} {567, A97}

\bibitem[\protect\citeauthoryear{{Sandin}}{{Sandin}}{2015}]{Sandin15}
{Sandin} C.,  2015, \mn@doi [\aap] {10.1051/0004-6361/201425168}, \href
  {http://adsabs.harvard.edu/abs/2015A%26A...577A.106S} {577, A106}

\bibitem[\protect\citeauthoryear{{S{\'e}rsic}}{{S{\'e}rsic}}{1968}]{Sersic68}
{S{\'e}rsic} J.~L.,  1968, {Atlas de galaxias australes}

\bibitem[\protect\citeauthoryear{{Sheth} et~al.,}{{Sheth}
  et~al.}{2010}]{Sheth+10}
{Sheth} K.,  et~al., 2010, \mn@doi [\pasp] {10.1086/657638}, \href
  {http://adsabs.harvard.edu/abs/2010PASP..122.1397S} {122, 1397}

\bibitem[\protect\citeauthoryear{{Smercina}, {Bell}, {Slater}, {Price},
  {Bailin}  \& {Monachesi}}{{Smercina} et~al.}{2017}]{Smercina+17}
{Smercina} A.,  {Bell} E.~F.,  {Slater} C.~T.,  {Price} P.~A.,  {Bailin} J.,
  {Monachesi} A.,  2017, \mn@doi [\apjl] {10.3847/2041-8213/aa78fa}, \href
  {http://adsabs.harvard.edu/abs/2017ApJ...843L...6S} {843, L6}

\bibitem[\protect\citeauthoryear{{Staudaher}, {Dale}, {van Zee}, {Barnes}  \&
  {Cook}}{{Staudaher} et~al.}{2015}]{Staudaher+15}
{Staudaher} S.~M.,  {Dale} D.~A.,  {van Zee} L.,  {Barnes} K.~L.,   {Cook}
  D.~O.,  2015, \mn@doi [\mnras] {10.1093/mnras/stv2064}, \href
  {http://adsabs.harvard.edu/abs/2015MNRAS.454.3613S} {454, 3613}

\bibitem[\protect\citeauthoryear{{Trujillo} \& {Fliri}}{{Trujillo} \&
  {Fliri}}{2016}]{Trujillo+16}
{Trujillo} I.,  {Fliri} J.,  2016, \mn@doi [\apj]
  {10.3847/0004-637X/823/2/123}, \href
  {http://adsabs.harvard.edu/abs/2016ApJ...823..123T} {823, 123}

\bibitem[\protect\citeauthoryear{{Tully}, {Rizzi}, {Shaya}, {Courtois},
  {Makarov}  \& {Jacobs}}{{Tully} et~al.}{2009}]{Tully+09}
{Tully} R.~B.,  {Rizzi} L.,  {Shaya} E.~J.,  {Courtois} H.~M.,  {Makarov}
  D.~I.,   {Jacobs} B.~A.,  2009, \mn@doi [\aj] {10.1088/0004-6256/138/2/323},
  \href {http://adsabs.harvard.edu/abs/2009AJ....138..323T} {138, 323}

\bibitem[\protect\citeauthoryear{{Wetzel}, {Hopkins}, {Kim},
  {Faucher-Gigu{\`e}re}, {Kere{\v s}}  \& {Quataert}}{{Wetzel}
  et~al.}{2016}]{Wetzel+16}
{Wetzel} A.~R.,  {Hopkins} P.~F.,  {Kim} J.-h.,  {Faucher-Gigu{\`e}re} C.-A.,
  {Kere{\v s}} D.,   {Quataert} E.,  2016, \mn@doi [\apjl]
  {10.3847/2041-8205/827/2/L23}, \href
  {http://adsabs.harvard.edu/abs/2016ApJ...827L..23W} {827, L23}

\bibitem[\protect\citeauthoryear{{Zaritsky}, {Smith}, {Frenk}  \&
  {White}}{{Zaritsky} et~al.}{1993}]{Zaritsky+93}
{Zaritsky} D.,  {Smith} R.,  {Frenk} C.,   {White} S.~D.~M.,  1993, \mn@doi
  [\apj] {10.1086/172379}, \href
  {http://adsabs.harvard.edu/abs/1993ApJ...405..464Z} {405, 464}

\bibitem[\protect\citeauthoryear{{Zucker} et~al.,}{{Zucker}
  et~al.}{2006}]{Zucker+06}
{Zucker} D.~B.,  et~al., 2006, \mn@doi [\apjl] {10.1086/505216}, \href
  {http://adsabs.harvard.edu/abs/2006ApJ...643L.103Z} {643, L103}

\bibitem[\protect\citeauthoryear{{Zucker} et~al.,}{{Zucker}
  et~al.}{2007}]{Zucker+07}
{Zucker} D.~B.,  et~al., 2007, \mn@doi [\apjl] {10.1086/516748}, \href
  {http://adsabs.harvard.edu/abs/2007ApJ...659L..21Z} {659, L21}

\bibitem[\protect\citeauthoryear{{de Vaucouleurs}, {de Vaucouleurs}, {Corwin},
  {Buta}, {Paturel}  \& {Fouqu{\'e}}}{{de Vaucouleurs}
  et~al.}{1991}]{DeVaucouleurs+91}
{de Vaucouleurs} G.,  {de Vaucouleurs} A.,  {Corwin} Jr. H.~G.,  {Buta} R.~J.,
  {Paturel} G.,   {Fouqu{\'e}} P.,  1991, {Third Reference Catalogue of Bright
  Galaxies. Volume I: Explanations and references. Volume II: Data for galaxies
  between 0$^{h}$ and 12$^{h}$. Volume III: Data for galaxies between 12$^{h}$
  and 24$^{h}$.}

\bibitem[\protect\citeauthoryear{{van Dokkum}, {Abraham}  \& {Merritt}}{{van
  Dokkum} et~al.}{2014}]{vanDokkum+14}
{van Dokkum} P.~G.,  {Abraham} R.,   {Merritt} A.,  2014, \mn@doi [\apjl]
  {10.1088/2041-8205/782/2/L24}, \href
  {http://adsabs.harvard.edu/abs/2014ApJ...782L..24V} {782, L24}

\makeatother
\end{thebibliography}

\bsp    % typesetting comment
\label{lastpage}
\end{document}